\documentclass[11pt,a4paper]{article}
\pdfoutput=1

\usepackage{jheppub}
\usepackage{mathtools}
\usepackage{amsthm}
\usepackage{marvosym}
\usepackage{tikz-cd}
\usepackage{hyperref}
\usepackage{empheq}
\usepackage[T1]{fontenc} 
\usepackage{soul}
\usepackage{dsfont}
\usepackage{graphicx}
\usepackage{mwe}
\usepackage{amsmath,amssymb,amsfonts}
\usepackage[mathscr]{eucal}

\DeclareMathOperator{\tr}{tr} % trace
\DeclareMathOperator{\Spin}{Spin} % Spin group
\DeclareMathOperator{\GL}{GL} % GL group
\DeclareMathOperator{\SL}{SL} % SL group
\DeclareMathOperator{\PSL}{PSL} % SL group
\DeclareMathOperator{\Hom}{Hom} % Hom functor
\newcommand\QGaussSum[2]{{\chi_{{#2}}({#1})}} % Quadratic Gauss Sum
\newcommand\DualQGaussSum[2]{{\varrho_{{#2}}({#1})}} % Quadratic Gauss Sum

\newcommand*\widefbox[1]{\fbox{\hspace{2em}#1\hspace{2em}}}

\DeclareMathOperator{\sgn}{sgn} % sign
\def\sig{{\sigma}}

\title{
Double-Janus Linear Sigma Models
and Generalized Reciprocity for Gauss Sums}

\author{Ori~J.~Ganor,}
\author{Hao-Yu~Sun,}
\author{Nesty~R.~Torres-Chicon}

\affiliation{
Department of Physics,
  University of California,\\
Berkeley, CA 94720, U.S.A.}

\emailAdd{ganor@berkeley.edu}
\emailAdd{hkdavidsun@berkeley.edu}
\emailAdd{ntorres@berkeley.edu}

\abstract{
We study the supersymmetric partition function of a 2d linear $\sigma$-model whose target space is a torus with a complex structure that varies along one worldsheet direction and a K\"ahler modulus that varies along the other.
This setup is inspired by the dimensional reduction of a Janus configuration of 4d $\mathcal{N}=4$ $U(1)$ Super-Yang-Mills theory compactified on a mapping torus ($T^2$ fibered over $S^1$) times a circle with an $\SL(2,\mathbb{Z})$ duality wall inserted on $S^1$, but our setup has minimal supersymmetry.
The partition function depends on two independent elements of $\SL(2,\mathbb{Z})$, one describing the duality twist, and the other describing the geometry of the mapping torus.
It is topological and can be written as a multivariate quadratic Gauss sum. By calculating the partition function in two different ways, we obtain identities relating different quadratic Gauss sums, generalizing the {\it Landsberg-Schaar} relation. These identities are a subset of a collection of identities discovered by F. Deloup. Each identity contains a phase which is an eighth root of unity, and we show how it arises as a Berry phase in the supersymmetric Janus-like configuration. Supersymmetry requires the complex structure to vary along a semicircle in the upper half-plane, as shown by Gaiotto and Witten in a related context, and that semicircle plays an important role in reproducing the correct Berry phase.
}
\begin{document}
\maketitle
\flushbottom

% ==============================================================
% ==============================================================
\newcommand{\secref}[1]{\S\ref{#1}}
\newcommand{\figref}[1]{Figure~\ref{#1}}
\newcommand{\appref}[1]{Appendix~\ref{#1}}
\newcommand{\footref}[1]{Footnote~\ref{#1}}

\newcommand\scalemath[2]{\scalebox{#1}{\mbox{\ensuremath{\displaystyle #2}}}}

\newcommand{\apprefrange}[2]{Appendices~\ref{#1}-\ref{#2}}
\newcommand{\tabref}[1]{Table~\ref{#1}}

\newcommand\rep[1]{{\bf {#1}}} % representation

\newcommand\SUSY[1]{{${\mathcal{N}}={#1}$}}  % supersymmetry
\newcommand\px[1]{{\partial_{#1}}}

\def\be{\begin{equation}}
\def\ee{\end{equation}}
\def\bear{\begin{eqnarray}}
\def\eear{\end{eqnarray}}
\def\nn{\nonumber}

\newcommand\bra[1]{{\left\langle{#1}\right\rvert}} % bra
\newcommand\ket[1]{{\left\lvert{#1}\right\rangle}} % ket

\newcommand{\C}{\mathbb{C}}
\newcommand{\R}{\mathbb{R}}
\newcommand{\Z}{\mathbb{Z}}
\newcommand{\CP}{\mathbb{CP}}

\newcommand\putRect[4]{
  \multiput(#1,#2)(0,#4){2}{\line(1,0){#3}}
  \multiput(#1,#2)(#3,0){2}{\line(0,1){#4}}
}

\def\Id{{\mathbb{I}}} % Identity matrix

\def\SO{{SO}}
\def\SU{{SU}}

\def\wfF{{\varphi}} % wavefunction 
\def\ABerry{{{\mathcal A}}} % Berry connection
\def\btau{{\overline{\tau}}}

\def\GWa{{a}} % Gaiotto-Witten variable a
\def\GWD{{D}} % Gaiotto-Witten variable D
\def\GWe{{\mathbf{e}}} % Gaiotto-Witten variable e
\def\GWpsi{{\psi}} % Gaiotto-Witten variable psi

\def\tGWpsi{{\tilde{\GWpsi}}} % \SL(2,Z) dual of $\GWpsi$

\def\Mst{M_{\text{st}}} % string scale
\def\gst{g_{\text{st}}} % string coupling constant
\def\gIIB{g_{\text{IIB}}} % IIB string coupling constant
\def\lst{\ell_{\text{st}}} % string length.
\def\lP{\ell_{\text{P}}} % Planck length.

\def\xa{{\mathbf{a}}} % \SL(2,Z) variable
\def\xb{{\mathbf{b}}} % \SL(2,Z) variable
\def\xc{{\mathbf{c}}} % \SL(2,Z) variable
\def\xd{{\mathbf{d}}} % \SL(2,Z) variable
\def\matMx{{M}} % \SL(2,Z) matrix acting on tau
\def\evMx{{\varsigma}} % eigenvalue of the above matrix

\def\Mf{{\mathbf{M}}} % 3D Melvin manifold
\def\lvk{{\mathbf{k}_{\tiny\rm cs}}} % the 'level' k
\def\xR{{R}}

\def\gYM{g_{\text{ym}}}

\def\ten{{\natural}} % 10 in Dirac matrices

\def\elDirac{{\gamma}} % eleven dimensional Dirac matrices
\def\tenDirac{{\Gamma}} % ten dimensional Dirac matrices

\def\xJJ{{\mathbf{y}}} % coordinate along Janus circle
\def\xB{{\mathbf{x}}} % coordinate along Base of mapping torus

\def\tauYM{{\tau}} % Yang-Mills coupling
\def\vCS{{\tau}} % complex structure of sigma-model target space
\def\vKS{{\tilde{\tau}}} % K\"ahler structure of sigma-model target space

\def\fArea{{\rho}} % area of $T^2$

\def\sTF{{\mathcal{M}_{\text{TF}}}} % torus fibration

\def\sCF{{\mathcal{M}_{\text{CF}}}} % circle fibration

\def\sMT{{\mathcal{M}}} % mapping torus

\def\gJ{{\mathfrak{g}}} % \SL(2,Z) element
\def\xT{{\xi}} % coordinate on fiber

\def\gB{{\tilde{\gJ}}} % \SL(2,Z) element of second mapping torus
\def\xS{{\eta}} % coordinate on fiber of second mapping torus

\def\ya{{\tilde{\xa}}} % \SL(2,Z) variable
\def\yb{{\tilde{\xb}}} % \SL(2,Z) variable
\def\yc{{\tilde{\xc}}} % \SL(2,Z) variable
\def\yd{{\tilde{\xd}}} % \SL(2,Z) variable
\def\matMy{{\widetilde{\matMx}}} % \SL(2,Z) matrix acting on rho
\def\evMy{{\widetilde{\varsigma}}} % eigenvalue of the above matrix

\def\ua{{\tilde{\alpha}}} % \SL(2,R) variable
\def\ub{{\tilde{\beta}}} % \SL(2,R) variable
\def\uc{{\tilde{\gamma}}} % \SL(2,R) variable
\def\ud{{\tilde{\delta}}} % \SL(2,R) variable
\def\matMu{{\widetilde{\mathfrak{M}}}} % \SL(2,R) matrix acting on rho

\def\PF{{\mathcal {Z}}} % partition function
\def\FPF{{\Delta}} % $1$-loop determinants

% SIGMA MODEL FIELDS
\def\bpartial{{\overline{\partial}}}
\def\fPsi{{\Psi}} % Sigma model fermionic field (L-moving)
\def\bfPsi{{\overline{\fPsi}}} % Sigma model fermionic field (R-moving)
\def\fX{{\mathbf{X}}} % Sigma model bosonic field
\def\cG{{G}} % Sigma model target space metric
\def\cB{{B}} % Sigma model target space 2-form
\def\cE{{E}} % Sigma model G+B
\def\cK{{\mathcal{K}}} % coupling between fermions (L-moving)
\def\bcK{{\overline{\cK}}} % coupling between fermions (R-moving)
\def\cW{{\mathcal{W}}} % coupling between fermions (L-R)
\def\bcW{{\overline{\cW}}} % coupling between fermions (L-R)

\def\fZ{{\mathbf{Z}}} % component fields of $\vZ$

% TARGET SPACE INDICES
%\def\xI{{\mathcal{I}}}
%\def\xJ{{\mathcal{J}}}
%\def\xK{{\mathcal{K}}}
%\def\xL{{\mathcal{L}}}
%\def\xM{{\mathcal{M}}}
%\def\xP{{\mathcal{P}}}

\def\xI{{{I}}}
\def\xJ{{{J}}}
\def\xK{{{K}}}
\def\xL{{{L}}}
\def\xM{{{M}}}
\def\xP{{{P}}}

\def\csT{{\widehat{\mathcal{T}}}}
\def\csS{{\widehat{\mathcal{S}}}}
\def\slR{{\bf{R}}}
\def\csR{{\hat{\slR}}}

\def\tr{{{\mbox{Tr}}}}

\def\slM{{\bf{M}}} % \SL(2,Z) matrix

\def\slT{{\bf{T}}}
\def\slS{{\bf{S}}}

\def\gcd{{\mbox{gcd}}}

\def\csM{{\hat{\slM}}} % operator on Hilbert space

\def\nV{{N}}
\def\rVar{{r}}

\def\nEq{{J}}
\def\yVar{{u}}

\def\Aco{{A}}
\def\Cco{{C}}
\def\Dco{{D}}

\def\xTorus{{x}}
\def\yTorus{{y}}

\def\XEq{{\mathbf X}} % polynomial equation
\def\zWind{{\xi}} % winding direction of target space

\def\GWpsi{{\psi}} % Gaiotto-Witten variable psi

\def\xf{{\bf{\xi}}} % 
\def\wfF{{\Psi}}

\def\ABerry{{{\mathcal A}}} % Berry connection
\def\BerryPhase{{\gamma_b}}
\def\btau{{\overline{\tau}}}

% SUSY TRANSFORMATION MATRICES
\newcommand\ASUSY[2]{{{{\mathcal{A}}^{#1}}_{#2}}} % parameters in SUSY transf.
\newcommand\bASUSY[2]{{{{\overline{\mathcal{A}}}^{#1}}_{#2}}} % parameters in SUSY transf.
\newcommand\BSUSY[2]{{{{\mathcal{C}}^{#1}}_{#2}}} % parameters in SUSY transf.
\newcommand\bBSUSY[2]{{{{\overline{\mathcal{C}}}^{#1}}_{#2}}} % parameters in SUSY transf.

\def\tauCS{{\tau}} % complex structure
\def\rhoKM{{\rho}} % Kahler modulus
\def\btauCS{{\overline{\tauCS}}}
\def\brhoKM{{\overline{\rhoKM}}}

\def\dtauCS{\tau'} % derivative of complex structure w.r.t. $\xWS^1$
\def\drhoKM{\dot{\rho}} % derivative of Kahler modulus w.r.t. $\xWS^2$

\def\ddtauCS{\tau''} % derivative of complex structure w.r.t. $\xWS^1$
\def\ddrhoKM{\ddot{\rho}} % derivative of Kahler modulus w.r.t. $\xWS^2$

\def\VtauCS{{\breve{\tau}_2}} % complex structure
\def\VrhoKM{{\breve{\rho}_2}} % Kahler modulus

\def\dVtauCS{{\breve{\tau}_2'}} % derivative of complex structure w.r.t. $\xWS^1$
\def\dVrhoKM{{\dot{\breve{\rho}}_2}} % derivative of Kahler modulus w.r.t. $\xWS^2$

\def\matB{{\cB}}
\def\matG{{\cG}}
\def\matE{{\cE}}
\def\matASUSY{{\mathcal{A}}}
\def\bmatASUSY{{\overline{\matASUSY}}}
\def\matK{{\cK}}
\def\bmatK{{\bcK}}
\def\matW{{\cW}}

% WORLDSHEET COORDINATES
\def\xWS{{\sigma}} % Worldsheet real coordinate
\def\zWS{{z}} % Worldsheet complex coordinate
\def\bzWS{{\overline{\zWS}}} % c.c. of Worldsheet complex coordinate

\def\dimTS{{N}} % target space dimensions

\def\delSUSY{{\widetilde{\delta}}} % new SUSY transformation
\def\pSUSY{{\eta}} % anti-commuting SUSY parameter

% FERMION METRIC
\def\matSV{{\Sigma}}
\def\bmatSV{{\overline{\Sigma}}}

% VIELBEIN FORMULATION
\def\matF{{V}} % vielbein
\def\matOf{{\Omega}} % gauge transformation
\def\matX{{\mathcal A}}
\def\bmatX{{\overline{\matX}}}

\def\matY{{\mathcal Y}}
\def\bmatY{{\overline{\matY}}}

\def\matU{{U}}
\def\bmatU{{\overline{\matU}}}

\def\matS{{S}}
\def\bmatS{{\overline{\matS}}}

% TRANSFORMATIONS
\def\matP{{P}} % \SL(2,R) matrix

\def\xRa{{\alpha}} % real \SL(2,R) element
\def\xRb{{\beta}} % real \SL(2,R) element
\def\xRc{{\gamma}} % real \SL(2,R) element
\def\xRd{{\delta}} % real \SL(2,R) element

\def\matGn{{\matG^{(new)}}}
\def\tauCSn{\tau^{(new)}}
\def\matWn{{\matW^{(new)}}}
\def\matASUSYn{{\matASUSY^{(new)}}}

% Vector variables for $T^2$ target space

\def\vX{{X}} % 2d vector of X-coordinates
\def\vPsi{{\Psi}} % 2d vector of Psi-coordinates
\def\bvPsi{{\overline{\Psi}}} % 2d vector of bPsi-coordinates
\def\vN{{N}} % 2d vector of integers signifying the winding
\def\vZ{{Z}} % oscillator part of $\vX$

\def\vY{{Y}} % 2d vector of X-coordinates at sigma=0 or sigma=1
\def\vK{{K}} % 2d vector of integers signifying a shift in $\vX$
\def\vKF{{\mathbf{K}}} % 2d vector of integers signifying a shift in $\vx$

\def\GrSt{{\Xi_0}}

\def\genT{{\overline{T}}} % generator of \SL(2,Z) (generic)
\def\genS{{\overline{S}}} % generator of \SL(2,Z)  (generic)
\def\tpow{{\overline{k}}} % power of T (generic)
\def\tpowNum{{\overline{r}}} % number of power of T (generic)

\def\matEps{{\varepsilon}}

\def\Ir{{\mathfrak{g}}} % segment counter
\def\Nr{{\mathfrak{n}}} % segment counter
\def\matMpow{{\nu}}

\def\matR{{{R}}}
\def\bmatR{{\overline{\matR}}}
\def\matL{{{L}}}
\def\bmatL{{\overline{\matL}}}

\def\matH{{{H}}}
\def\bmatH{{\overline{\matH}}}

\def\matRi{{{J}}}
\def\bmatRi{{\overline{\matRi}}}
\def\matLi{{{Q}}}

\def\matRz{{H}}
\def\bmatRz{{\overline{\matRz}}}
\def\matLz{{{O}}}

\def\bXi{{\overline{\Xi}}}

\def\hA{{\mathbf{A}}} % holonomies
\def\hB{{\mathbf{B}}} % holonomies

\def\matEABCD{{\mathfrak{Y}}}
\def\matGAGA{{\mathfrak{H}}}

\def\modeX{{\mathfrak{X}}}
\def\modeP{{\mathfrak{P}}}
\def\cD{{\mathcal{D}}} % path-integral measure

\def\varf{{\mathfrak{f}}}

\def\evxg{{\breve{\lambda}}}
\def\evyg{{\breve{\mu}}}

\def\xal{{\alpha}} % \SL(2,R) variable
\def\xbl{{\beta}} % \SL(2,R) variable
\def\xcl{{\gamma}} % \SL(2,R) variable
\def\xdl{{\delta}} % \SL(2,R) variable
\def\yal{{\tilde{\xal}}} % \SL(2,R) variable
\def\ybl{{\tilde{\xbl}}} % \SL(2,R) variable
\def\ycl{{\tilde{\xcl}}} % \SL(2,R) variable
\def\ydl{{\tilde{\xdl}}} % \SL(2,R) variable

\def\xPg{{\Omega}}
\def\yPg{{\tilde{\xPg}}}

\def\NPsi{{\Upsilon}} % new fermionic variable
\def\bNPsi{{\overline{\NPsi}}} % new fermionic variable

\def\UPsi{{\Theta}} % new fermionic variable
\def\bUPsi{{\overline{\UPsi}}} % new fermionic variable

\def\muF{{\mu}}

\def\UMode{{\mathbf{B}}} % fermionic mode
\def\bUMode{{\overline{\UMode}}} % fermionic mode

\def\dUMode{{\dot{\UMode}}} % time derivative of fermionic mode
\def\dbUMode{{\dot{\bUMode}}} % time derivative of fermionic mode

%trace calculation
\def\be{\begin{equation}}
\def\ee{\end{equation}}
\def\bear{\begin{eqnarray}}
\def\eear{\end{eqnarray}}
\def\nn{\nonumber}

\newcommand{\Q}{\mathbb{Q}}

\def\Id{{\mathbb{I}}} % Identity matrix

\def\xa{{\mathbf{a}}} % \SL(2,Z) variable
\def\xb{{\mathbf{b}}} % \SL(2,Z) variable
\def\xc{{\mathbf{c}}} % \SL(2,Z) variable
\def\xd{{\mathbf{d}}} % \SL(2,Z) variable
\def\matMx{{M}} % \SL(2,Z) matrix acting on tau

\def\Mf{{\mathbf{M}}} % 3D Melvin manifold
\def\xR{{R}}

\def\gYM{g_{\text{ym}}}

\def\ya{{\tilde{\xa}}} % \SL(2,Z) variable
\def\yb{{\tilde{\xb}}} % \SL(2,Z) variable
\def\yc{{\tilde{\xc}}} % \SL(2,Z) variable
\def\yd{{\tilde{\xd}}} % \SL(2,Z) variable
\def\matMy{{\widetilde{\matMx}}} % \SL(2,Z) matrix acting on rho

\def\xg{{\matMx}}
\def\yg{{\matMy}}

\def\vN{{\mathcal{N}}} % 2d vector of integers signifying the winding

\def\genTx{{T}} % generator of \SL(2,Z) acting on $\tauCS$
\def\genSx{{S}} % generator of \SL(2,Z) acting on $\tauCS$
\def\tpowx{{l}} % power of T in $\matMx$
\def\tpowNumx{{s}} % number of power of T in $\matMx$
\def\thalfpowx{{v}} % half the power of T in $\matMx$

\def\genTy{{\widetilde{T}}} % generator of \SL(2,Z) acting on $\rhoKM$
\def\genSy{{\widetilde{S}}} % generator of \SL(2,Z) acting on $\rhoKM$
\def\tpowy{{\mathbf{k}}} % power of T in $\matMy$
\def\tpowNumy{{r}} % number of power of T in $\matMy$
\def\thalfpowy{{u}} % half the power of T in $\matMx$

\def\matKcpl{{K}} % coupling constant matrix
\def\matKcplx{{K}} % coupling constant matrix related to $\matMx$
\def\matKcply{{\widetilde{K}}} % coupling constant matrix related to $\matMy$
\def\iKy{{\mathbf{i}}} % counter for $\matKcply$ elements
\def\jKy{{\mathbf{j}}} % counter for $\matKcply$ elements

\def\smithP{{P}}
\def\smithQ{{Q}}

\def\xV{{v}}
\def\xU{{u}}

\def\yW{{\widetilde{w}}}
\def\yV{{\widetilde{v}}}
\def\yU{{\widetilde{u}}}

\def\latZv{{\Lambda}}
\def\latWv{{\Lambda'}}

\def\latAG{{\widetilde{\Lambda}}}

\def\tvphi{{\widetilde{\varphi}}} % map between lattices

\def\matJ{{\mathcal J}}

\def\xB{{\theta}} % coordinate on $S^1$
\def\vxF{{\mathbf{x}}} % coordinate on $T^2$

\def\matEps{{\epsilon}}

\newcommand\Hilb[1]{{\mathcal{H}_{({#1})}}} % Hilbert subspace

\def\Reqv{{\mathcal{R}}} % equivalence relation
\newcommand\trHp[2]{{\tr^{({#2})}_{{#1}}}}
\def\LSterm{{\mathcal{X}}}

\def\opO{{\hat{\mathcal{O}}}} % some operator

\def\matYS{{\mathfrak Y}} % change of fermionic variable matrix
\def\bmatYS{{\overline{\mathfrak Y}}} % change of fermionic variable matrix
\def\vPsiN{{\widehat{\vPsi}}} % change of fermionic variables to normalized kinetic term
\def\bvPsiN{{\widehat{\bvPsi}}} % change of fermionic variables to normalized kinetic term

\def\tI{{\tilde{I}}} % discretized action

\def\vZNN{{\widetilde{\vZ}}}
\def\fZNN{{\widetilde{\fZ}}}
\def\vPsiNN{{\widetilde{\vPsi}}} % change of fermionic variables to normalized kinetic term
\def\bvPsiNN{{\widetilde{\bvPsi}}} % change of fermionic variables to normalized kinetic term
\newcommand\fPsiNN[1]{{\widetilde{\fPsi^{#1}}}} % change of fermionic variables to normalized kinetic term
\newcommand\bfPsiNN[1]{{\widetilde{\bfPsi^{#1}}}} % change of fermionic variables to normalized kinetic

\def\vPsiNNN{{\widetilde{\widetilde{\vPsi}}}} % change of fermionic variables to normalized kinetic term
\def\bvPsiNNN{{\widetilde{\widetilde{\bvPsi}}}} % change of fermionic variables to normalized kinetic term

\def\matMMb{{\mathfrak{M}}} % bosonic mass matrix (to be squared)
\def\evMMb{{\mathfrak{m}}} % bosonic mass matrix eigenvalue (to be squared)

\def\VSchKM{{\mathbf{V}}} % 1d effective potential for 1d Schrodinger problem
\def\VSchCSa{{\mathbf{U}_1}} % 1d effective potential for 1d Schrodinger problem
\def\VSchCSb{{\mathbf{U}_2}} % 1d effective potential for 1d Schrodinger problem
\def\evSchKM{{\mu}} % eigenvalues of the Schrodinger problem in rhoKM direction (2)
\def\evSchCSa{{\varepsilon}} % eigenvalues of the Schrodinger problem in tauCS direction (1)
\def\evSchCSb{{\vartheta}} % eigenvalues of the Schrodinger problem in tauCS direction (1)
\def\xSch{{\mathbf{x}}} % Schrodiner variable
\def\effmass{{m}} % effective mass
\def\WSch{{\mathbf{W}}} % superpotential for Schrodinger problem
\def\QSch{{q}} % supercharge for Schrodinger problem

\def\Nx{{\mathbf{x}}} % component of $\vN$
\def\Ny{{\mathbf{y}}} % component of $\vN$
\def\Nz{{\mathbf{z}}} % $\Nx+\Ny$

\def\Kx{{\mathbf{m}}} % component of $\vK$
\def\Ky{{\mathbf{n}}} % component of $\vK$

\def\phaseT{{\varphi}} % phase.
\def\vNCS{{\overline{\mathcal{N}}}} % vector of integers for CS

\def\vXR{{\xi}}
\def\levely{{{q}}} % level of geometric quantization
\def\massL{{m}} % mass for Landau problem
\def\tauLL{{\tau}} % tau for Landau problem discussion
\def\btauLL{{\overline{\tau}}} % tau for Landau problem discussion
\def\tauGW{{\tau}} % tau for Gaiotto-Witten discussion

\def\matSf{{F}} % for holomorphic solutions section
\def\bmatSf{{\overline{F}}} % for holomorphic solutions section
\def\matCf{{\varkappa}}  % for holomorphic solutions section
\def\bmatCf{{\overline{\varkappa}}} % for holomorphic solutions section
\def\matSfX{{\varpi}} % for holomorphic solutions section
\def\bmatSfX{{\overline{\varpi}}} % for holomorphic solutions section

\def\Pauli{{\boldsymbol{\sigma}}} % Pauli matrix

\def\nS{{\mathbf{n}}} % counter for Smith normal form lattice
\def\aS{{\mathbf{a}}} % counter for Smith normal form lattice

\def\matADeloup{{\mathbf{A}}} % symmetric matrix related to Deloup's generalization
\def\matBDeloup{{\mathbf{B}}} % symmetric matrix related to Deloup's generalization
\def\xDeloup{{\mathbf{x}}} % vector related to Deloup's quadratic form
\def\WuClass{{\mathbf{w}}} % Wu class representative

\def\mcS{{{\mathcal S}}}
\def\mcT{{{\mathcal T}}}

\def\AG{{{\mathcal A}}} % abelian group
\def\qf{{\mathfrak{q}}} % quadratic form
\def\qBf{{\mathfrak{b}}} % bilinear form
\def\qfL{{q}} % quadratic form on lattice
\def\qBL{{\mathbf{B}}} % bilinear form on lattice

\def\mcW{{\mathbf{w}}}
\def\mcV{{\mathbf{v}}}
\def\mcU{{\mathbf{u}}}

\def\mcVRep{{\mathbf{V}}} % representative of v
\def\mcURep{{\mathbf{U}}} % representative of u
\def\mcWRep{{\mathbf{W}}} % representative of u

\def\isomMTCS{{\widetilde{\varphi}}} % isomorphism between MT and CS
\def\qfMT{{\widetilde{\qf}}} % quadratic form on $\GrSt$
\def\qBfMT{{\widetilde{\qBf}}} % bilinear form for MT

\def\phaseW{{\phi}} % phase in S for Weil Rep

% ==============================================================

% ==============================================================
\section{Introduction}
\label{sec:Intro}
In these notes, we study the torus partition function of a supersymmetric 2d linear $\sigma$-model with $T^2$ target space (a free theory) whose complex structure varies along one of the worldsheet directions (parametrized by $0\le\xWS_1<1$) and whose K\"ahler modulus varies along the other direction (parametrized by $0\le\xWS_2<1$). The periodic boundary conditions can be twisted both with an element in the mapping class group (MCG) of the target $T^2$ (we choose to do that twist along the $\xWS_1$ direction) and with a T-duality transformation (along $\xWS_2$). 
%We choose to twist with an MCG element along the $\xWS_1$-direction and with a T-duality along the $\xWS_2$-direction.
The partition function thus depends on (the conjugacy classes of) $\xg,\yg\in\SL(2,\mathbb{Z})$ that respectively describe the MCG ``geometrical'' element and the T-duality transformation. We represent the complex structure of the target space by $\tauCS$ (taking values in the upper half-plane $\mathbb{H}$), which is allowed to vary as a function of $\xWS_1$. More concretely, as we complete a loop around the first cycle of the worldsheet (by varying $\xWS_1$ from $0$ to $1$), the variable $\tauCS$ may undergo a $\PSL(2,\Z)$ transformation 
$
\tauCS\rightarrow \left(\xa\tauCS+\xb\right)/\left(\xc\tauCS+\xd\right),
$
and in order to prevent a discontinuity at $\xWS_1=0$ we need to impose boundary conditions (connecting the fields at $\xWS_1=0$ to the fields at $\xWS_1=1$) that involve an element of the MCG of the target space, encoded by 
$$\xg=\begin{pmatrix}\xa & \xb \\ \xc & \xd \\ \end{pmatrix}\in \SL(2,\Z).$$
Similarly, the K\"ahler structure of the target space is represented by $\rhoKM=\rhoKM_1+i\rhoKM_2$ on the upper half-plane (with $\rhoKM_2$ proportional to the area of the $T^2$ target and $\rhoKM_1$ proportional to the Kalb-Ramond flux). As $\xWS_2$ varies from $0$ to $1$, the variable $\rho$ may undergo a $\PSL(2,\Z)$ transformation $\rho\rightarrow \left.\left(\ya\rho+\yb\right)\right/\left(\yc\rho+\yd\right)$, and in order to prevent a discontinuity at $\xWS_2=0$ we need to impose boundary conditions (connecting the fields at $\xWS_2=0$ to the fields at $\xWS_2=1$) that involve a ``T-duality wall'' labeled by 
$$\yg=\begin{pmatrix}\ya & \yb \\ \yc & \yd \\ \end{pmatrix}\in \SL(2,\Z).$$

The motivation for considering such a setup is that it arises as a limit of a sort of ``double-Janus'' configuration of 4d \SUSY{4} Super-Yang-Mills (SYM) theory. 
Janus configurations are 4d SYM theories with a coupling constant that varies along one direction of space. They were first introduced by Bak, Gutperle and Hirano \cite{Bak:2003jk} in a non-supersymmetric dilatonic deformation of AdS$_5$, an exact solution to the Type IIB SUGRA equations, as a way to create a discontinuous jump in the {\it real} Yang-Mills coupling constant (which is related to the asymptotic boundary value of the dilaton according to the standard AdS/CFT dictionary) across a codimension-1 interface.
A supersymmetric Janus configuration was introduced in \cite{Clark:2005te}, and later, the jump was ``smoothed out'' in \cite{D'Hoker:2006uv}, but still without a $\theta$-angle. Subsequently, Gaiotto and Witten presented \cite{Gaiotto:2008sd} the action of a deformation of \SUSY{4} SYM that preserves half of the supersymmetries with a {\it complex} coupling constant $\tau=\frac{4\pi i}{\gYM^2}+\frac{\theta}{2\pi}$ that varies as a function of one spatial direction, say $\tau(x_3)$. Janus configurations have been further explored in \cite{Kim:2008dj,Kim:2009wv} and have been introduced into sphere partition functions in \cite{Terashima:2011qi,Drukker:2010jp}. They can also be constructed for theories in other dimensions \cite{Minahan:2018kme}. Configurations for 2d $\sigma$-models have been proposed earlier in \cite{Gaiotto:2009fs} and sphere partition functions with Janus configurations have been calculated in \cite{Drukker:2010jp,Goto:2018bci}.

Taking the gauge group to be $U(n)$, the Gaiotto-Witten action allows us to smoothly introduce an $\SL(2,\Z)$ duality twist (sometimes referred to as a ``duality wall'' or ``S-fold'' and studied in various string theory and gauge theory contexts, for example, in \cite{Dabholkar:1998kv,Ganor:1998ze,Hellerman:2002ax,Dabholkar:2002sy,Dabholkar:2005ve,Gadde:2014wma}) in an $S^1$ compactification of 4d SYM. Parameterizing $S^1$ by $0\le x_3<1$, the duality twist is an unconventional boundary condition that sets 
\begin{equation}
\label{eq:sdual}
\displaystyle{\tau(1)=\frac{\xa\tau(0)+\xb}{\xc\tau(0)+\xd}}\quad \text{for}\,\, \xg\in\SL(2,\Z),
\end{equation}
together with the implied electric-magnetic duality action on the fields. We will refer to this theory as a {\it closed Janus configuration.} This setup was studied in \cite{Ganor:2014pha} for $U(1)$ gauge group, where the low energy limit is easily shown to be described by a Chern-Simons action with an abelian gauge group, determined (not uniquely) by a decomposition of $\xg$ into $$T=\begin{pmatrix} 1 & 1 \\ 0 & 1 \\ \end{pmatrix}\quad \text{and}\quad S=\begin{pmatrix} 0 & -1 \\ 1 & 0 \\ \end{pmatrix}$$ 
generators. Furthermore, it was shown in \cite{Ganor:2014pha} that a T-dual string theory background provides a geometrical interpretation for the quantum algebra of Wilson loops. A similar setup was also studied in \cite{Assel:2018vtq}, from the perspective of the holographic dual. Duality walls have also been studied in \cite{Martucci:2014ema}, and recently a gravitational anomaly was discovered \cite{Hsieh:2019iba} in such compactifications.

The Gaiotto-Witten Janus configuration can be constructed as a limit of a compactification of the 6d $(2,0)$-theory on a $T^2$-fibration over $\R$ when the area of the $T^2$ fiber shrinks to zero. We will review this construction in \appref{app:JanusAndMT}. To construct the closed Janus configuration requires us to first take the limit where the area of the $T^2$ fiber shrinks to zero, since otherwise the area will be discontinuous along $S^1$ (as we will explain in \appref{app:JanusAndMT}), and so we are back to \SUSY{4} SYM with the S-fold twist $\xg$. In the context of 3d-3d correspondence \cite{Dimofte:2011ju}, this configuration is related to the 6d $(2,0)$ theory on an (auxiliary) 3d mapping torus studied in \cite{Dimofte:2011py,Gang:2015wya,Gang:2015bwa,Chun:2019mal}, where a higher-genus fiber is also considered.
The setup for the present paper is derived from an abelian double-Janus configuration that is a prelude to the study of a nonabelian theory. It can be obtained as a limit of a Gaiotto-Witten Janus configuration by compactifying on a small torus and allowing its complex structure parameter to vary as a function of time. In other words, we compactify the closed 4d Janus configuration on another (auxiliary) mapping torus labeled by another $\SL(2,\Z)$ element $\yg$. We are interested in the partition function $\PF(\xg,\yg)$ which is a function of the two duality twists. Here $\xg$ is the S-duality element that acts on the $\mathcal{N}=4$ SYM coupling constant, while $\yg$ is the MCG element of the $T^2$ in the {\it geometrical} mapping torus. 
To preserve SUSY, it is again convenient to take the limit where the area of the $T^2$ fiber shrinks to zero first. Thus, we first reduce the 4d gauge theory to 2d, which for $U(1)$ gauge group becomes a $\sigma$-model with a $T^2\times\R^6$ target space. The $\R^6$ factor does not play much of a role in what follows, so we ignore it. We are thus led to study ``double-Janus'' configurations for a linear $\sigma$-model with $T^2$ target space where the complex structure varies in one direction and the K\"ahler structure (i.e., the complexified area) varies in the other direction, with $\xg$- and $\yg$-boundary conditions respectively [where $\xg\in \SL(2,\Z_\tau)$ and $\yg\in \SL(2,\Z_\rho)$]. We will show that $\PF(\xg,\yg)$ is essentially a finite sum over certain roots of unity, and by calculating it in two different limits, respectively corresponding to different limits of the shape of the (physical) $T^2$ target, we arrive at a number-theoretic identity known as the {\it Landsberg-Schaar} identity.

The rest of this paper is organized as follows.
We begin in \secref{sec:QRbasics} with a brief review of Quadratic Reciprocity, Gauss sums, and the Landsberg-Schaar identity. We then construct the double-Janus $\sigma$-model in the 2d bulk in \secref{sec:sigmaM}, where we present the general constraints from supersymmetry, and various solutions. In \secref{sec:DTwists} we compactify the double-Janus solution on a torus and introduce the twisted boundary conditions (with both geometrical and T-duality twists, $M$ and $\widetilde{M}$, respectively). In \secref{sec:PF} we calculate the double-Janus partition function, including both bosonic and fermionic one-loop determinants, which cancel each other out, leaving only a number-theoretic quadratic Gauss sum.
In \secref{sec:CSMT} we discuss connections with abelian Chern-Simons theory partition functions and the dual ``strings on mapping tori'' introduced in \cite{Ganor:2014pha} in the context of 4d closed Janus configurations. We use this dual formulation to determine the precise normalization of the partition function. In \secref{sec:QRDJSM} we show how the basic Landsberg-Schaar relation follows with a Berry phase factor included, and in \secref{sec:Generalizations} we derive its multivariate generalizations, along with a comparison against known generalizations in the mathematical literature. Finally we conclude in \secref{sec:Discussion}.

% ==============================================================

\section{Quadratic Reciprocity}
\label{sec:QRbasics}
%This section is a review of basic facts from elementary number theory, and more details could be found in the references \cite{LS,NumberTheory}. 

{\it Quadratic Reciprocity} is a classic duality in elementary number theory. In order to introduce it, we first set up the context and review a few related concepts.
If $p$ is an odd prime number and $q$ is an integer, then $q$ is called a {\it quadratic residue} (mod $p$) if $x^2\equiv q$ (mod $p$) has integer solutions $x$. The information on solutions to this equation is packaged inside the {\it Legendre symbol} $\displaystyle{\left(\frac{q}{p}\right)}$ defined as follows:

$$
{\left(\frac{q}{p}\right)} \equiv \left\{\begin{array}{cc}
{-1} & \text{if $\sqrt{q}$ (mod $p$) doesn't exist;} \\&  \\
{0} & \text{if $q\equiv 0$ (mod $p$);} \\ & \\ 
{1} & \text{if $\pm\sqrt{q}$ (mod $p$) exist and are distinct.} \\
\end{array}\right.
$$

The {\it Law of Quadratic Reciprocity} states that if $p$ and $q$ are odd primes, then
\begin{equation}
\label{eqn:QR}
\displaystyle{\left(\frac{p}{q}\right)}\displaystyle{\left(\frac{q}{p}\right)}=(-1)^{\left(\frac{p-1}{2}\right)\left(\frac{q-1}{2}\right)}.
\end{equation}
It is a nontrivial statement that connects the existence of solutions to $x^2\equiv q$ (mod $p$) with the existence of solutions to the dual equation $x^2\equiv p$ (mod $q$).
Quadratic Reciprocity was originally conjectured by Euler and Legendre in the late 18$^{\text{th}}$ century, and then proved by Gauss in 1801. It was a catalyst for subsequent modern developments in Algebraic Number Theory, such as Artin's Reciprocity Law \cite{Artin:1927}.

A {\it Quadratic Gauss Sum} is the discrete Fourier transform of the Legendre symbol:
\begin{equation}\label{eqn:QGS}
\QGaussSum{a}{p} \equiv \sum_{b=0}^{p-1} e^{2\pi i a b/p}\displaystyle{\left(\frac{b}{p}\right)}
=\sum_{n=0}^{p-1}e^{2\pi i a n^2/p}\,,\qquad a{\not\equiv}0\pmod{p}\,,
\end{equation}
where the last equality follows from the identity $\displaystyle{\sum_{b=0}^{p-1}e^{2\pi i a b/p}=0}$ for $a{\not\equiv}0$ (mod $p$). 
%If $a\equiv 0$ (mod $p$), we define $\QGaussSum{a}{p}=p$.
It is not hard to prove that $\displaystyle{\QGaussSum{a}{p}=\displaystyle{\left(\frac{a}{p}\right)}\QGaussSum{1}{p}}$, so the quadratic Gauss sum is proportional to the quadratic residue.
It can also be shown that $\QGaussSum{1}{p} = \sqrt{p}$ if $p\equiv 1$ (mod $4$) and $\QGaussSum{1}{p} = i\sqrt{p}$ if $p\equiv 3$ (mod $4$) (see \cite{IrelandRosen}). Recent results on closed forms of some quadratic Gauss sum include \cite{Milgram:2014}.
Quadratic Reciprocity is then a statement about the relation between quadratic Gauss sums.
For example, if both $p$ and $q$ are $1$ (mod $4$) then $\QGaussSum{q}{p}/\sqrt{p} = \QGaussSum{p}{q}/\sqrt{q}$.

A convenient way to prove the Quadratic Reciprocity \eqref{eqn:QR} is via the Landsberg-Schaar identity \cite{LS1,LS2}
\begin{equation}
\label{eqn:LSrel}
\frac{e^{\pi i/4}}{\sqrt{2p}}\sum_{n=0}^{2p-1}e^{-\pi i n^2q/2p}
=\frac{1}{\sqrt{q}}\sum_{n=0}^{q-1}e^{2\pi i n^2 p/q}\,,
\qquad
1\le p,q\in\Z.
\end{equation}
The identity can easily be proved using the modular transformation properties of the Jacobi theta function $\displaystyle{\vartheta\left(0;\tau\right) \equiv \sum_{n=-\infty}^\infty \exp\left(\pi i\tau n^2\right)}$ evaluated at zero\footnote{In Riemann and Mumford's notations, $\vartheta(z;\tau)\equiv\vartheta_{00}(z;\tau)$ is denoted as $\theta_3(z;q)$, where $q\equiv e^{2\pi i\tau}$.} and its asymptotic behavior as the argument $\tau$ vertically approaches the real axis \cite{Analysis,Analysis2}.\footnote{The Landsberg-Schaar identity was also recently proved using a non-analytical method \cite{newLS}. For the history of development of various proofs other than Gauss's after 1801, see \cite{Bulletin}.}
In this paper, we will describe a physical system with a finite number of quantum ground states that reproduces (\ref{eqn:LSrel}) directly.

If $p$ and $q$ are both even, say $p=2p'$ and $q=2q'$ then it is easy to see that \eqref{eqn:LSrel} reduces to
$$
\frac{e^{\pi i/4}}{2\sqrt{p'}}\sum_{n=0}^{2p'-1}e^{-\pi i n^2q'/2p'}
=\frac{1}{\sqrt{2q'}}\sum_{n=0}^{q-1}e^{2\pi i n^2 p'/q'}\,.
$$
(This is seen by first summing the terms with $n$ and $n+2p'$ on the LHS and $n$ and $n+q'$ on the RHS.)

Before proceeding to the physical system, we note that for odd primes $p$ and $q$, the Landsberg-Schaar identity is a slightly modified version of Quadratic Reciprocity. To see this, define
\begin{equation}
\DualQGaussSum{a}{p} \equiv\sum_{n=0}^{2p-1}e^{\pi i a n^2/2p}\,,
\end{equation}
which is periodic in $a$ with period $4p$, and satisfies $\DualQGaussSum{a+2p}{p}=(-i)^{ap}\DualQGaussSum{a}{p}$ (as can easily be seen by relabeling $n\rightarrow n+p$).
%$$
%\DualQGaussSum{a+2p}{p}=
%\sum_{n=0}^{2p-1}(-1)^{an}e^{\pi i a n^2/2p}
%=\sum_{n=0}^{2p-1}(-1)^{an-ap}e^{\pi i a(n^2-2pn+p^2)/2p}
%=(-i)^{ap}\DualQGaussSum{a}{p}
%$$
%We  note that $2\QGaussSum{a}{p}=\DualQGaussSum{4a}{p}$. Since $4$ is a quadratic residue, we have $\QGaussSum{a}{p}=\QGaussSum{4a}{p}$.
In fact, it is elementary to check (by splitting the sum over $n$ into sums over odd and even numbers) that
\begin{equation}
\DualQGaussSum{a}{p} =\left(1+i^{p a}\right)\QGaussSum{a}{p}\,.
\end{equation}
It then follows from Quadratic Reciprocity and the results quoted above for quadratic Gauss sums that if $p\neq q$ are odd primes, 
%$$
%\frac{1}{\sqrt{2p}}\DualQGaussSum{q}{p}
%=\frac{1}{\sqrt{q}}(-1)^{(q-1)/2}e^{i\pi/4}\QGaussSum{p}{q}\,.
%$$
\begin{equation}\label{eqn:QDualQ}
\frac{1}{\sqrt{2p}}\DualQGaussSum{q}{p}
=\frac{1}{\sqrt{q}}e^{\frac{1}{4}(2q-1)i\pi}\QGaussSum{p}{q}\,.
\end{equation}
Taking the complex conjugate and noting the known result $\displaystyle{\left(\frac{-1}{q}\right)}=(-1)^{(q-1)/2}$, the Landsberg-Schaar identity (\ref{eqn:LSrel}) follows in the form
\begin{equation}\label{eqn:LSS}
\frac{1}{\sqrt{2p}}\DualQGaussSum{-q}{p}
=\frac{1}{\sqrt{q}}e^{-\frac{1}{4}i\pi}\QGaussSum{p}{q}\,.
\end{equation}
We will now construct a $(p,q)$-dependent quantum field theory whose partition function can be calculated in two different ways; one gives an expression proportional to $\QGaussSum{p}{q}$ while the other gives an expression proportional to $\DualQGaussSum{-q}{p}$. 

In the next sections we will show how \eqref{eqn:LSS} and generalizations of it arise from studying the partition function on $T^2$ of supersymmetric 2d $\sigma$-models whose target space is (also) a $T^2$ with complex structure $\tauCS$ and K\"ahler modulus $\rhoKM$ that vary along the $T^2$ worldsheet. We will include boundary conditions with duality twists along two independent cycles of the worldsheet, acting as 
$$
\tauCS\rightarrow\frac{\xa\tauCS+\xb}{\xc\tauCS+\xd}
\,,\qquad\text{and}\qquad
\rhoKM\rightarrow\frac{\ya\rhoKM+\yb}{\yc\rhoKM+\yd}
\,,
$$
So that the partition function $\PF(\matMx,\matMy)$ will depend on two $\SL(2,\Z)$ elements
$$
\matMx\equiv\begin{pmatrix}\xa & \xb \\ \xc & \xd \\ \end{pmatrix}\in \SL(2,\Z)_{\tauCS}
\,,\qquad
\matMy\equiv\begin{pmatrix}\ya & \yb \\ \yc & \yd \\ \end{pmatrix}\in \SL(2,\Z)_{\rhoKM}
\,.
$$
The basic Landsberg-Schaar relation \eqref{eqn:LSS} will be recovered in the special case where
\be\label{eqn:MyMxSpecial}
\matMx=\begin{pmatrix}
0 & -1 \\ 1 & q+2
\end{pmatrix},\qquad
\matMy=\begin{pmatrix}
0 & -1 \\ 1 & 2p+2
\end{pmatrix}.
\ee
We now present the details.

% ==============================================================

\section{Double-Janus \texorpdfstring{$\sigma$}{}-models}
\label{sec:sigmaM}
In this section we construct a 2d $\sigma$-model whose target space is a $T^2$ with complex structure $\tauCS$ and K\"ahler modulus $\rhoKM$ that vary along the worldsheet so as to preserve some amount of supersymmetry.
We use the terms ``worldsheet'' and ``target-space'', borrowed from string-theory, but we emphasize that we are dealing only with a 2d CFT. 
In particular, we will be working with a fixed metric
$$
ds^2 = (d\xWS_1)^2 + (d\xWS_2)^2 = d\zWS d\bzWS\,,\qquad
$$
where we denote the (Euclidean) worldsheet coordinates by $(\xWS_1,\xWS_2)$, and we set
$$
\zWS\equiv\xWS_1 + i\xWS_2\,,\quad
\bzWS\equiv\xWS_1 - i\xWS_2\,,\quad
\partial\equiv\frac{\partial}{\partial\zWS}=\frac{1}{2}(\partial_1-i\partial_2)\,,\quad
\bpartial\equiv\frac{\partial}{\partial\bzWS}=\frac{1}{2}\left(\partial_1+i\partial_2\right)\,.
$$
Dimensional reduction of the Gaiotto-Witten action \cite{Gaiotto:2008sd}, in the special case of a $U(1)$ gauge group, is one way to obtain such a double-Janus Lagrangian, with both $\tauCS$ and $\rhoKM$ varying along the worldsheet. 
%In this model, the Yang-Mills coupling constant is taken to be $\rhoKM(\xWS_1)$, which varies along direction $\xWS_1$, and the remaining $3$ directions are taken to be in the form of a small $T^2$ of complex structure $\tauCS(\xWS_2)$ fibered over $\R$ in the direction of $\xWS_2$, with an R-symmetry connection that matches the holonomy of this $3$-manifold. The holonomy of the $3$-manifold can be designed to be only $SO(2)$. [The $T^2$ can be taken to be of the form of a product of two circles, $S^1_a\times S^1_b$, with $S^1_a$ of a fixed radius and $S^1_b$ of a variable radius, or more generally, we can take such a metric and act on it with $\SL(2,\R)$.] 
In this way we can preserve $4$ supersymmetries, but we also get additional massive noncompact scalar fields. We will describe this model briefly in \appref{app:JanusAndMT}, but for our purposes it will be sufficient to work with a minimally supersymmetric model which we describe in \secref{subsec:MinSUSYDJ}.

% --------------------------------------------------------------
\subsection{Minimally supersymmetric double-Janus model}
\label{subsec:MinSUSYDJ}

Our starting point is a supersymmetric $\sigma$-model with target-space $T^\dimTS$ (later we will set $\dimTS=2$).
We let $\xI,\xJ=1,\dots,\dimTS$ label target-space coordinates. The scalar 
fields will be denoted by $\fX^\xI$, the left-moving fermionic fields will be denoted 
by $\fPsi^\xI$, and the right-moving fermionic fields will be denoted by $\bfPsi^\xI$. 
We combine them into $(2\dimTS)$-component column vectors, denoted by $\vX$, $\vPsi$ and $\bvPsi$, with transposed row-vectors denoted by $\vX^t$, $\vPsi^t$, and $\bvPsi^t$.
We assume that $\fX^\xI\sim\fX^\xI+2\pi$ is a periodic field.
All fields are functions of $(\xWS_1,\xWS_2)$.

The components of the target-space metric and Kalb-Ramond field will be denoted by $\cG_{\xI\xJ}$ and $\cB_{\xI\xJ}$, respectively, and combined into the $(2\dimTS)\times(2\dimTS)$ matrices $\matG$ and $\matB$, with $\matG^t=\matG$ and $\matB^t=-\matB$. We denote
$$
\matE\equiv\matG+\matB.
$$
This matrix will appear in the kinetic term of the scalar fields $\vX$, and we allow $\matE$ to vary with $(\xWS_1,\xWS_2)$, in a predetermined way, as in a Janus configuration.
$\matE$ is thus a $(\xWS_1,\xWS_2)$-dependent background.
To preserve supersymmetry, the action requires additional ``mass terms'' and ``background gauge field terms'' and takes the general form:
\be\label{eqn:sigmaI}
I =
\frac{1}{\pi}\int\left(
\partial\vX^t\matE\bpartial\vX
+i\vPsi^t\matSV\bpartial\vPsi
+i\bvPsi^t\bmatSV\partial\bvPsi
+i\vPsi^t\matK\vPsi
+i\bvPsi^t\bmatK\bvPsi
+\bvPsi^t\matW\vPsi
\right)d^2\xWS
\,.
\ee
$\matSV$, $\bmatSV$, $\matK$, $\bmatK$, and $\matW$ are background $(2\dimTS)\times(2\dimTS)$ matrices with, possibly, $(\xWS_1,\xWS_2)$-dependent elements.
$\matSV=\matSV^t$ and $\bmatSV=\bmatSV^t$ are symmetric matrices that define the fermionic kinetic terms, $\matK=-\matK^t$ and $\bmatK=-\bmatK^t$ are antisymmetric and enter in effective ``background $SO(2\dimTS)$ gauge field terms'', and $\matW$ is an effective ``mass term''.

The supersymmetry transformations take the form
\be\label{eqn:delSUSY}
\delSUSY\vX = \pSUSY\left(\vPsi-\bvPsi\right)\,,
\qquad
\delSUSY\vPsi = i\pSUSY\matSV^{-1}\matG\partial\vX\,,
\qquad
\delSUSY\bvPsi = -i\pSUSY\bmatSV^{-1}\matG\bpartial\vX\,,
\ee
where $\pSUSY$ is an anticommuting parameter.
$\pSUSY$ is real, and complex conjugation acts as
\be\label{eqn:cc}
\pSUSY^\star=\pSUSY,\qquad
\vPsi^\star=\bvPsi,\qquad
\bvPsi^\star=\vPsi,\qquad
\left(\pSUSY\vPsi\right)^\star=\vPsi^\star\pSUSY^\star=-\pSUSY\bvPsi,\qquad
\left(\fPsi^\xI\fPsi^\xJ\right)^\star=\bfPsi^\xJ\bfPsi^\xI.
\ee
Note that invariance of the action under the $(\cdots)^\star$ operation requires $\matB=0$, $\matSV^\star=\bmatSV$, $\matK^\star=\bmatK$ and $\matW^\star=\matW^t$ (but we will not require this except in special cases below).
Invariance of the action under the SUSY transformations \eqref{eqn:delSUSY} requires the following relations among $\matG$, $\matE$, $\matSV$ and $\bmatSV$:
\bear
0 &=& 
2\bpartial\matSV-\bpartial\matE\matG^{-1}\matSV-\matSV\matG^{-1}\bpartial\matE^t,
\label{eqn:bpSV}\\
0 &=&  
2\partial\bmatSV-\partial\matE^t\matG^{-1}\bmatSV-\bmatSV\matG^{-1}\partial\matE,
\label{eqn:pbSV}\\
0 &=&
\matSV\matG^{-1}\bpartial\matE +\partial\matE\matG^{-1}\bmatSV.
\label{eqn:SVbSV}
\eear
Then, $\matK$, $\bmatK$, and $\matW$ are determined in terms of $\matE$, $\matSV$, and $\bmatSV$ by
\bear
\matK &=& 
\frac{1}{2}\left(\bpartial\matSV-\bpartial\matE\matG^{-1}\matSV\right)
=
\frac{1}{4}\matSV\matG^{-1}\bpartial\matE^t
-\frac{1}{4}\bpartial\matE\matG^{-1}\matSV,
\label{eqn:matK}\\
\bmatK &=& 
\frac{1}{2}\left(\partial\bmatSV-\partial\matE^t\matG^{-1}\bmatSV\right)
=
\frac{1}{4}\bmatSV\matG^{-1}\partial\matE
-\frac{1}{4}\partial\matE^t\matG^{-1}\bmatSV,
\label{eqn:bmatK}\\
\matW &=&i\bmatSV\matG^{-1}\partial\matE^t =
-i\bpartial\matE^t\matG^{-1}\matSV.
\label{eqn:matW}
\eear
For future reference, we list the linear equations of motion that follow from \eqref{eqn:sigmaI}. The bosonic equations are
\be\label{eqn:bosonicEOM}
0=\bpartial\left(\matE^t\partial\vX\right)+\partial\left(\matE\bpartial\vX\right)
=2\matG\bpartial\partial\vX
+\left(\bpartial\matE^t\right)\partial\vX
+\left(\partial\matE\right)\bpartial\vX,
\ee
and the fermionic equations of motion are
\be\label{eqn:fermionicEOM}
\bpartial\vPsi =
-\frac{1}{2}\matG^{-1}\bpartial\matE\bvPsi
-\matSV^{-1}\left(\matK+\frac{1}{2}\bpartial\matSV\right)\vPsi
\,,\qquad
\partial\bvPsi =
-\frac{1}{2}\matG^{-1}\partial\matE^t\vPsi
-\bmatSV^{-1}\left(\bmatK+\frac{1}{2}\partial\bmatSV\right)\bvPsi\,,
\ee
which can be simplified, using \eqref{eqn:matK}-\eqref{eqn:matW}, to
\be\label{eqn:fermionicEOMsimplified}
0 =
2\matG\bpartial\vPsi+\left(\bpartial\matE\right)\bvPsi+\left(\bpartial\matE^t\right)\vPsi
\,,\qquad
0 =
2\matG\partial\bvPsi+\left(\partial\matE^t\right)\vPsi+\left(\partial\matE\right)\bvPsi
\,.
\ee
The supersymmetry constraint equations \eqref{eqn:bpSV}-\eqref{eqn:SVbSV} simplify somewhat when $\matG$ is expressed in terms of a ``vielbein'' $\matF$ as
\be\label{eqn:Vielbein}
\matG=\matF^t\matF.
\ee
Here $\matF$ is a $(2\dimTS)\times(2\dimTS)$ matrix, which is not uniquely defined in terms of $\matG$, but different choices differ by $\matF\rightarrow\matOf\matF$, where $\matOf$ is $O(2\dimTS)$-valued. We also define
\be\label{eqn:matSdef}
\matS\equiv\left(\matF^t\right)^{-1}\matSV\matF^{-1}
\qquad
\bmatS\equiv\left(\matF^t\right)^{-1}\bmatSV\matF^{-1},
\ee
and
\bear
\matX_\zWS&\equiv&
\frac{1}{2}\left[\left(\matF^t\right)^{-1}\partial\matF^t-\partial\matF\matF^{-1}\right]
+\frac{1}{2}\left(\matF^t\right)^{-1}\partial\matB\matF^{-1},
\label{eqn:matXdef}\\
\bmatX_\bzWS&\equiv&
\frac{1}{2}\left[\left(\matF^t\right)^{-1}\bpartial\matF^t-\bpartial\matF\matF^{-1}\right]
-\frac{1}{2}\left(\matF^t\right)^{-1}\bpartial\matB\matF^{-1}.
\label{eqn:bmatXdef}
\eear
Then,
$\matS=\matS^t$ and $\bmatS=\bmatS^t$ are symmetric matrices transforming under $\matF\rightarrow\matOf\matF$ as $\matS\rightarrow\matOf\matS\matOf^t$ and $\bmatS\rightarrow\matOf\bmatS\matOf^t$, while $\matX_\zWS$ and $\bmatX_\bzWS$ are antisymmetric and transform like a gauge field:
$$
\matX_\zWS\rightarrow
\matOf\matX_\zWS\matOf^t+\matOf\partial\matOf^t,\qquad
\bmatX_\bzWS\rightarrow
\matOf\bmatX_\bzWS\matOf^t+\matOf\bpartial\matOf^t.
$$
The constraints \eqref{eqn:bpSV}-\eqref{eqn:pbSV} can then be written as
\be\label{eqn:pSbpbS}
0=\bpartial\matS+\left[\bmatX_\bzWS,\matS\right],
\qquad
0=\partial\bmatS+\left[\matX_\zWS,\bmatS\right].
\ee
Next, we define the symmetric matrices
\bear
\matY &\equiv& \left(\matF^t\right)^{-1}\left(\partial\matE^t\right)\matF^{-1} =
\left(\matF^t\right)^{-1}\left(\partial\matF^t\right)
+\left(\partial\matF\right)\matF^{-1}
-\left(\matF^t\right)^{-1}\left(\partial\matB\right)\matF^{-1}
\,,\label{eqn:matYdef}\\
\bmatY &\equiv& \left(\matF^t\right)^{-1}\left(\bpartial\matE^t\right)\matF^{-1} =
\left(\matF^t\right)^{-1}\left(\bpartial\matF^t\right)
+\left(\bpartial\matF\right)\matF^{-1}
-\left(\matF^t\right)^{-1}\left(\bpartial\matB\right)\matF^{-1}
\,.\label{eqn:bmatYdef}
\eear
Then, \eqref{eqn:SVbSV} can be written as
\be\label{eqn:YS}
0 = \bmatS\matY+\bmatY\matS.
\ee
Thus, given $\matF$ and $\matB$, we can find $\matS$ and $\bmatS$ by solving \eqref{eqn:pSbpbS}, and then \eqref{eqn:YS} yields a constraint on $\matF$ and $\matB$.
We do not know the general solution $\left(\matG,\matB,\matS,\bmatS\right)$ to \eqref{eqn:bpSV}-\eqref{eqn:SVbSV}, but below we will discuss a special class of solutions that will suffice for our needs. We restrict to $\dimTS=1$ and begin with a simple class of solutions where $\matG$ is diagonal and $\matB=0$. We then apply solution generating transformations (i.e., T-duality) to obtain a wider class of solutions.

% --------------------------------------------------------------
\subsection{\texorpdfstring{$\xWS_1$}{}-independent solutions}
\label{subsec:indepsol}

A simple way to satisfy \eqref{eqn:pSbpbS} is to set
$$
\matS=\bmatS=\Id,
$$
where $\Id$ is the identity matrix.
Then, \eqref{eqn:YS} becomes $\matY=\bmatY$, which by \eqref{eqn:matYdef}-\eqref{eqn:bmatYdef} requires $0=\partial\matE+\bpartial\matE=\partial_1\matE$. Thus, $\matE$, and hence $\matG$ and $\matB$ are functions of $\xWS_2$ only. How they vary with $\xWS_2$ is arbitrary, and any pair of $\matG(\xWS_2)$ and $\matB(\xWS_2)$ determines a supersymmetric action \eqref{eqn:sigmaI}.
The remaining couplings in \eqref{eqn:sigmaI} can then easily be calculated from \eqref{eqn:matSdef} and \eqref{eqn:matK}-\eqref{eqn:matW}:
$$
\matSV=\bmatSV=\matG,\qquad
\matK=\bmatK=-\frac{i}{4}\partial_2\matB,\qquad
\matW=\frac{1}{2}\partial_2\matE^t.
$$
By setting $\matS=-\bmatS=i\Id$ we similarly get a solution with couplings that are functions of $\xWS_1$ only.
In this paper, however, we are more interested in solutions where $\matG$ and $\matB$ vary nontrivially with both $\xWS_1$ and $\xWS_2$.

% --------------------------------------------------------------
\subsection{Holomorphic solutions}
\label{subsec:HoloSols}

For the $T^2$ target space, another class of solutions can be obtained by requiring $\rhoKM$ to be constant and allowing $\tauCS$ to vary holomorphically with $\zWS$. The setup is then an elliptic fibration, reminiscent of F-theory \cite{Vafa:1996xn}.
We can take\footnote{The minus sign in front of $\tauCS_1$ is in order for equations \eqref{eqn:vPsiMZ}-\eqref{eqn:vXMZ} to be simpler.}
$$
\matF=\rhoKM_2^{\frac{1}{2}}\tauCS_2^{-\frac{1}{2}}
\begin{pmatrix} 1 & -\tauCS_1 \\
 0 & \tauCS_2 \\
 \end{pmatrix},
$$
and using \eqref{eqn:matXdef}-\eqref{eqn:bmatXdef}, we calculate
$$
\matX_\zWS=-\frac{\partial\tauCS_1}{2\tauCS_2}\matEps
\,,\qquad
\bmatX_\bzWS=-\frac{\bpartial\tauCS_1}{2\tauCS_2}\matEps
\,,
$$
where
\be
\label{eqn:matEpsDEF}
\matEps\equiv\begin{pmatrix} 0 & -1 \\ 1 & 0 \\ \end{pmatrix}.
\ee
Since $\bpartial\tauCS=0$, we can substitute $\partial\tauCS_1=i\partial\tauCS_2$ and $\bpartial\tauCS_1=-i\bpartial\tauCS_2$ and write
$$
\matX_\zWS=-i\frac{\partial\tauCS_2}{2\tauCS_2}\matEps
\,,\qquad
\bmatX_\bzWS=i\frac{\bpartial\tauCS_2}{2\tauCS_2}\matEps.
$$
Then, \eqref{eqn:pSbpbS} has the general solution
\be\label{eqn:matSfromSf}
\matS=e^{-\frac{i}{2}\left(\log\tauCS_2\right)\matEps}
\matSf(\zWS)
e^{\frac{i}{2}(\log\tauCS_2)\matEps},
\qquad
\bmatS=e^{\frac{i}{2}(\log\tauCS_2)\matEps}
\bmatSf(\bzWS)
e^{-\frac{i}{2}(\log\tauCS_2)\matEps},
\ee
where $\matSf(\zWS)$ and $\bmatSf(\bzWS)$ are arbitrary holomoprhic and antiholomorphic (symmetric) matrix-valued functions on the worldsheet.
We also calculate from \eqref{eqn:matYdef}-\eqref{eqn:bmatYdef},
$$
\matY=
\frac{1}{\tauCS_2}
\begin{pmatrix}
-\partial\tauCS_2 & -\partial\tauCS_1 \\
-\partial\tauCS_1 & \partial\tauCS_2 \\
\end{pmatrix}
=
\frac{\partial\tauCS_2}{\tauCS_2}
\begin{pmatrix}
-1 & -i \\
-i & 1 \\
\end{pmatrix}
\,,\qquad
\bmatY=
\frac{1}{\tauCS_2}
\begin{pmatrix}
-\bpartial\tauCS_2 & -\bpartial\tauCS_1 \\
-\bpartial\tauCS_1 & \bpartial\tauCS_2 \\
\end{pmatrix}
=
\frac{\bpartial\tauCS_2}{\tauCS_2}
\begin{pmatrix}
-1 & i \\
i & 1 \\
\end{pmatrix}\,.
$$
We set
$\matCf\equiv
\begin{pmatrix}
-1 & -i \\
-i & 1 \\
\end{pmatrix}$,
$\bmatCf\equiv
\begin{pmatrix}
-1 & i \\
i & 1 \\
\end{pmatrix}$,
and noting that
$\matCf\matEps=-\matEps\matCf =i\matCf$
and
$\matEps\bmatCf=-\bmatCf\matEps=i\bmatCf$ imply
$e^{-\frac{i}{2}(\log\tauCS_2)\matEps}\matCf e^{-\frac{i}{2}(\log\tauCS_2)\matEps}
=\matCf$ and $e^{-\frac{i}{2}(\log\tauCS_2)\matEps}\bmatCf e^{-\frac{i}{2}(\log\tauCS_2)\matEps}=\bmatCf$,
we find that \eqref{eqn:YS} yields
\be\label{eqn:tauSfCf}
0 =
\left(\partial\tauCS_2\right)\bmatSf\matCf
+\left(\bpartial\tauCS_2\right)\bmatCf\matSf.
\ee
Since, for a holomorphic $\tauCS$ we have
$\partial\tauCS_2 = -\frac{i}{2}\partial\tauCS$, which is also holomorphic, while
$\bpartial\tauCS_2=\frac{i}{2}\bpartial\btauCS$ is antiholomoprhic, we see that \eqref{eqn:tauSfCf} is possible only if 
\be\label{eqn:matSfFromSfX}
\matSf=\left(\partial\tauCS_2\right)\matSfX,\qquad
\bmatSf=\left(\bpartial\tauCS_2\right)\bmatSfX,
\ee
for some constant symmetric matrices $\matSfX$, $\bmatSfX$ that satisfy
\be\label{eqn:SfCf}
0 = \bmatSfX\matCf+\bmatCf\matSfX.
\ee
Since $\matCf$ and $\bmatCf$ are nilpotent ($\matCf^2=\bmatCf^2=0$), \eqref{eqn:SfCf} requires $\bmatCf\bmatSfX\matCf=\bmatCf\matSfX\matCf=0$ which is equivalent to 
\be\label{eqn:trSfXcond}
\text{Tr}\,\matSfX=\text{Tr}\,\bmatSfX=0.
\ee
If we also require $\bmatSfX$ to be the complex conjugate of $\matSfX$ then \eqref{eqn:SfCf} is satisfied provided 
\be\label{eqn:ReSfXcond}
\text{Re}\,\matSfX_{12}=-\text{Im}\,\matSfX_{11}.
\ee
Once we choose a $2\times 2$ matrix $\matSfX$ (and its complex conjugate $\bmatSfX$) that satisfies \eqref{eqn:trSfXcond} and \eqref{eqn:ReSfXcond}, we calculate $\matSf$ and $\bmatSf$ from \eqref{eqn:matSfFromSfX}, and then we calculate $\matS$ and $\bmatS$ from \eqref{eqn:matSfromSf}. We then find $\matSV$ and $\bmatSV$ from \eqref{eqn:matSdef}.

These holomorphic solutions might be interesting to explore, but we will not discuss them further in the present paper, since our focus is the closed double-Janus solutions to be described in \secref{subsec:MoreDJSols}. We will construct these solutions by first finding solutions where $\matE$ is diagonal.

% --------------------------------------------------------------
\subsection{Diagonal solutions}
\label{subsec:DiagSols}

We get another simple class of solutions to \eqref{eqn:bpSV}-\eqref{eqn:SVbSV} [or, equivalently, to \eqref{eqn:pSbpbS} and \eqref{eqn:YS}] by setting $\matB=0$ and requiring $\matG$ to be diagonal.
We can then choose a diagonal $\matF=\matG^{1/2}$ in \eqref{eqn:Vielbein} and we calculate from \eqref{eqn:matXdef}-\eqref{eqn:bmatXdef} that $\matX_\zWS=\bmatX_\bzWS=0$, and from \eqref{eqn:matYdef}-\eqref{eqn:bmatYdef} we calculate $\matY=\partial\log\matG$ and $\bmatY=\bpartial\log\matG$. Then, \eqref{eqn:pSbpbS} and \eqref{eqn:YS} require
$$
0 = \partial\bmatS = \bpartial\matS = 
\left(\bpartial\log\matG\right)\matS+\bmatS\left(\partial\log\matG\right)
\,.
$$
We will make the further assumption that $\matS$ and $\bmatS$ are constant matrices, as is the case if the worldsheet is $\C$ and we require $\matS$ and $\bmatS$ to be bounded, or if the worldsheet is $T^2$ with periodic boundary conditions. 
We also assume that $\bmatS$ is the complex conjugate of $\matS$, and for simplicity we proceed to analyze the case $\dimTS=2$, i.e., a $T^2$ target space.

If $\matS$ is also diagonal, we proceed as follows.
We first assume without loss of generality that $\matS_{11}>0$ (since we can always rotate $\zWS$ by a phase to make it so). Then
$
0=\left(\bpartial\log\matG_{11}\right)\matS_{11}+\bmatS_{11}\left(\partial\log\matG_{11}\right)
$
says that $\matG_{11}$ is a function of $\xWS_2$ only. If $\matS_{22}$ is also real then a similar conclusion about $\matG_{22}$ shows that we have a special case of \secref{subsec:indepsol}. If $\matS_{22}$ is not real, then $\matG_{22}$ is a function of a different linear combination of $\xWS_1$ and $\xWS_2$. In any case, this seems to be a rather restricted system, and we will not pursue this further in this paper.

If $\matS$ is not diagonal, we can assume that $\bmatS_{12}=\matS_{12}>0$ (again achieved by rotating $\zWS$ if necessary). The off-diagonal components of
\be\label{eqn:plogGS}
0=\left(\bpartial\log\matG\right)\matS+\bmatS\left(\partial\log\matG\right)
\ee
yield
\be\label{eqn:plogGbplogG}
\bpartial\log\matG_{11}=-\partial\log\matG_{22}\qquad\text{and}\qquad
\bpartial\log\matG_{22}=-\partial\log\matG_{11},
\ee
from which it follows that $\partial^2\log\matG_{11}=\bpartial^2\log\matG_{11}$ and similarly for $\matG_{22}$. Since $\partial^2-\bpartial^2=i\partial_1\partial_2$, and combined with \eqref{eqn:plogGbplogG}, we find that $\matG$ takes the form
$$
\matG=\begin{pmatrix}
\frac{\rhoKM_2}{\tauCS_2} & 0 \\
0 & \rhoKM_2\tauCS_2 \\
\end{pmatrix}\,,\qquad
\tauCS_2=\tauCS_2(\xWS_1),\qquad
\rhoKM_2=\rhoKM_2(\xWS_2),
$$
where we have written the metric in terms of the K\"ahler modulus $\rhoKM=\rhoKM_1+i\rhoKM_2$ (with $\rhoKM_1=0$ since $\matB=0$) and the complex structure $\tauCS=\tauCS_1+i\tauCS_2$ (with $\tauCS_1=0$ since $\matG$ is diagonal), and the analysis above shows that (for $\matS_{12}>0$) $\tauCS_2=\tauCS_2(\xWS_1)$ is a function of $\xWS_1$ only, and $\rhoKM_2=\rhoKM_2(\xWS_2)$ is a function of $\xWS_2$ only.
For the future, we denote
$$
\dtauCS_2\equiv\frac{d\tauCS_2}{d\xWS_1},\qquad
\drhoKM_2\equiv\frac{d\rhoKM_2}{d\xWS_2}.
$$
Then, since we assume that $\matS$ and $\bmatS$ are constant, the diagonal components of \eqref{eqn:plogGS} imply that either $\matS_{11}=\matS_{22}=0$, or both $\dtauCS_2/\tauCS_2$ and $\drhoKM_2/\rhoKM_2$ are constant. We will first consider the more general case that $\tauCS_2$ and $\rhoKM_2$ are arbitrary functions of $\xWS_1$ and $\xWS_2$ and we therefore assume that $\matS_{11}=\matS_{22}=0$. Although later on we will discuss the case that $\log\tauCS_2$ and $\log\rhoKM_2$ are linear functions of $\xWS_1$ and $\xWS_2$, respectively, we will still keep the assumption $\matS_{11}=\matS_{22}=0$, since it leads to the simpliest model.
We can take $\matS_{12}=1$ without loss of generality (by rescaling $\zWS$ if necessary).
The action \eqref{eqn:sigmaI} then takes the form
\bear
I &=&
\frac{1}{\pi}\int\left[
\frac{\rhoKM_2}{\tauCS_2}\partial\fX^1\bpartial\fX^1
+\rhoKM_2\tauCS_2\partial\fX^2\bpartial\fX^2
+i\rhoKM_2\fPsi^1\bpartial\fPsi^2
+i\rhoKM_2\fPsi^2\bpartial\fPsi^1
+i\rhoKM_2\bfPsi^1\partial\bfPsi^2
+i\rhoKM_2\bfPsi^2\partial\bfPsi^1
\nn\right.\\ &&
\left.+\frac{i\rhoKM_2\tauCS_2'}{2\tauCS_2}\fPsi^1\fPsi^2
+\frac{i\rhoKM_2\tauCS_2'}{2\tauCS_2}\bfPsi^1\bfPsi^2
+\left(\frac{\dot{\rhoKM}_2}{2}+\frac{i\rhoKM_2\tauCS_2'}{2\tauCS_2}\right)\bfPsi^1\fPsi^2
+\left(\frac{\dot{\rhoKM}_2}{2}-\frac{i\rhoKM_2\tauCS_2'}{2\tauCS_2}\right)\bfPsi^2\fPsi^1
\right] d^2\xWS.
\label{eqn:sigmaIdiag}
\eear
Note that this action is real, in the sense defined below \eqref{eqn:cc}.
The fermionic part of the action \eqref{eqn:sigmaIdiag} can be simplified with a change of variables, but we will defer that to \secref{subsec:Psidet}.

% --------------------------------------------------------------
\subsection{\texorpdfstring{$\SL(2,\R)$}{}-generated double-Janus solutions}
\label{subsec:MoreDJSols}

The action \eqref{eqn:sigmaIdiag} describes a special solution to \eqref{eqn:bpSV}-\eqref{eqn:SVbSV} [substituted into \eqref{eqn:sigmaI}] where the complex structure $\tauCS=\tauCS_1+i\tauCS_2$ and K\"ahler parameter $\rhoKM_1+i\rhoKM_2$ of the $T^2$ target space are each allowed to vary along the imaginary axes of their respective upper half-planes.
We will now apply fractional linear transformations, acting as
$$
\tauCS\rightarrow
\frac{\xal\tauCS+\xbl}{\xcl\tauCS+\xdl},\qquad
\rhoKM\rightarrow
\frac{\yal\rhoKM+\ybl}{\ycl\rhoKM+\ydl},
\qquad\text{for some}\quad
\begin{pmatrix}
\xal & \xbl \\
\xcl & \xdl \\
\end{pmatrix},
\begin{pmatrix}
\yal & \ybl \\
\ycl & \ydl \\
\end{pmatrix}
\in \SL(2,\R),
$$
to generate new solutions.
In general, we expect an $O(n,n,\R)$ group of transformations acting on the parameters $\left(\matE,\matSV,\bmatSV\right)$, and converting a solution of \eqref{eqn:bpSV}-\eqref{eqn:SVbSV} to a new solution $\left(\matE',\matSV',\bmatSV'\right)$.
This is an extension of the well-known $O(n,n,\R)$ action on $\sigma$-models with $T^n$ target space, parametrized by a Narain lattice in the isotropic (i.e., non-Janus) case. (See \cite{Giveon:1994fu} for a review.)
Such transformations are genererated by linear coordinate reparametrizations (which modify the boundary conditions since the new coordinates are still required to obey the same $2\pi$ periodicity conditions as the old ones) and T-dualities. The geometrical transformations act as 
$$
\vX'=\matP\vX,\qquad
\vPsi'=\matP\vPsi,\qquad
\bvPsi'=\matP\bvPsi\,,
$$
and therefore
\be\label{eqn:Ptransf}
\matE'=\left(\matP^{-1}\right)^t\matE\matP^{-1},\qquad
\matSV'=\left(\matP^{-1}\right)^t\matSV\matP^{-1},\qquad
\bmatSV'=\left(\matP^{-1}\right)^t\bmatSV\matP^{-1},
\ee
where $\matP\in \GL(n,\R)$ is a constant matrix. (Later, we will restrict to the case $n=2$.)

T-duality on all directions acts as
\be\label{eqn:TdualityAll}
\matE' =\matE^{-1},\qquad
\matSV' = \matE^{-1}\matSV\left(\matE^t\right)^{-1},\qquad
\bmatSV' = \left(\matE^t\right)^{-1}\bmatSV\matE^{-1}.
\ee
The terms in the action \eqref{eqn:sigmaIdiag} can then easily be calculated from \eqref{eqn:matK}-\eqref{eqn:matW}. For example, after some algebra, we get
\bear
\bmatK' &=&
\left(\matE^t\right)^{-1}\left[\bmatK
+\frac{1}{2}\partial\matE^t\left(\matE^t\right)^{-1}\bmatSV
-\frac{1}{2}\bmatSV\matE^{-1}\partial\matE\right]\matE^{-1}
,\nn\\
\matK' &=&
\matE^{-1}\left[\matK
+\frac{1}{2}\left(\bpartial\matE\right)\matE^{-1}\matSV
-\frac{1}{2}\matSV\left(\matE^t\right)^{-1}\bpartial\matE^t\right]\left(\matE^t\right)^{-1}
.\nn
\eear
For the case of $T^2$ target space, we can define two commuting $\PSL(2,\R)$ actions.
We take the target space metric to be 
$$
\cG_{\xI\xJ}d\fX^\xI d\fX^\xJ
=\frac{\rhoKM_2}{\tauCS_2}\left|\tauCS d\fX^2-d\fX^1\right|^2,
$$
so that the geometrical $\PSL(2,\R)_\tauCS$ acts as
$$
\matE'=\left(\matP^{-1}\right)^t\matE\matP^{-1},\quad
\matSV'=\left(\matP^{-1}\right)^t\matSV\matP^{-1},\quad
\bmatSV'=\left(\matP^{-1}\right)^t\bmatSV\matP^{-1},\qquad
\tauCS\rightarrow\tauCS'=
\frac{\xal\tauCS+\xbl}{\xcl\tauCS+\xdl},
$$
where 
\be\label{eqn:xabcdl}
\matP=\begin{pmatrix}
\xal & \xbl \\
\xcl & \xdl \\
\end{pmatrix}\in \SL(2,\R).
\ee
The other group, $\PSL(2,\R)_\rhoKM$, acts as
$$
\rhoKM\rightarrow
\frac{\yal\rhoKM+\ybl}{\ycl\rhoKM+\ydl},
$$
combined with
\bear
\matEps\matE' &=& \left(\ycl\matEps\matE+\ydl\Id\right)^{-1}\left(\yal\matEps\matE+\ybl\Id\right)
\,,\label{eqn:matEp}\\
%\matEps\matW' &=& (\ycl\matEps\matE+\ydl\Id)^{-1}\matEps\matW(\ycl\matEps\matE+\ydl\Id)^{-1}
%\,,\label{eqn:matWp}\\
%\matG' &=& [(\ycl\matEps\matE+\ydl\Id)^t]^{-1}\matG(\ycl\matEps\matE+\ydl\Id)^{-1}
%\,.\label{eqn:matGp}\\
\matSV' &=& 
\left(\ycl\matE\matEps+\ydl\Id\right)^{-1}\matSV\left[\left(\ycl\matE\matEps+\ydl\Id\right)^{-1}\right]^t
\,,\label{eqn:matSVp}\\
\bmatSV' &=&
\left[\left(\ycl\matEps\matE+\ydl\Id\right)^{-1}\right]^t\bmatSV\left(\ycl\matEps\matE+\ydl\Id\right)^{-1}
\,,\label{eqn:bmatSVp}
\eear
where $\matEps$ is the antisymmetric matrix defined in \eqref{eqn:matEpsDEF}, and
\be\label{eqn:yabcdl}
\begin{pmatrix}
\yal & \ybl \\
\ycl & \ydl \\
\end{pmatrix}
\in\SL(2,\R).
\ee
Applying these transformations to the solution \eqref{eqn:sigmaIdiag}, we get a new solution with $\tauCS(\xWS_1)$ taking values on a semicircle of radius $1/2|\xcl\xdl|$ in the upper half-plane that intersects the real axis at the points $\xal/\xcl$ and $\xbl/\xdl$, and similarly, $\rhoKM(\xWS_2)$ takes values on a semicircle of radius $1/2|\ycl\ydl|$ that intersects the real axis at the points $\yal/\ycl$ and $\ybl/\ydl$. This behavior of the modular parameters is similar to that derived by Gaiotto and Witten in \cite{Gaiotto:2008sd} for the supersymmetric Janus configurations of $\mathcal{N}=4$ SYM. 

The solutions discussed above are more suitable for our needs, since for suitably chosen parameters, the semicircles will be invariant under some $\SL(2,\Z)$ duality transformations. The worldsheet of such models can then be compactified on a torus, thus making $\xWS_1$ and $\xWS_2$ periodic, and the periodicities $\xWS_1\rightarrow\xWS_1+1$ and $\xWS_2\rightarrow\xWS_2+1$ are accompanied by duality twists. In the context of 4d \SUSY{4} SYM, such a setup has been used in 
\cite{Ganor:2014pha,Assel:2018vtq} to compactify the Gaiotto-Witten solution to 3d. A generic $\SL(2,\Z)$ matrix $$
\matMx=\begin{pmatrix}
\xa & \xb \\ 
\xc & \xd \\
\end{pmatrix}
$$
with $|\xa+\xd|>2$ (a hyperbolic element\footnote{Or equivalently, a {\it pseudo-Anosov homeomorphism} of $T^2$, the fiber of the non-geometric mapping torus, which turns out to be hyperbolic.}) preserves the semicircle of radius $\sqrt{
(\xa+\xd)^2-4}/2|\xc|$ that intersects the real axis at 
$$
\tauCS=\frac{1}{2\xc}\left(
\xa-\xd\pm
\sqrt{
(\xa+\xd)^2-4
}
\right)
.
$$
We will now discuss in detail how to incorporate the duality twists, and the ``interaction'' between the $\tauCS$ and $\rhoKM$ twists.

% ==============================================================
\section{Duality twists}
\label{sec:DTwists}
At the end of \secref{subsec:MoreDJSols} we obtained a model with $\tauCS$ varying as a function of $\xWS_1$ while taking values on a semicircle that is preserved by
$$
\matMx = \begin{pmatrix}
\xa & \xb \\ \xc & \xd \\
\end{pmatrix},
$$
and with $\rhoKM$ varying as a function of $\xWS_2$ while taking values on another semicircle that is preserved by 
$$
\matMy = \begin{pmatrix}
\ya & \yb \\ \yc & \yd \\
\end{pmatrix}.
$$
We will now compactify the worldsheet.
To begin with, we put the theory on a noncompact rectangular worldsheet with
$$
0\le\xWS_1,\xWS_2< 1\,.
$$

% --------------------------------------------------------------
\subsection{Geometrical twist}
\label{subsec:GeomTwist}

Let us first discuss the boundary conditions relating $\xWS_1=0$ to $\xWS_1=1$.
The complex structure $\tauCS$ of the target space $T^2$ depends on $\xWS_1$, and if $\tauCS(0)\neq\tauCS(1)$, we can insert an MCG element that acts nontrivially, but geometrically, on $\vX$.
We then require $\tauCS(0)$ to be related to $\tauCS(1)$ by the element $\matMx\in \SL(2,\Z)$ so that
\be\label{eqn:tauCS01}
\tauCS(1) = \frac{\xa\tauCS(0)+\xb}{\xc\tauCS(0)+\xd}\,.
\ee
The boundary conditions on the fermionic fields are
\be\label{eqn:vPsiMZ}
\vPsi(1,\xWS_2)=\matMx\vPsi(0,\xWS_2)\,,
\qquad
\bvPsi(1,\xWS_2)=\matMx\bvPsi(0,\xWS_2)\,.
\ee
For the bosonic fields, we require $\vX(1,\xWS_2)$ to be related to $\matMx\vX(0,\xWS_2)$, up to a vector whose components are integer multiples of $2\pi$: 
\be\label{eqn:vXMZ}
\vX(1,\xWS_2)-\matMx\vX(0,\xWS_2)\in 2\pi\Z^2\,.
\ee
Denote
\be\label{eqn:vNdef}
\vN\equiv\frac{1}{2\pi}\left[\vX(1,\xWS_2)-\matMx\vX(0,\xWS_2)\right]\in\Z^2\,,
\ee
which can be thought of as a vector of ``winding numbers'', and being a vector of integers, it is independent of $\xWS_2$.
We can construct a vector with a simpler boundary condition by removing a constant piece from $\vX$, 
\be\label{eqn:vZdef}
\vZ\equiv\vX-2\pi\left(\Id-\matMx\right)^{-1}\vN
\,,
\ee
which satisfies the periodicity condition
\be\label{eqn:vZbc}
\vZ(1,\xWS_2)=\matMx\vZ(0,\xWS_2)\,.
\ee
Note that $\vZ$ is also invariant under the discrete translations that are parametrized by a vector of integers $\vK$ as:
\be\label{eqn:vXvKshift}
\vX\rightarrow\vX+2\pi\vK\,,\quad
\vN\rightarrow\vN+(\Id-\matMx)\vK\,,
\qquad(\vK\in\Z^2)
\,.
\ee

In Minkowski (worldsheet) signature, the minimum energy configuration would be $\vZ=0$ which corresponds to $\vX=2\pi(\Id-\matMx)^{-1}\vN$. These are the fixed points of the action $\vX\rightarrow\matMx\vX$ (acting on the $T^2$ target).
Moreover, the equivalence
$$
\vX\sim\vX+2\pi\vK\,,\qquad
(\vK\in\Z^2)\,,
$$
which is required to describe a $T^2$ target space (instead of $\R^2$), acts on $\vN$ as
\be\label{eqn:vNi}
\vN\sim\vN+(\Id-\matMx)\vK\,.
\ee
The lattice $\Z^2$ subject to the identification \eqref{eqn:vNi} is a finite abelian group with $\det(\Id-\matMx)=|\xa+\xd-2|$ elements. (This abelian group played an important role in \cite{Ganor:2014pha}, where it was related to a group of symmetry operators in a related context.) We denote this group by
\be
\label{eqn:GrStDef}
\GrSt\equiv\left\{\text{$\vN$ subject to $\vN\sim\vN+(\Id-\matMx)\vK$ for all $\vK\in\Z^2$}\right\}
=(\Z^2)/(\Id-\matMx)(\Z^2).
\ee
The path integral over field configurations $\vX$ (for any worldsheet signature) subject to the periodicity condition \eqref{eqn:vXMZ} is equivalent to a path integral over $\vZ$, subject to the boundary condition \eqref{eqn:vZbc}, and a sum over the finite group $\GrSt$.

% --------------------------------------------------------------
\subsection{T-duality twist}
\label{subsec:Ttwist}

Now, we introduce an $\SL(2,\Z)$ duality twist in the $\xWS_2$ direction. This twist acts on the $\rhoKM(\xWS_2)$ parameter by an element $\matMy\in \SL(2,\Z)$ so that
\be\label{eqn:rhoKM01}
\rhoKM(1) = \frac{\ya\rhoKM(0)+\yb}{\yc\rhoKM(0)+\yd}
\,,\qquad
\matMy = \begin{pmatrix}
\ya & \yb \\ \yc & \yd \\
\end{pmatrix}
\,,
\ee
resulting in an asymmetric orbifold \cite{Orbifold}.
The $\matMy$-twist induces certain nontrivial boundary conditions for the fields $\vX$, $\vPsi$ and $\bvPsi$. To describe them, we denote
\be\label{eqn:sub1notation}
\vX_1(\xWS_1,\xWS_2)\equiv\vX(\xWS_1,1+\xWS_2),\quad
\vPsi_1(\xWS_1,\xWS_2)\equiv\vPsi(\xWS_1,1+\xWS_2),\quad
\bvPsi_1(\xWS_1,\xWS_2)\equiv\bvPsi(\xWS_1,1+\xWS_2).
\ee
These fields will be used to describe the fields in the vicinity of $\xWS_2=1$.
In the same vein, we also denote
\be\label{eqn:sub0notation}
\vX_0\equiv\vX,\quad
\vPsi_0\equiv\vPsi,\quad
\bvPsi_0\equiv\bvPsi,
\ee
when referring to fields in the vicinity of $\xWS_2=0$.
(Note that $\vX_0$ is identical to $\vX$, but the subscript ``$0$'' is useful for clarity when we couple $\vX_0$ to $\vX_1$ and restrict the fields to $\xWS_2=0$.)
We use a similar subscript notation for the background fields, i.e.,  $\matE_0\equiv\matE$, $\matE_1(\xWS_1,\xWS_2)\equiv\matE(\xWS_1,1+\xWS_2)$, etc.
On the level of equations of motion, the boundary conditions on the bosonic fields are
\be\label{eqn:vXbcTdl}
\bpartial\vX_1
=\left(\yc\matEps\matE_0+\yd\Id\right)\bpartial\vX_0
,\qquad
\partial\vX_1
=\left(\yc\matE_0\matEps+\yd\Id\right)^t\partial\vX_0.
\ee
This is a special case of  B\"acklund transformation (for Cauchy-Riemann-like equations), whereby the equations of motion for the fields at $\xWS_2=1$ get mapped to the integrability conditions for the fields at $\xWS_2=0$:
$$
\partial\left(\bpartial\vX_1\right)-\bpartial\left(\partial\vX_1\right)
=0=\yc\matEps\left[\bpartial\left(\matE_0^t\partial\vX_0\right)+\partial\left(\matE_0\bpartial\vX_0\right)\right],
$$
which vanishes thanks to the equation of motion \eqref{eqn:bosonicEOM}.

The fermionic boundary conditions can be derived from the SUSY transformations \eqref{eqn:delSUSY}, together with the boundary conditions on the background parameters \eqref{eqn:matEp}-\eqref{eqn:bmatSVp}, and they are
\be\label{eqn:vPsiTdl}
%\bvPsi_1=\bmatSV_1^{-1}\matG_1(\yc\matEps\matE_0+\yd\Id)\matG_0^{-1}\bmatSV_0\bvPsi_0
\bvPsi_1=\left(\yc\matEps\matE_0+\yd\Id\right)\bvPsi_0
,\qquad
%\vPsi_1=-\matSV_1^{-1}\matG_1(\yc\matEps\matE_0^t-\yd\Id)\matG_0^{-1}\matSV_0\vPsi_0.
\vPsi_1=\left(\yc\matE_0\matEps+\yd\Id\right)^t\vPsi_0.
\ee
When applied to \eqref{eqn:vPsiTdl}, the SUSY variation \eqref{eqn:delSUSY} gives the bosonic boundary conditions \eqref{eqn:vXbcTdl}, and using the fermionic equations of motion \eqref{eqn:fermionicEOMsimplified}, one can easily check that the SUSY variation of \eqref{eqn:vXbcTdl} is also satisfied.
In deriving \eqref{eqn:vPsiTdl} we also used the relations
\bear
\matEps\matE_1 &=& \left(\yc\matEps\matE_0+\yd\Id\right)^{-1}\left(\ya\matEps\matE_0+\yb\Id\right)
\,,\label{eqn:matE1}\\
\matSV_1 &=& 
\left(\yc\matE_0\matEps+\yd\Id\right)^{-1}\matSV_0\left[\left(\yc\matE_0\matEps+\yd\Id\right)^{-1}\right]^t
\,,\label{eqn:matSV1}\\
\bmatSV_1 &=&
\left[\left(\yc\matEps\matE_0+\yd\Id\right)^{-1}\right]^t\bmatSV_0\left(\yc\matEps\matE_0+\yd\Id\right)^{-1}
\,,\label{eqn:bmatSV1}
\eear
which mirror \eqref{eqn:matEp}-\eqref{eqn:bmatSVp}, but with integers $\ya,\dots,\yd$ instead of real numbers $\yal,\dots,\ydl$,
and we also used the identity
$$
\matG_1 =
\left[\left(\yc\matEps\matE_0+\yd\Id\right)^t\right]^{-1}\matG_0\left(\yc\matEps\matE_0+\yd\Id\right)^{-1}
 = \left(\yc\matE_0\matEps+\yd\Id\right)^{-1}\matG_0\left[\left(\yc\matE_0\matEps+\yd\Id\right)^t\right]^{-1},
$$
which easily follows from $2\matG=\matE+\matE^t$.

Quantum mechanically, the boundary conditions \eqref{eqn:vXbcTdl} are introduced by inserting a ``duality wall'' (see for instance \cite{Kapustin:2009av} and the supersymmetric case discussed in \cite{Ganor:2012ek}). The T-duality wall can be thought of as the dimensional reduction of the $T(U(1))$ theory,\footnote{$T(U(1))$, or more generally $T(G)$, is the 3d action introduced in \cite{Gaiotto:2008ak} to capture the action of S-duality in the $\mathcal{N}=4$ Super-Yang-Mills theory, and see also \cite{Lozano:1995aq,Ganor:1996pe,Witten:2003ya,Ganor:2008hd,Ganor:2010md,Martucci:2014ema} for related ideas.} and in general, to incorporate the $\matMy$ twist into the action we have to decompose $\matMy$ in terms of the generators
$$
\genTy=\begin{pmatrix}
1 & 1 \\ 0 & 1 \\
\end{pmatrix}
\,,\qquad
\genSy=\begin{pmatrix}
0 & -1 \\ 1 & 0 \\
\end{pmatrix}
\,,
$$
as
\be
\label{eq:Tdual}
\matMy = \genSy\genTy^{\tpowy_1}\cdots\genSy\genTy^{\tpowy_\tpowNumy},
\ee
where $\tpowy_1,\dots,\tpowy_\tpowNumy$ are integers.
The decomposition \eqref{eqn:matMyST} is not unique, but for our purposes it will be sufficient to consider the case $\tpowNumy=1$, so we will just set
\be\label{eqn:matMyST}
\matMy = \genSy\genTy^\tpowy
\,.
\ee
(We will comment on the general case in \secref{subsec:MoreGens}.)
%It will also be necessary to assume that $\tpowy$ is even ($\tpowy\in 2\Z$).\footnote{As we will see later, the low-energy reduction of the geometrically twisted $\xWS_1$ direction leads to a system that is similar to a geometrically quantized $T^2$ at level $(\tpowy-2)$. If the level is odd, there is an ambiguity in the definition of the action of the generator $T$ of the MCG, which is needed for the $\matMx$ twist.}
The $\matMy$-twist can then be inserted by adding a $1$-dimensional term $I'$ and a $0$-dimensional term $I''$ to the action as follows.
$I'$ is defined as the integral over $\xWS_1$ at $\xWS_2=0$, and is a sum
$$
I'=I_b'+I_f'
$$
of bosonic and fermionic terms.
The bosonic term is given by
\be\label{eqn:Ibprime}
I_b'=
\left.-\frac{i\tpowy}{4\pi}\int_0^1\vX_0^t\matEps d\vX_0\right\rvert_{\xWS_2=0}
+\left.\frac{i}{2\pi}\int_0^1\vX_1^t\matEps d\vX_0\right\rvert_{\xWS_2=0}\,.
\ee
The first term in \eqref{eqn:Ibprime} implements the $\genTy^\tpowy$ component of \eqref{eqn:matMyST} while the second term couples the fields at $\xWS_2=1$ to the fields at $\xWS_2=0$ and implements the $\tilde{S}$ component of $\matMy$.
From now own, any field with a subscript ``0'' or ``1'' will be implicitly understood to be evaluated at $\xWS_2=0$.

The fermionic $1$-dimensional action $I_f'$ can be designed so that if we denote by $I_f$ the fermionic part of the bulk action $I$ given in \eqref{eqn:sigmaI}, then the equations of motion derived from $I_f+I_f'$ at $\xWS_2=0$ and $\xWS_2=1$ will be equivalent to the boundary conditions \eqref{eqn:vPsiTdl}.
We take the ansatz\footnote{
In $I_f'$ we did not include mixed chirality terms of the form $\bvPsi_1^t\matL\vPsi_0$, $\vPsi_1^t\bmatL\bvPsi_0$, $\bvPsi_1^t\matLi\vPsi_1$, $\bvPsi_0^t\matLz\vPsi_0$ (with $2\times 2$ matrices $\matL$, $\bmatL$, $\matLi$ and $\matLz$) since they are not necessary.
}
\be\label{eqn:Ifprime}
I_f' =
-\frac{i}{2\pi}\int\left(
\bvPsi_1^t\matR\bvPsi_0 +\vPsi_1^t\bmatR\vPsi_0 
+\bvPsi_1^t\matRi\bvPsi_1 +\vPsi_1^t\bmatRi\vPsi_1 
+\bvPsi_0^t\matRz\bvPsi_0 +\vPsi_0^t\bmatRz\vPsi_0 
%+\bvPsi_1^t\matL\vPsi_0+\vPsi_1^t\bmatL\bvPsi_0
%+\bvPsi_1^t\matLi\vPsi_1+\bvPsi_0^t\matLz\vPsi_0
\right) d\xWS_1\,,
\ee
where $\matR$, $\bmatR$, $\matRi$, $\bmatRi$, $\matRz$ and $\bmatRz$ are $2\times 2$ antisymmetric matrices:
\be\label{eqn:matRmatRt}
\matRi^t=-\matRi\,,\qquad
\bmatRi^t=-\bmatRi\,,\qquad
\matRz^t=-\matRz\,,\qquad
\bmatRz^t=-\bmatRz\,.
\ee
To be compatible with \eqref{eqn:vPsiTdl}, the following relations must hold:
\bear
2\bmatRi &=&
-i\matSV_1-\bmatR\left(-\matEps\matE_0^t+\tpowy\Id\right)^{-1}
\,,\label{eqn:bmatRifromR}\\
2\matRi &=&
i\bmatSV_1 -\matR\left(\matEps\matE_0+\tpowy\Id\right)^{-1}
\,,\label{eqn:matRifromR}\\
2\bmatRz &=&
i\matSV_0
+\bmatR^t\left(-\matEps\matE_0^t+\tpowy\Id\right)
\,,\label{eqn:bmatRzfromR}\\
2\matRz &=&
-i\bmatSV_0
+\matR^t\left(\matEps\matE_0+\tpowy\Id\right)
\,.\label{eqn:matRzfromR}
\eear
Requiring $\matRz$ and $\bmatRz$ to be antisymmetric [as in \eqref{eqn:matRmatRt}], and using \eqref{eqn:matSV1}-\eqref{eqn:bmatSV1}] we get two equations for $\matR$ and $\bmatR$:
\bear
-2i\matSV_0 &=&
(\matE_0\matEps+\tpowy\Id)\bmatR
+\bmatR^t\left(-\matEps\matE_0^t+\tpowy\Id\right)
\,,\label{eqn:matSVbmatR}\\
2i\bmatSV_0 &=& 
\left(-\matE_0^t\matEps+\tpowy\Id\right)\matR
+\matR^t\left(\matEps\matE_0+\tpowy\Id\right)
\,.
\label{eqn:bmatSVmatR}
\eear
It can easily be checked that if \eqref{eqn:matSVbmatR}-\eqref{eqn:bmatSVmatR} are satisfied, $\matRi$ and $\bmatRi$ as given by \eqref{eqn:bmatRifromR}-\eqref{eqn:matRzfromR} are also antisymmetric.

In order to solve \eqref{eqn:matSVbmatR}-\eqref{eqn:bmatSVmatR} it is convenient to change variables as follows. Since $\matSV$ and $\bmatSV$ are symmetric and $\matG$ is symmetric and nondegenerate (as we assume), we can find matrices $\matYS$ and $\bmatYS$ so that 
\be\label{eqn:YSbYS}
\matSV=\matYS^t\matG\matYS
\,,\qquad
\bmatSV=\bmatYS^t\matG\bmatYS.
\ee
We can also require the boundary conditions
\be\label{eqn:YSbYStransf}
\matYS_1 = 
(\matE_0\matEps+\tpowy\Id)^t\matYS_0
\left[\left(\matE_0\matEps+\tpowy\Id\right)^{-1}\right]^t
\,,\qquad
\bmatYS_1 = 
(\matEps\matE_0+\tpowy\Id)
\bmatYS_0
(\matEps\matE_0+\tpowy\Id)^{-1},
\ee
which are compatible with \eqref{eqn:matSV1}-\eqref{eqn:bmatSV1}.
Then, it is not hard to check that
\bear
\bmatR &=&
i\matYS_1^t\matEps\matYS_0
=
i(\matE_0\matEps+\tpowy\Id)^{-1}
\matYS_0^t(\matE_0\matEps+\tpowy\Id)
\matEps\matYS_0
\,,\label{eqn:bmatRsol}\\
\matR &=&
i\bmatYS_1^t\matEps\bmatYS_0
=
i\left[(\matEps\matE_0+\tpowy\Id)^{-1}\right]^t
\bmatYS_0^t
\left(\matEps\matE_0+\tpowy\Id\right)^t
\matEps\bmatYS_0
\label{eqn:matRsol}
\eear
are solutions to \eqref{eqn:matSVbmatR}-\eqref{eqn:bmatSVmatR}.

The meaning of \eqref{eqn:bmatRsol}-\eqref{eqn:matRsol} becomes clearer if we change field variables to
$$
\vPsiN\equiv\matYS\vPsi,\qquad
\bvPsiN\equiv\bmatYS\bvPsiN.
$$
In terms of the new variables, the kinetic terms 
$i\vPsi^t\matSV\bpartial\vPsi$ and $i\bvPsi^t\bmatSV\partial\bvPsi$ of \eqref{eqn:sigmaI} become $i\vPsi^t\matG\bpartial\vPsi+(\cdots)$ and $i\bvPsi^t\matG\partial\bvPsi+(\cdots)$, where $(\cdots)$ are corrections to the $\vPsi^t\matK\vPsi$ and $\bvPsi^t\bmatK\bvPsi$ terms.
In terms of the new field variables, \eqref{eqn:Ifprime} can be written as
%$$
%\vPsi_1^t\bmatR\vPsi_0 =
%i\vPsiN_1^t\matYS_1^t\bmatR\matYS_0\vPsiN_0
%=i\vPsiN_1^t\matEps\vPsiN_0.
%$$
\bear
I_f' &=&
\frac{1}{2\pi}\int\left(
\bvPsiN_1^t\matEps\bvPsiN_0
+\vPsiN_1^t\matEps\vPsiN_0
-\frac{1}{2}\bvPsiN_1^t\matB_1\bvPsiN_1
-\frac{1}{2}\vPsiN_1^t\matB_1\vPsiN_1
\nn\right.\\ &&
\left.+\frac{1}{2}\bvPsiN_0^t\matB_0\bvPsiN_0
-\frac{1}{2}\tpowy\bvPsiN_0^t\matEps\bvPsiN_0
+\frac{1}{2}\vPsiN_0^t\matB_0\vPsiN_0
-\frac{1}{2}\tpowy\vPsiN_0^t\matEps\vPsiN_0
\right) d\xWS_1.
\label{eqn:IfPrimeN}
\eear
Thus, the coupling between $\vPsiN_1$ and $\vPsiN_0$ is given by the constant matrix $\matEps$. However, the disadvantage of the new variables is that formulae \eqref{eqn:delSUSY}, \eqref{eqn:bpSV}-\eqref{eqn:SVbSV} and \eqref{eqn:matK}-\eqref{eqn:matW} become more cumbersome.

We also note that the total action $I+I_b'+I_f'$ is not invariant under the SUSY transformation \eqref{eqn:delSUSY}, but this is expected with a duality wall, and a similar problem occurs with translations. Consider, for example, an $\genSy$-duality wall with $\rhoKM=i$ (which does not require a Janus configuration to match the field values at $\xWS_2=0$ to those at $\xWS_2=1$). In that case, the system is translationally invariant under $\xWS_2\rightarrow\xWS_2+\varepsilon$, but not manifestly so, because under a translation the duality wall is moved from $\xWS_2=0$ to $\xWS_2=\varepsilon$, and demonstrating invariance requires an additional duality transformation within the strip $0<\xWS_2<\varepsilon$.

Finally, to complete the action we need to add a $0$-dimensional term that couples the field $\vX(0,0)$ to $\vX(1,0)$. It takes the form
\be\label{eqn:Ipp}
I''=
\frac{i}{4\pi}(\tpowy-2)\vX(1,0)^t\matEps\matMx\vX(0,0).
\ee
%This can also be written as
%\be
%I''=
%-\frac{i}{4\pi}(\tpowy-2)\vX(0,0)^t\matEps\matMx^{-1}\vX(1,0).
%\ee
This term is necessary to reproduce the correct equations of motion at the intersection $(0,0)$ of the two duality walls. Indeed, if we denote by $I_b$ the bosonic part of $I$ from \eqref{eqn:sigmaI}, then $I_b+I_b'+I''$ leads to the bosonic boundary conditions \eqref{eqn:vXbcTdl}. Note that $\vX(1,0)$ is related to $\vX(0,0)$ by \eqref{eqn:vNdef}.
Without $I''$, the variation of the action $I_b+I_b'$ will have an unwanted term $\frac{i}{2}(\tpowy-2)\delta\vX(0,0)^t\matEps\matMx^{-1}\vN$, which would be too restrictive (leading to $\vN=0$).
Thanks to \eqref{eqn:vNdef} and the identity $\matMx^t\matEps\matMx=\matEps$ (which follows from $\det\matMx=1$), we can rewrite $I''$ as
\be\label{eqn:IppAlt}
I''=
\frac{i}{2}(\tpowy-2)\vN^t\matEps\matMx\vX(0,0).
\ee
If $\tpowy$ is odd then $I_b'+I''$ might not be invariant under \eqref{eqn:vXvKshift}. This is clearer after rewriting $I_b'+I''$ as
\be\label{eqn:IbpIpp}
I_b'+I''=
\left.-\frac{i\tpowy}{4\pi}\int_0^1\vZ_0^t\matEps d\vZ_0\right\rvert_{\xWS_2=0}
+\left.\frac{i}{2\pi}\int_0^1\vZ_1^t\matEps d\vZ_0\right\rvert_{\xWS_2=0}
+i\pi(\tpowy-2)\vN^t\matEps(\Id-\matMx)^{-1}\vN,
%-i\pi(2-\tpowy)\vN^t[(\Id-\matMx^{-1})^t]^{-1}\matEps\vN,
\ee
where $\vZ$ and $\vN$ where defined in \eqref{eqn:vZdef} and \eqref{eqn:vNdef}, and subscripts of $\vZ_0$ and $\vZ_1$ are defined similarly to those of $\vX_0$ and $\vX_1$ in \eqref{eqn:sub0notation} and \eqref{eqn:sub1notation}.
$\vZ$ is invariant under the discrete shift \eqref{eqn:vXvKshift}, but the last term on the RHS of \eqref{eqn:IbpIpp} changes by
$i\pi(\tpowy-2)\left(\vN^t-\vK^t\right)\matEps(\Id-\matMx)\vK$
$\pmod{2\pi i\Z}$,
which might be an odd multiple of $i\pi$ if $\tpowy$ is odd. We therefore require $\tpowy$ to be even.\footnote{We can, in fact, allow $\tpowy$ to be odd if we require $\matMx\in\Gamma(2)$, where $\Gamma(2)$ is the {\it principal congruence subgroup} of $\SL(2,\Z)$ with $\xa\equiv\xd\equiv 1\pmod{2}$ and $\xb\equiv\xc\equiv 0\pmod{2}$, and it has index 6 inside $\SL(2,\Z)$.
\label{foot:GammaTwo}} 
(We will see another aspect of this requirement in \secref{subsec:MT}.)

At this point one may wonder if an additional term quadratic in $\vN$ can be added to the action. This, however, will violate locality, since $\vN$ can only be calculated by continuously following the value of $\vX$ from $\xWS_1=0$ to $\xWS_1=1$.
We will address the issue of locality in more details in \appref{app:Locality}, where we will also argue that $I''$ is necessary to incorporate the discrete target space periodicity \eqref{eqn:vXvKshift} in a local way, at least for even $\tpowy$.

% --------------------------------------------------------------
\subsection{The partition function}
\label{subsec:PFwithSqrt}

Thus, we have completed the construction of the action, which is given by the sum of 
\eqref{eqn:sigmaI}, \eqref{eqn:Ibprime}, \eqref{eqn:Ifprime} and \eqref{eqn:Ipp}:
$$
I_{\text{tot}} = I+I_b'+I_f'+I'',
$$
where $I$ is obtained by $\SL(2,\R)_\rhoKM\times \SL(2,\R)_\tauCS$ transformations on the diagonal model \eqref{eqn:sigmaIdiag}.
The model we have constructed consists of the bulk 2d action \eqref{eqn:sigmaI} and two duality walls inserted between $\xWS_1=0$ and $\xWS_1=1$ and between $\xWS_2=0$ and $\xWS_2=1$. The wall at $\xWS_1=0$ is given by nontrivial boundary conditions \eqref{eqn:vPsiMZ}-\eqref{eqn:vXMZ}, and the wall at $\xWS_2=0$ is described by the 1d action $I'$ given by \eqref{eqn:Ibprime}-\eqref{eqn:Ifprime}. At the intersection of the walls we have the additional 0d term $I''$ given by \eqref{eqn:Ipp}. This is depicted in \figref{fig:BulkAndWalls}.
% ----------------------------
\begin{figure}[t]
\begin{picture}(400,160)
\put(50,30){\begin{picture}(0,0)
\thinlines
\putRect{0}{0}{300}{100}

\put(300,0){\vector(1,0){20}}\put(322,-1){$\xWS_1$}
\put(0,100){\vector(0,1){20}}\put(-2,122){$\xWS_2$}
\thicklines
\put(0,1){\circle*{6}}
\put(16,16){\vector(-1,-1){12}}\put(18,13){$I''$}

\put(0,2){\line(1,0){300}}
\multiput(2,0)(2,0){149}{\line(0,1){2}}

\put(70,60){Bulk: $I=\int(\cdots)d\xWS_1 d\xWS_2$}

\put(70,33){Duality wall:}
\put(70,18){$I'=I_b'+I_f' = \int(\cdots)d\xWS_1$}
\qbezier(190,20)(195,15)(195,2)
\put(195,2){\vector(0,-1){0}}

\put(50,-10){$\tauCS(\xWS_1)$}\put(77,-7){\vector(1,0){60}}
\put(-30,20){$\rhoKM(\xWS_2)$}\put(-25,27){\vector(0,1){32}}

% duality connection $\matMy$
\qbezier(250,0)(260,-20)(270,0)
\qbezier(270,0)(280,20)(280,50)
\put(280,50){\vector(0,1){0}}
\put(282,42){$\matMy$}
\qbezier(250,100)(260,120)(270,100)
\qbezier(270,100)(280,80)(280,50)

% duality connection $\matMx$
\qbezier(300,50)(320,60)(300,70)
\qbezier(300,70)(280,80)(150,80)
\put(150,80){\vector(1,0){0}}
\put(142,84){$\matMx$}
\qbezier(0,50)(-20,60)(0,70)
\qbezier(0,70)(20,80)(150,80)

\thicklines
\end{picture}}
\end{picture}
\caption{
Our field theory is defined on a $T^2$ parametrized by $0\le\xWS_1,\xWS_2\le 1$,
with $\tauCS$ varying as a function of $\xWS_1$, and $\rhoKM$ varying as a function of $\xWS_2$. Duality walls connect $\xWS_1=0$ to $\xWS_1=1$ [with the geometrical $\matMx\in \SL(2,\Z)$] and $\xWS_2=0$ to $\xWS_2=1$ [with the T-duality $\matMy\in \SL(2,\Z)$], and their intersection supports a 0d action $I''$.
}
\label{fig:BulkAndWalls}
\end{figure}
% ----------------------------

The partition function is defined as
\be\label{eqn:PFdefWithSqrt}
\PF=\frac{1}{\sqrt{|\GrSt|}}\int
e^{-I-I_b'-I_f'-I''}\cD\vX\cD\vPsi\cD\bvPsi.
\ee
The prefactor $\left.1\right/\sqrt{|\GrSt|}$ is necessary in order to have a properly normalized T-duality wall at $\xWS_2=1$. This can be argued by considering $\xWS_2$ as the time direction and realizing the T-duality wall at $\xWS_2=1$ as a unitary transformation on the Hilbert space. We will outline this approach in more detail in the \secref{sec:CSMT}, but first we will calculate $\PF$ by evaluating the one-loop determinants in the action.

% ==============================================================

\section{Calculating the partition function}
\label{sec:PF}

We will now calculate the partition function for the model we have constructed in \secref{subsec:GeomTwist}-\secref{subsec:Ttwist}.
First, it is convenient to express the bosonic 1d and 0d terms $I_b'+I''$ in terms of $\vZ$ and $\vN$ as in \eqref{eqn:IbpIpp}.
The advantage is that $\vZ$ satisfies linear boundary conditions \eqref{eqn:vXMZ} at $\xWS_1$, unlike $\vX$ whose boundary conditions include an affine term \eqref{eqn:vNdef}.
The discrete $\vN$ decouples and the partition function now takes the form
\be\label{eqn:PFFPFsum}
\PF=
\frac{1}{\sqrt{|\GrSt|}}\left(\frac{\FPF_f}{\FPF_b}\right)
\sum_{\vN\in\GrSt}
\exp\left[
i\pi(2-\tpowy)\vN^t\matEps(\Id-\matMx)^{-1}\vN
\right],
\ee
where the finite abelian group $\GrSt$ was defined in \eqref{eqn:GrStDef}, and where
$\FPF_b^{-1}$ is the result of integrating over fluctuations of $\vZ$ and $\FPF_f$ is the result of integrating over $\vPsi$ and $\bvPsi$.
%Schematically,
%$$
%\FPF_b=\det\lbrack
%2\matG\bpartial\partial
%+(\bpartial\matE^t)\partial
%+(\partial\matE)\bpartial
%\rbrack^{1/2}
%$$
%and
%$$
%\FPF_f=\Pfaffian\begin{pmatrix}
%-i(\matSV\bpartial+\matK+\tfrac{1}{2}\bpartial\matSV) & -\tfrac{1}{2}\matW \\
%\tfrac{1}{2}\matW^t & -i(\bmatSV\partial+\bmatK+\tfrac{1}{2}\partial\bmatSV) \\
%\end{pmatrix}
%$$
%[See \eqref{eqn:sigmaI} and \eqref{eqn:bosonicEOM}.] 

% --------------------------------------------------------------
\subsection{The bosonic one-loop determinant}
\label{subsec:Zdet}

Let us discuss the bosonic part first.
To compute $\FPF_b$ we need a metric on the space of fluctuations, which we take to be
\be\label{eqn:deltaZmetric}
\|\delta\vZ\|^2=
\int \delta\vZ^t\matG\delta\vZ d^2\xWS
\,.
\ee
Given a choice of vielbein $\matF$ as in \eqref{eqn:Vielbein}, we can change variables to 
$\vZNN\equiv\matF\vZ$, which brings the metric $\|\delta\vZ\|^2$ into the normal form. We will now show how to calculate $\FPF_b$ for the model whose 2d bulk is obtained by acting with $\SL(2,\R)_\tauCS\times \SL(2,\R)_\rhoKM$ transformations, defined in \secref{subsec:MoreDJSols}, on the diagonal model of \eqref{eqn:sigmaIdiag}. Since $\vZ$, $\vPsi$ and $\bvPsi$ are unaware of the periodicity of the target space $T^2$, their boundary conditions do not require the transformations to be restricted to $\SL(2,\Z)$, and so we can return to the diagonal model and calculate the fluctuations there.

In this way the model constructed at the end of \secref{subsec:MoreDJSols}, with $\tauCS$ and $\rhoKM$ varying along semicircles that are invariant under $\matMx$ and $\matMy$, respectively, gets converted to a model in which $\tauCS$ and $\rhoKM$ vary along the imaginary axis. In this model, whose bulk action is given by \eqref{eqn:sigmaIdiag}, the boundary conditions along the $\xWS_1$ and $\xWS_2$ directions are determined by the eigenvalues of $\matMx$ and $\matMy$,
$$
\matMx=\begin{pmatrix}
\xa & \xb \\ 
\xc & \xd \\
\end{pmatrix}
\rightarrow\begin{pmatrix}
e^{\evMx} & 0 \\
0 & e^{-\evMx} \\
\end{pmatrix}
\,,\qquad
\matMy=\begin{pmatrix}
\ya & \yb \\ 
\yc & \yd \\
\end{pmatrix}
\rightarrow\begin{pmatrix}
e^{\evMy} & 0 \\
0 & e^{-\evMy} \\
\end{pmatrix}\,,
$$
where
$$
e^{\evMx}=\frac{1}{2}\left(\xa+\xd+\sqrt{(\xa+\xd)^2-4}\right)
\,,\qquad
e^{\evMy}=\frac{1}{2}\left(\ya+\yd+\sqrt{\left(\ya+\yd\right)^2-4}\right)
\,.
$$
Then, \eqref{eqn:tauCS01} and \eqref{eqn:rhoKM01} become
\be\label{eqn:bcCSKMevM}
\tauCS(1) = e^{2\evMx}\tauCS(0)
\,,\qquad
\rhoKM(1) = e^{2\evMy}\rhoKM(0)
\,,
\ee
and the boundary conditions \eqref{eqn:vZbc} become
\be\label{eqn:bcfZevMx}
\fZ^1(0,\xWS_2)=e^{\evMx}\fZ^1(1,\xWS_2)\,,\qquad
\fZ^2(0,\xWS_2)=e^{-\evMx}\fZ^2(1,\xWS_2)\,,
\ee
where $\fZ^1$ and $\fZ^2$ are the components of $\vZ$,
while \eqref{eqn:vXbcTdl} becomes
\be\label{eqn:bcvZevMy}
\bpartial\vZ(\xWS_1,1)=e^{-\evMy}\bpartial\vZ(\xWS_1,0)
\,,\qquad
\partial\vZ(\xWS_1,1)
=e^{-\evMy}\partial\vZ(\xWS_1,0),
\ee
which we can solve simply by requiring
$$
\vZ(\xWS_1,1)=e^{-\evMy}\vZ(\xWS_1,0).
$$
In \eqref{eqn:sigmaIdiag}, we can take the vielbein to be diagonal as well, so that the change of variables from $\vZ$ to the normalized $\vZNN$ is given by
\be\label{eqn:fZNN12}
\fZNN^1=\rhoKM_2^{\frac{1}{2}}\tauCS_2^{-\frac{1}{2}}\fZ^1\,,
\qquad
\fZNN^2=\rhoKM_2^{\frac{1}{2}}\tauCS_2^{\frac{1}{2}}\fZ^2\,.
\ee
The boundary conditions \eqref{eqn:bcCSKMevM}, \eqref{eqn:bcfZevMx} and \eqref{eqn:bcvZevMy} imply that $\vZNN$ is periodic in $\xWS_1$ and $\xWS_2$.
Substituting \eqref{eqn:fZNN12} into the bosonic part of \eqref{eqn:sigmaIdiag}, and integrating by parts, we get the bosonic part of the action in the form
\bear
I_b &=&
\frac{1}{\pi}\int\left(
\partial\vZ^t\bpartial\vZ
+\tfrac{1}{4}\vZ^t\matMMb^2\vZ
\right)d^2\xWS
\,,
\label{eqn:IbMMb}
\eear
where the effective mass matrix (squared) is defined as
\bear
\matMMb^2 = \begin{pmatrix}
\evMMb_1^2 & 0 \\
0 & \evMMb_2^2 \\
\end{pmatrix}
\eear
with the eigenvalues given in terms of  $\tauCS_2(\xWS_1)$ and $\rhoKM_2(\xWS_2)$ by
\bear
\evMMb_1^2 &=&
\frac{\ddrhoKM_2}{2\rhoKM_2}
-\frac{\drhoKM_2^2}{4\rhoKM_2^2}
-\frac{\ddtauCS_2}{2\tauCS_2}
+\frac{3{\dtauCS_2}^2}{4\tauCS_2^2}
\,,\label{eqn:evMMb1}\\
\evMMb_2^2 &=&
\frac{\ddrhoKM_2}{2\rhoKM_2}
-\frac{\drhoKM_2^2}{4\rhoKM_2^2}
+\frac{\ddtauCS_2}{2\tauCS_2}
-\frac{{\dtauCS_2}^2}{4\tauCS_2^2}
\,.\label{eqn:evMMb2}
\eear
The contribution of the fluctuations of $\vZ$ to the partition function are now expressed in terms of the product of the eigenvalues of the operator $-2\partial\bpartial+\matMMb^2$.
These take the form of sums of two 1d Schr\"odinger problems as follows.
Define
\be\label{eqn:defVSch}
\VSchKM(\xWS_2)\equiv
\frac{\ddot{\rhoKM}_2}{2\rhoKM_2}
-\frac{\drhoKM_2^2}{4\rhoKM_2^2}
\,,\qquad
\VSchCSa(\xWS_1)\equiv
-\frac{\tauCS_2''}{2\tauCS_2}
+\frac{3{\tauCS_2'}^2}{4\tauCS_2^2}
\,,\qquad
\VSchCSb(\xWS_1)\equiv
\frac{\tauCS_2''}{2\tauCS_2}
-\frac{{\tauCS_2'}^2}{4\tauCS_2^2}
\,.
\ee
The boundary conditions \eqref{eqn:bcCSKMevM} ensure that $\VSchKM$, $\VSchCSa$ and $\VSchCSb$ are periodic.
We now need to solve three separate 1d Schr\"odinger problems with the following periodic potentials defined on the circle parametrized by the periodic coordinate $0\le\xSch<1$:
\be\label{eqn:evSchDef}
\begin{aligned}
-\frac{d}{d\xSch^2}+\VSchKM(x) &\,\,
\text{with eigenvalues $\evSchKM_0, \evSchKM_1, \evSchKM_2, \ldots$} \\
-\frac{d}{d\xSch^2}+\VSchCSa(x) &\,\,
\text{with eigenvalues $\evSchCSa_0, \evSchCSa_1, \evSchCSa_2, \ldots$} \\
-\frac{d}{d\xSch^2}+\VSchCSb(x) &\,\,
\text{with eigenvalues $\evSchCSb_0, \evSchCSb_1, \evSchCSb_2, \ldots$} \\
\end{aligned}
\ee
Then, $\evSchKM_i+\evSchCSa_j$ are the eigenvalues of the fluctuations of $\fZNN^1$, and $\evSchKM_i+\evSchCSb_j$  are the eigenvalues of the fluctuations of $\fZNN^2$ (with $0\le i,j<\infty$).
We will see below that the eigenvalues of $\VSchCSa$ and $\VSchCSb$ are the same, $\evSchCSa_j=\evSchCSb_j$, so that, formally, we can write
\be\label{eqn:FPFbEval}
\FPF_b = 
\prod_{i,j}\left\{(\evSchKM_i+\evSchCSa_j)(\evSchKM_i+\evSchCSb_j)\right\}^{1/2}
=\prod_{i,j}(\evSchKM_i+\evSchCSa_j).
\ee
The eigenvalues $\evSchKM_j$, $\evSchCSa_j=\evSchCSb_j$ are all positive.
That the eigenvalues are nonnegative is a consequence of supersymmetry, and the fact that the $\xWS_1$-Schr\"odinger problem and $\xWS_2$-Schr\"odinger problem are decoupled. Thus, to show that $\evSchKM_j\ge 0$ we can look at the model with constant $\tauCS_2$ and $\matMx=\Id$. This model is translationally invariant in $\xWS_1$, which we can identify as time. Because of supersymmetry, all energy states are nonnegative, and therefore all single-particle eigenvalues $\evSchKM_j$ must be nonnegative. Similarly, to argue that $\evSchCSa_j$ and $\evSchCSb_j$ are nonnegative we take $\rhoKM_2$ to be constant and $\matMy=\Id$ and identify $\xWS_2$ as the time direction. We can also argue more directly that the eigenvalues are positive, by noting that each of the potentials $\VSchKM$, $\VSchCSa$ and $\VSchCSb$ can be written in terms of a superpotential $\WSch$ as $\WSch^2+d\WSch/d\xSch$, with 
\be\label{eqn:WSchValues}
\WSch\rightarrow
\text{
$\frac{\drhoKM_2}{2\rhoKM_2}$, or $\pm\frac{\dtauCS_2}{2\tauCS_2}$.
}
\ee
The Hamiltonian can then be expressed in terms of the superpotential as
\cite{Witten:1981nf,Cooper:1982dm,Cooper:1994eh}
$$
-\frac{d}{d\xSch^2}+\WSch^2+\frac{d\WSch}{d\xSch}=\QSch^\dagger\QSch,\qquad
\text{where}\qquad
\QSch\equiv\frac{d}{d\xSch}-\WSch(\xSch),\quad
\QSch^\dagger =-\frac{d}{d\xSch}-\WSch(\xSch)\,.
$$
It is now clear that the eigenvalues are nonnegative, and it is also clear that a zero eigenvalue would require a nontrivial kernel for either $\QSch$ or $\QSch^\dagger$, but for $\evMx$ and $\evMy$ both nonzero, and for $\WSch$ given in \eqref{eqn:WSchValues}, there are no periodic zero eigenvalues for $\QSch$ or $\QSch^\dagger$.
This 1d supersymmetry is also the reason why $\evSchCSa_j=\evSchCSb_j$.
Setting
$$
\WSch=\frac{\dtauCS_2}{2\tauCS_2},
$$
we find
$$
\QSch\QSch^\dagger = -\frac{d}{d\xSch^2}+\VSchCSa(x)
\,,\qquad
\QSch^\dagger\QSch = -\frac{d}{d\xSch^2}+\VSchCSb(x),
$$
and since we have established that the kernels of both $\QSch$ and $\QSch^\dagger$ are trivial, it follows that the eigenvalues of $\QSch\QSch^\dagger$ and $\QSch^\dagger\QSch$ are equal, as is well-known.
Thus, the double-Janus $\sigma$-model realizes a second quantized version of supersymmetric quantum mechanics.

We note that there is a special case where the three effective Schr\"odinger potentials are positive constants.
For this we take
\be\label{eqn:taurhoexp}
\tauCS(\xWS_1)=e^{2\evMx\xWS_1},\qquad
\rhoKM(\xWS_2)=e^{2\evMy\xWS_2},
\ee
which satisfies the boundary conditions \eqref{eqn:bcCSKMevM} and leads to
$$
\VSchKM=\evMy^2,
\qquad
\VSchCSa=\VSchCSb=\evMx^2,
$$
and the action \eqref{eqn:IbMMb} becomes that of two free, translationally invariant, massive bosons:
\bear
I_b &=&
\frac{1}{\pi}\int\left(
\partial\vZ^t\bpartial\vZ
+\frac{1}{4}\effmass^2\vZ^t\vZ
\right)d^2\xWS
\,,
\label{eqn:IbMMbC}
\eear
with constant mass squared given in terms of the eigenvalues of the $\SL(2,\Z)$ twist matrices $\matMx$ and $\matMy$ by
\be\label{eqn:effm2ev}
\effmass^2=\evMy^2+\evMx^2.
\ee
Up to an unimportant constant, then
\be\label{eqn:FPFbeffm}
\FPF_b=
%\propto
\det\left(-\partial\bpartial+\frac{1}{4}\effmass^2\right).
\ee
% {\bf (CHECK LITERATURE ON FREE MASSIVE BOSON!)}
%\be\label{eqn:FPFbeffm}
%\FPF_b\propto
%\det(-\partial\bpartial+\frac{\effmass^2}{4})\propto
%\prod_{n_1,n_2=-\infty}^\infty
%(n_1^2+n_2^2+\frac{\effmass^2}{4\pi^2})
%\rightarrow
%\prod_{n=-\infty}^\infty\frac{1}{2\sinh(\frac{1}{2}\sqrt{4\pi^2 n^2+\effmass^2})}
%\,.
%\ee
%where the last product is obtained by a standard Hilbert space approach to the partition %function of a free massive boson.

% --------------------------------------------------------------
\subsection{The fermionic one-loop determinant}
\label{subsec:Psidet}

To define the fermionic determinant we again need a metric on the space of fluctuations, similarly to \eqref{eqn:deltaZmetric}.
We choose it to be given in terms of $\matSV$ and $\bmatSV$, for $\vPsi$ and $\bvPsi$, respectively. Then, using \eqref{eqn:YSbYS}, and the vielbein, we can redefine
$$
\vPsiNNN\equiv\matF\matYS\vPsi,
\qquad
\bvPsiNNN\equiv\matF\bmatYS\bvPsi,
$$
for which the kinetic term in the bulk action is normalized.
As in \secref{subsec:Zdet}, we can then change variables to the diagonal model \eqref{eqn:sigmaIdiag}.
It is convenient to use, instead of $\vPsiNNN$ which has kinetic term $i\vPsiNNN^t\bpartial\vPsiNNN$, another set of variables
\be\label{eqn:PsiNNdef}
\fPsiNN{1}\equiv\rhoKM_2^{1/2}\tauCS_2^{-1/2}\fPsi^1,\quad
\fPsiNN{2}\equiv\rhoKM_2^{1/2}\tauCS_2^{1/2}\fPsi^2,\qquad
\bfPsiNN{1}\equiv\rhoKM_2^{1/2}\tauCS_2^{-1/2}\bfPsi^1,\quad
\bfPsiNN{2}\equiv\rhoKM_2^{1/2}\tauCS_2^{1/2}\bfPsi^2,
\ee
similarly to \eqref{eqn:fZNN12}.
These new fields are periodic in $\xWS_1$ and $\xWS_2$ and the fermionic part of \eqref{eqn:sigmaIdiag} is given in terms of them by
%\bear
%I_f &=&
%\tfrac{1}{\pi}\int\Bigl\lbrack
%2i\fPsiNN{1}\bpartial\fPsiNN{2}
%+2i\bfPsiNN{1}\partial\bfPsiNN{2}
%+(\frac{\drhoKM_2}{2\rhoKM_2}+\frac{i\tauCS_2'}{2\tauCS_2})\bfPsiNN{1}\fPsiNN{2}
%+(\frac{\drhoKM_2}{2\rhoKM_2}-\frac{i\tauCS_2'}{2\tauCS_2})\bfPsiNN{2}\fPsiNN{1}
%\Bigr\rbrack d^2\xWS\,.
%\nn\\ &&
%\label{eqn:IfNN}
%\eear
\bear
I_f &=&
\frac{1}{\pi}\int\left[
2i\fPsiNN{1}\bpartial\fPsiNN{2}
+2i\bfPsiNN{1}\partial\bfPsiNN{2}
+\left(\frac{i\dtauCS_2}{2\tauCS_2}+\frac{\drhoKM_2}{2\rhoKM_2}\right)\bfPsiNN{1}\fPsiNN{2}
+\left(\frac{i\dtauCS_2}{2\tauCS_2}-\frac{\drhoKM_2}{2\rhoKM_2}\right)\fPsiNN{1}\bfPsiNN{2}
\right] d^2\xWS\,.
\nn\\ &&
\label{eqn:IfNN}
\eear

The fermionic determinant is now expressed, formally, as
%$$
%\FPF_f=\det\begin{pmatrix}
%(\frac{\drhoKM_2}{4\rhoKM_2}-\frac{i\tauCS_2'}{4\tauCS_2}) & i\bpartial \\
%-i\partial  & (\frac{\drhoKM_2}{4\rhoKM_2}+\frac{i\tauCS_2'}{4\tauCS_2})\\
%\end{pmatrix}.
%$$
%Or
\be\label{eqn:FPFfdet}
\FPF_f=\det\begin{pmatrix}
 i\bpartial & \frac{i\dtauCS_2}{4\tauCS_2}-\frac{\drhoKM_2}{4\rhoKM_2}\\
\frac{i\dtauCS_2}{4\tauCS_2}+\frac{\drhoKM_2}{4\rhoKM_2} & i\partial\\
\end{pmatrix}.
\ee
To calculate the determinant, we use the well known identity for the determinant of a $(2N)\times(2N)$ matrix given in $4$ blocks of $N\times N$ matrices $A,B,C,D$, with $C$ invertible, as
%\footnote{It can easily be proven by multiplying the matrix on the left by $\begin{pmatrix} \Id & -A^{-1}C \\ 0 & \Id \\ \end{pmatrix}$.}
\be\label{eqn:detABCD}
\det\begin{pmatrix} A & B \\ C & D \\ \end{pmatrix}
=\det\left(A C^{-1} D C - B C\right)
\xrightarrow{\text{if $CD=DC$}}
\det(A D - C B).
\ee
By choosing a different representation of 2d Dirac matrices\footnote{
The matrix in \eqref{eqn:FPFfdet} can be written in terms of Pauli matrices as $\frac{i}{2}\Id\partial_1-\frac{1}{2}\Pauli_3\partial_2+i\Pauli_1\frac{\dtauCS_2}{4\tauCS_2}-i\Pauli_2\frac{\drhoKM_2}{4\rhoKM_2}$ and we got \eqref{eqn:FPFfdetAlt} by rotating $\Pauli_1\rightarrow\Pauli_3$ and $\Pauli_3\rightarrow-\Pauli_1$.} we rewrite \eqref{eqn:FPFfdet} as
\be\label{eqn:FPFfdetAlt}
\FPF_f=\det\begin{pmatrix}
\frac{i}{2}\partial_1+\frac{i\dtauCS_2}{4\tauCS_2}\quad & 
\frac{1}{2}\partial_2-\frac{\drhoKM_2}{4\rhoKM_2}
\\
\frac{1}{2}\partial_2+\frac{\drhoKM_2}{4\rhoKM_2}\quad & 
\frac{i}{2}\partial_1-\frac{i\dtauCS_2}{4\tauCS_2}\\
\end{pmatrix}.
\ee
Applying the identity \eqref{eqn:detABCD} to \eqref{eqn:FPFfdet} (with commuting $C\equiv\frac{1}{2}\partial_2+\frac{\drhoKM_2}{4\rhoKM_2}$ and $D\equiv\frac{i}{2}\partial_1-\frac{i\tauCS_2'}{4\tauCS_2}$), we get
\bear
\FPF_f &=&
\det\left(
-\frac{1}{4}\partial_1^2
-\frac{1}{4}\partial_2^2
+\frac{\ddrhoKM_2}{8\rhoKM_2}
+\frac{\ddtauCS_2}{8\tauCS_2}
-\frac{{\dtauCS_2}^2}{16\tauCS_2^2}
-\frac{\drhoKM_2^2}{16\rhoKM_2^2}
\right)
=\det\left(-\bpartial\partial+\frac{1}{4}\evMMb_2^2\right),
\label{eqn:FPFfDetpbp}
\eear
where $\evMMb_2^2$ was defined in \eqref{eqn:evMMb2}. Following the arguments at the end of \secref{subsec:Zdet}, and using \eqref{eqn:FPFbEval}, we find that
$$
\FPF_f=
\prod_{i,j}\left(\evSchKM_i+\evSchCSa_j\right)=\FPF_b.
$$
For example, in the special case \eqref{eqn:taurhoexp}, we get constant
$$
\frac{i\tauCS_2'}{4\tauCS_2}+\frac{\drhoKM_2}{4\rhoKM_2} = 
\frac{1}{2}\left(\evMy+i\evMx\right),
$$
and then
\be\label{eqn:FPFfeffm}
\FPF_f=\det\left(-\bpartial\partial+\frac{1}{4}\effmass^2\right),
\ee
with $\effmass^2$ given by \eqref{eqn:effm2ev}.
Comparing \eqref{eqn:FPFfeffm} to \eqref{eqn:FPFbeffm}, we find
$\FPF_f=\FPF_b$, as expected.

We conclude that the partition function \eqref{eqn:PFFPFsum} reduces to
\be\label{eqn:PFFPFsumRed}
\boxed{\PF=
\frac{1}{\sqrt{|\GrSt|}}
\sum_{\vN\in\GrSt}
\exp\left\lbrack
i\pi(2-\tpowy)\vN^t\matEps(\Id-\matMx)^{-1}\vN
\right\rbrack}.
\ee
This is essentially a quadratic Gauss sum, and we will now see how this double-Janus configuration naturally leads to the Quadratic Reciprocity.

% ==============================================================

\section{Connections with abelian Chern-Simons theory and strings on a mapping torus}
\label{sec:CSMT}

We now return to the matter of normalization of the partition function $\mathcal{Z}$.
In particular, we need to explain the prefactor $1/\sqrt{|\GrSt|}$ that appears in \eqref{eqn:PFdefWithSqrt} and \eqref{eqn:PFFPFsumRed}.
This factor will be necessary to reproduce the Landsberg-Schaar identity in \secref{subsec:RecoverLS}, and it can be argued by treating $\xWS_2$ and ``time'' and realizing the T-duality $\matMy$ in \eqref{eq:Tdual} as an operator on a Hilbert space of dimension $|\GrSt|$.
%We will see that the Hilbert space of ground states has $|\GrSt|$ states, and we will argue that $1/\sqrt{|\GrSt|}$ is required to normalize the T-duality operator $\matMy$. We begin by relating the problem of counting ground states of $I_{\text{tot}}$ to the geometry of a mapping torus.

We will argue that, up to a known phase, \eqref{eqn:PFFPFsumRed} can be recast as the trace of the operators representing $M=ST^\tpowy\in\SL(2,\Z)$ in a certain $|\GrSt|$-dimensional representation of $\SL(2,\Z)$. We will give two equivalent constructions and physical interpretations for this representation, as a Hilbert space of ground states, one in terms of low-energy strings on a mapping-torus target space, and another in terms of an effective abelian Chern-Simons theory on $T^2$. The first construction can be expressed in terms of the element in $\matMx$ explicitly, and it directly leads to \eqref{eqn:PFFPFsumRed}, but the construction of matrix elements of $\mathscr{S}$ and $\mathscr{T}$, representing $S$ and $T$ in $\SL(2,\Z)$ respectively, will only be given up to independent $(\pm)$ signs -- an ambiguity that has a physical origin. The second construction relies on a particular decomposition of $\matMx$ as
\be\label{eqn:MxST}
\matMx = 
\genSx\genTx^{\tpowx_1}
\genSx\genTx^{\tpowx_2}
\cdots
\genSx\genTx^{\tpowx_\tpowNumx}.
\ee
We note that this decomposition is unique if $\matMx$ belongs to the subgroup $\langle\genSx,\genTx^2\rangle$ generated by $\genSx$ and $\genTx^2$ [and isomorphic to the {\it Hecke congruence subgroup} $\Gamma_0(2)\subset\SL(2,\Z)$], as we will assume in \secref{subsec:MST2gen}.

The $\SL(2,\Z)$ representations that we will need belong to the subclass of representations that appear in the theory of {\it modular tensor category}\footnote{Mathematically it is equivalent defined as a {\it ribbon fusion category}, or a {\it braided fusion category} with a {\it spherical structure}, whose modular $\mathcal{S}$ matrix is invertible.}, i.e., the representations one obtains by studying Chern-Simons theory with a compact abelian gauge group.
The physical interpretation presented in this section is partly based on results from \cite{Ganor:2014pha}.
More recently a much deeper theory related to nonabelian Chern-Simons was developed in \cite{Gukov:2016njj,Cheng:2018vpl}, related to the study of the 6d $(2,0)$-theory on plumbed $3$-manifolds, including mapping tori, but our case is somewhat different.
We will begin with a few generalities about such representations, which are defined through a quadratic form on a finite abelian group. (In general, the classification of unimodular, symmetric quadratic forms on finite abelian groups \cite{Wall} is equivalent to a classification of {\it pointed modular tensor categories}, i.e., theories of ``abelian anyons'' in physical terminology \cite{Anyon, wang}. For additional recent insight on abelian Chern-Simons theory and abelian anyons, see \cite{Moore,Tachikawa,Gomis}.)

% -------------------------------------------------------------

\subsection{Representations of \texorpdfstring{$\SL(2,\Z)$}{} from quadratic forms on abelian groups}
\label{subsec:WeilReps}

In this subsection we will denote the standard $\SL(2,\Z)$ generators by $\mcS$ and $\mcT$, in order to distinguish the abstract discussion from the concrete $S,T$.
The construction, known as the {\it Weil representation}\footnote{Most generally it is defined for $Mp(2n)$, a double cover of $Sp(2n)$, over an arbitrary local or finite field.} \cite{Williams, Deloup2,Cheng:2018vpl} begins with a finite abelian group $\AG$ on which a quadratic form $\qf(\cdot)$ is defined.
In our case the abelian group is $\AG\simeq\GrSt$ and the complex vector space of the representation is the group algebra $\C[\AG]$, which we identify with the Hilbert space spanned by a basis of states of the form $\ket{\mcV}$, with $\mcV\in\AG$. The quadratic form $\qf(\cdot)$ is required to take values in $\Q/\Z$, i.e., $\qf(\mcV)$ is a rational number up to an undetermined integer part. Thus, the phase $\exp[2\pi i\qf(\mcV)]$ is well-defined for every element $\mcV\in\AG$. From $\qf(\cdot)$ we then construct a symmetric {\it bilinear form}\footnote{Conversely, given a bilinear form $\qBf(\cdot,\cdot)$ (not necessarily symmetric), if a function $\qf$ on $\AG$ satisfies $\qBf(\mcU,\mcV)=\qf(\mcU+\mcV)-\qf(\mcU)-\qf(\mcV)+\qf(0),\,\forall\, \mcU,\mcV\in\AG$, then $\qf$ is called a {\it quadratic refinement} of $\qBf(\cdot,\cdot)$ \cite{Moore,Tachikawa,Gomis}.}:
$$
\qBf(\mcU,\mcV)\equiv\qf(\mcU+\mcV)-\qf(\mcU)-\qf(\mcV),\qquad
\mcU,\mcV\in\AG.
$$
We assume that the abelian group $\AG$ is given as a quotient of a lattice by a sublattice $\AG=\latAG/\latAG'$ and $\qf$ descends from a quadratic form $\qfL$ on $\latAG$ that takes integer values on $\latAG'$ [so that for any basis $\{\mcVRep_i\}$ of $\latAG'$, the associated bilinear form $\qBL$ is represented by a symmetric matrix with integer elements and even integers on the diagonal].
Then, the Weil representation constructed from this data is given by the action of the $\SL(2,\Z)$ generators $\mcS$ and $\mcT$ on the basis vectors $\ket{\mcV}$ of Hilbert space as follows:
\be
\label{eqn:WeilR}
\mcT\ket{\mcV} = 
e^{-i\phaseW}e^{2\pi i\qf(\mcV)}\ket{\mcV},
\qquad
\mcS\ket{\mcV}
=
\frac{1}{\sqrt{|\AG|}}
\sum_{\mcU\in\AG} 
e^{-2\pi i\qBf(\mcV,\mcU)}
\ket{\mcU},
\ee
where $|\AG|$ is the number of elements in $\AG$ and the phase $\phaseW$ is given by $\frac{\pi}{12}\sig(\qf)$ \cite{wang, signature}, where $\sig(\qf)$ is the {\it signature} of the quadratic form (i.e., the difference between the number of positive and negative eigenvalues of the matrix representing the bilinear form $\qBL$ in any basis $\{\mcVRep_i\}$ of $\latAG'$). The phase is also given by the cubic root of the {\it Gauss-Milgram sum}: \cite{milgram,Williams,signature}:
$$
e^{-3i\phaseW}=e^{-\frac{1}{4}\pi i\sig(\qf)}
=\frac{1}{\sqrt{|\AG|}}\sum_{\mcV\in\AG}e^{-2\pi i\qf(\mcV)}\,.
$$

In our case, there are two ways to describe $\qf(\cdot)$ on $\AG$.
In the first, we decompose $\matMx\in\SL(2,\Z)$ into $\genSx$ and $\genTx$ generators as in \eqref{eqn:MxST}, and we assume that all the powers are even, so that $\tpowx_i=2\thalfpowx_i$ for $\thalfpowx_i\in\Z$.
We then construct the $\tpowNumx\times\tpowNumx$ symmetric matrix
\be
\label{eqn:KcplM}
\matKcpl\equiv
\begin{pmatrix}
\tpowx_1 & -1 & 0 &  & -1 \\
-1 & \ddots & \ddots & \ddots & \\
0 & \ddots & \ddots & \ddots & 0 \\
&\ddots  & \ddots & \ddots & -1 \\
-1 &  & 0 & -1 & \tpowx_\tpowNumx \\
\end{pmatrix}\,,
\ee
with $\{\tpowx_i\}$ along the diagonal, and $(-1)$'s on the $(i,i+1)$ and $(1,\tpowNumx)$ places.
For the special case $\tpowNumx=2$ we take instead
\be
\label{eqn:KcplM2}
\matKcpl\equiv
\begin{pmatrix}
\tpowx_1 & -2 \\
-2 &  \tpowx_2 \\
\end{pmatrix}\,,
\ee
and for $\tpowNumx=1$ we take $\matKcpl\equiv(\tpowx_1-2)$.

We then define $\latAG=\Z^\tpowNumx$ and $\latAG'\subset\latAG$ to be the sublattice generated by the columns of $\matKcpl$, i.e.,
$$
\latAG'=\{\matKcpl\mcWRep: \mcWRep\in\Z^\tpowNumx\}\equiv\matKcpl(\Z^\tpowNumx).
$$
One can show \cite{Ganor:2014pha} that $\latAG/\latAG'$ is isomorphic to $\GrSt$, and we therefore take $\AG=\latAG/\latAG'$, and define
\be\label{eqn:qfmcV}
\qf(\mcV)=\mcVRep^t\matKcpl^{-1}\mcVRep,\qquad
\text{for any representative $\mcVRep\in\latAG$ of $\mcV\in\latAG/\latAG'$.}
\ee
This quadratic form on $\AG$ then defines, through \eqref{eqn:WeilR}, a $|\AG|$-dimensional Weil representation of $\SL(2,\Z)$. We will elaborate on its physical interpretation (related to Chern-Simons theory with $U(1)^\tpowNumx$ gauge group) in \secref{subsec:CSconnection}.

The expression \eqref{eqn:qfmcV} and the associated Weil representation \eqref{eqn:WeilR} is not yet satisfactory for us, since it is not yet clear how it is related to \eqref{eqn:PFFPFsumRed}. We would like to argue that, up to a phase, \eqref{eqn:qfmcV} can be written as $\tr(\mcS\mcT^\tpowy)$ (corresponding to $\matMy=\genSy\genTy^\tpowy$) in the Weil representation, but to see this we will need to recast the expressions for the matrix elements of $\mcS$ and $\mcT$ directly in terms of $\matMx$. We will see that this can be achieved only up to $\pm$ signs on elements in $\mcS$ and $\mcT$. Nevertheless, these signs drop out of the expression for $\tr(\mcS\mcT^\tpowy)$. This leads us to the {\it second} way to describe $\qf(\cdot)$, which we now present.

First, it is convenient to introduce a new lattice $(\matMx-\Id)^{-1}(\Z^2)$, and for $\vN\in\GrSt=\Z^2/(\matMx-\Id)\Z^2$ [see \eqref{eqn:GrStDef}], we define
\be\label{eqn:xVfromvN}
\xV\equiv(\matMx-\Id)^{-1}\vN{\pmod{\Z^2}},
\qquad\text{so that $\xV\in(\matMx-\Id)^{-1}(\Z^2)/\Z^2$.}
\ee
The quotient $(\matMx-\Id)^{-1}(\Z^2)/\Z^2$ is canonically identified with $\GrSt$ in this way.
As mentioned above, there is an isomorphism $\isomMTCS:\GrSt\cong\latAG/\latAG'$ (which we will describe in detail in \secref{subsec:isoCSMT}). Using it, we can define a quadratic form on $\GrSt$ simply as
\be\label{eqn:qfMTisomqf}
\qfMT(\xV)\equiv\qf\left(\isomMTCS(\xV)\right),\qquad\text{for $\vN\in\GrSt$.}
\ee
Using $\qfMT$ we can then define the action of $\mcS$ and $\mcT$ on the states $\ket{\xV}$ in $\GrSt$ as
\be
\label{eqn:WeilRMT}
\mcT\ket{\xV} = 
e^{-i\phaseW}e^{2\pi i\qfMT(\xV)}\ket{\xV},
\qquad
\mcS\ket{\xV}
=
\frac{1}{\sqrt{|\GrSt|}}
\sum_{\xU\in\GrSt} 
e^{-2\pi i\qBfMT(\xV,\xU)}
\ket{\xU},\qquad
\text{for $\xV\in\GrSt$,}
\ee
where
$$
\qBfMT(\xU,\xV)\equiv\qfMT(\xU+\xV)-\qfMT(\xU)-\qfMT(\xV),\qquad
\xU,\xV\in\GrSt.
$$
We claim that (at least for $\tpowNumx\le 2$) the doubled quadratic form $2\qfMT(\cdot)$ has a simple expression in terms of $\matMx$:
\be
\label{eqn:2qfMT}
2\qfMT(\xV)=\xV^t\matEps\matMx\xV\quad{\pmod\Z},
\ee
and therefore
\be\label{eqn:mcT2xV}
\mcT^2\ket{\xV} = 
e^{-2i\phaseW}e^{2\pi i\xV^t\matEps\matMx\xV}\ket{\xV}.
\ee
We will prove \eqref{eqn:2qfMT} in \appref{app:proof1} for $\tpowNumx\le 2$.
It implies that the matrix elements of $\mcT$ and $\mcS$ can be expressed as follows (with undetermined $\pm$ signs that depend on $\xU,\xV\in\GrSt$, as well as other input data):
\be
\label{eqn:STmatrixElements}
\bra{\xU}\mcT\ket{\xV}=
\pm\delta_{\xU\xV}e^{-i\phaseW}e^{\pi i\xV^t\matEps\matMx\xV},\qquad
\bra{\xU}\mcS\ket{\xV}=
\pm
\frac{1}{\sqrt{|\GrSt|}}
e^{-\pi i(\xU^t\matEps\matMx\xV+\xV^t\matEps\matMx\xU)}.
\ee
(In Appendix \ref{app:proof2} we will show what the $\pm$ signs are in a particular example.)
Fortunately, for even $\tpowy=2\thalfpowy$, the trace $\tr\left(\mcS\mcT^{2\thalfpowy}\right)$ is independent of the unknown $\pm$ signs in \eqref{eqn:STmatrixElements}:
\be\label{eqn:trSTtpowy}
\tr\left(\mcS\mcT^{2\thalfpowy}\right)=
\frac{e^{-2i\thalfpowy\phaseW}}{\sqrt{|\GrSt|}}
\sum_{\xV}e^{2(\thalfpowy-1)\pi i\xV^t\matEps\matMx\xV}\,.
\ee
Using \eqref{eqn:xVfromvN}, we easily check\footnote{Using $\vN^t\matEps\vN=0$ and $\matEps\matMx=(\matMx^t)^{-1}\matEps$.} that \eqref{eqn:trSTtpowy} reproduces \eqref{eqn:PFFPFsumRed}, up to the phase $e^{-2i\thalfpowy\phaseW}$.

In \secref{subsec:MoreGens} we will also need to calculate $\tr(\mcS\mcT^{2\thalfpowy_1}\mcS\mcT^{2\thalfpowy_2})$, and luckily again, this is independent of the ambiguous $\pm$ signs in \eqref{eqn:STmatrixElements}:
\be\label{eqn:trSTtpowySTtpowy}
\tr\left(\mcS\mcT^{2\thalfpowy_1}\mcS\mcT^{2\thalfpowy_2}\right)=
\frac{e^{-2i(\thalfpowy_1+\thalfpowy_2)\phaseW}}{|\GrSt|}
\sum_{\xU,\xV}e^{2\pi i(\thalfpowy_2\xV^t\matEps\matMx\xV
+\thalfpowy_1\xU^t\matEps\matMx\xU
-\xU^t\matEps\matMx\xV-\xV^t\matEps\matMx\xU)}\,.
\ee
We will now offer a physical interpretation for \eqref{eqn:2qfMT}.

% --------------------------------------------------------------
\subsection{Low-energy strings on a mapping torus}
\label{subsec:MT}

There is a simple geometrical interpretation of \eqref{eqn:mcT2xV} in terms of the topology of (an auxiliary) {\it mapping torus} -- a manifold formed by fibering $T^2$ over $S^1$.
Let $\xB\in\R$ be a periodic coordinate on the $S^1$ base with periodicity 1, and let $\vxF\in\R^2$ be coordinates that will soon parametrize a torus, after additional periodicity conditions are imposed. The mapping torus is defined as the set of points $(\xB,\vxF)$ with identification, 
\be\label{eqn:MTdef}
(\xB,\vxF)\sim(\xB,\vxF+\vKF)\sim\left(\xB+1,\matMx^{-1}\vxF\right)\,,\qquad\forall\,\vKF\in\Z^2,
\ee
for a fixed $\matMx\in\SL(2,\Z)$, which acts as an MCG element on the fiber.
In our case, a mapping torus is formed by fixing $\xWS_2$ and considering the configuration space of the fields $\vX(\xWS_1,\xWS_2)$ as $\xWS_1$ (which we identified as $\xB$ above) varies from $0$ to $1$. A {\it ground state} of the string corresponds to a point on the $T^2$ fiber that is invariant modulo $\Z^2$ under the geometrical $\matMx$-twist. Representing this point by $\xV\in\R^2$, we thus require $\xV-\matMx\xV\in\Z^2$.
This has a discrete set of rational solutions $\xV\in\Q^2$, and we formally define a Hilbert space $\mathcal{H}_M$ of states with basis $\left\{\ket{\xV}\right\}$ comprising of states $\ket{\xV}$ such that
\be\label{eqn:MIxV}
(\matMx-\Id)\xV\in\Z^2
\ee
in which $\ket{\xV}=\ket{\xU}$ if $\xV-\xU\in\Z^2$.
Now we define the lattice
\be
\label{eqn:latZvdef}
\latZv\equiv(\matMx-\Id)^{-1}\left(\Z^2\right)=\left\{
\xV\in\Q^2: (\matMx-\Id)\xV\in\Z^2
\right\}\supset\Z^2\,.
\ee
Then a solution to \eqref{eqn:MIxV} with the identification ``$\xV\sim\xU$ whenever $\xV-\xU\in\Z^2$'' defines an element of the coset space $\latZv/\Z^2$, which map to a basis of the Hilbert space $\mathcal{H}_M$.
This coset space is a finite abelian group, i.e., the cokernal of $M-\Id$, which can be identified with isometries of the mapping torus. (See \cite{Ganor:2014pha} for more details.)
It is easy to see that $\latZv/\Z^2$ is isomorphic to $\GrSt$ [defined in \eqref{eqn:GrStDef}] and that the number of states is
$$
|\latZv/\Z^2| = |\det(\matMx-\Id)|=|\text{Tr}\,\matMx-2|.
$$
We need to know the action of $\SL(2,\Z)$, generated by $S$ and $T$, on those quantum states. It can be described as T-duality on the $T^2$ fiber and, as we argued in \eqref{eqn:2qfMT}, (at least for $\matMx=\genSx\genTx^{2\thalfpowx_1}$ or $\matMx=\genSx\genTx^{2\thalfpowx_1}\genSx\genTx^{2\thalfpowx_2}$) is given by
\be
\label{eqn:SyMT}
\mathscr{S}\ket{\xV} = \frac{1}{\sqrt{|\latZv/\Z^2|}}\sum_{\xU\in\latZv/\Z^2}(\pm)
e^{-\pi i(\xV^t\matEps\matMx\xU+\xU^t\matEps\matMx\xV)}\ket{\xU},
\ee
\be
\label{eqn:TyMT}
\mathscr{T}^\tpow\ket{\xV}=(\pm) e^{-i\tpow\phaseW}e^{\tpow\pi i\xV^t\matEps\matMx\xV}\ket{\xV}\,,
\ee
where all the $(\pm)$ signs in \eqref{eqn:SyMT} and \eqref{eqn:TyMT} are independent of $u$ and $v$ respectively, and where $\phaseW$ in \eqref{eqn:TyMT} has been defined in \secref{subsec:WeilReps} [e.g., \eqref{eqn:WeilR}] and is a constant phase chosen so that $(\mathscr{S}\mathscr{T})^3=\mathscr{S}^2$ equals the charge conjugation operator [which represents the $-\Id\in\SL(2,\Z)$], which acts as
$$
\mathscr{S}^2\ket{\xV}=\ket{-\xV}.
$$
$\mathscr{T}^\tpow$ is ill-defined for odd $\tpow$ [unless $\matMx$ is such that $\xV^t\matEps\matMx\xV$ is an even integer for all $\xV\in\Z^2$, which is when $\matMx\in\Gamma(2)$, as mentioned in \footref{foot:GammaTwo}], so to avoid extra complications we assume $\tpow\in 2\Z$.
Note that the definitions \eqref{eqn:SyMT} and \eqref{eqn:TyMT} are independent of the representatives $\xV$ and $\xU$, because for $\vKF\in\Z^2$ we have $\vKF^t\matEps(\matMx-\Id)\xU\in\Z$ when $(\matMx-\Id)\xV\in\Z^2$, and also $\xU^t\matEps(\matMx-\Id)\vKF =
%\xU^t(\matMx^{-1})^t\matEps\vKF-\xU^t\matEps\vKF =
%\xU^t(\Id-\matMx^t)(\matMx^{-1})^t\matEps\vKF =
[(\Id-\matMx)\xU]^t\matEps\matMx\vKF\in\Z$.
Note also that since $\matEps$ is antisymmetric as in \eqref{eqn:matEpsDEF}, we have $\xV^t\matEps\matMx\xV = \xV^t\matEps(\matMx-\Id)\xV$.

The phase $\exp\left(-\tpow\pi i\xV^t\matEps\matMx\xV\right)$ has a nice geometrical interpretation (analogous to the one discussed in \S3.7 of \cite{Ganor:2010md} for $\matMx=\genSx$).
The expression 
$$
-\frac{1}{2}\xV^t\matEps\matMx\xV=\frac{1}{2}\xV^t\matEps\matMx^{-1}\xV
$$ 
is the area of a triangle in $\R^2$ with sides given by the vectors $\xV$ and $\matMx^{-1}\xV$. To see how this is related to $\mathscr{T}$, consider a string worldsheet (i.e., an $\vX$ field configuration) that interpolates between the states $\ket{0}$ (a string at $\xV=0$ for say $\xWS_2=0$) and $\ket{\xV}$ (for $\xV\neq 0$ and $\xWS_2>0$). We can realize it by constructing a section of the mapping torus with $\vxF=\zeta\xV$ and then letting $\zeta\in[0,1]$ and $\xB\in(0,1)$ be the coordinates of the worldsheet (i.e., identify $\xB$ with $\xWS_1$, and $\zeta$ with $\xWS_2$). If we attach to this surface $\Sigma$ the triangle with vertices $\left\{0,\xV,\matMx^{-1}\xV\right\}$, we obtain a surface whose boundary is the union of three loops: the loop corresponding to string state $\ket{0}$, the loop corresponding to string state $\ket{\xV}$, and the loop from $(0,\xV)$ to $(0,\matMx^{-1}\xV)\sim(1,\xV)$ at constant $\xB=0$, which is a closed loop thanks to \eqref{eqn:MTdef} and \eqref{eqn:MIxV}.
If we now consider the scattering amplitude of an inelastic scattering process with two string states $\ket{0}$ going into two final string states $\ket{\xV}$ and $\ket{-\xV}$:
\be\label{eqn:scattering}
\ket{0}\otimes\ket{0}\rightarrow\ket{\xV}\otimes\ket{-\xV},
\ee
then it is calculated in string theory by a path integral over worldsheets $\Sigma$ with four boundary components corresponding to the four string states $\ket{\xV}$, $\ket{-\xV}$ and $\ket{0}$'s (wrapped twice with opposite orientation). Since the duality operation $\mathscr{T}^\tpow$ acts on the Kalb-Ramond field $B$ as $B\rightarrow B+\pi\tpow d\vxF^t\wedge\matEps d\vxF$, it multiplies the scattering amplitude by the phase 
$$
\exp\left( i\int_\Sigma B\right).
$$
The construction above shows that this phase is $4\pi\tpow$ times the area of the triangle with vertices $\left\{0,\xV,\matMx^{-1}\xV\right\}$, which corresponds to a wavefunction normalization of each of the $\ket{\pm\xV}$ states by $\exp\left(\tpow\pi i\xV^t\matEps\matMx^{-1}\xV\right)$, as required by \eqref{eqn:TyMT}.
This explains the physical origin of the $\pm$ sign ambiguity in \eqref{eqn:STmatrixElements}, since the process \eqref{eqn:scattering} requires two states for charge conservation.\footnote{As explained in \cite{Ganor:2014pha}, the ``charges'' correspond to the first homology group $H_1(\Z)$ of the mapping torus, which is isomorphic to $\GrSt$, and $\ket{\xV}$ has charge $\xV$.}

We can now recover the partition function \eqref{eqn:PFFPFsumRed} by calculating
\be\label{eqn:trSTk}
\text{Tr}_{\mathcal{H}_M}\left(\mathscr{S}\mathscr{T}^\tpow\right) =
\sum_\xV\bra{\xV}\mathscr{S}\mathscr{T}^\tpow\ket{\xV}
=
\frac{e^{-i\tpow\phaseW}}{\sqrt{|\latZv/\Z^2|}}
\sum_\xV
e^{\left(2-\tpow\right)\pi i\xV^t\matEps(\matMx-\Id)\xV}.
\ee
Up to the phase $\phaseW$, this equals \eqref{eqn:PFFPFsumRed} after the substitution
$$
\vN=(\Id-\matMx)\xV.
$$

% --------------------------------------------------------------
\subsection{Connection with \texorpdfstring{$U(1)^{\tpowNumx}$}{} Chern-Simons theory}
\label{subsec:CSconnection}

The quadratic Gauss sum \eqref{eqn:PFFPFsumRed} can also be expressed as a trace similar to \eqref{eqn:trSTk}, but with $\mathscr{S}$ and $\mathscr{T}$ defined as MCG representation generators acting on the Hilbert space of an abelian Chern-Simons theory placed on $T^2$ \cite{Ganor:2014pha}. To see this, let us first consider Chern-Simons theory at level $\lvk\in\mathbb{Z}$ with a $U(1)$ gauge group, compactified on $T^2\times\R$, where $\R$ is the Euclidean time direction and $T^2$ is a torus parametrized by periodic coordinates $0\le x_1, x_2<1.$ The action is 
$$
I = \frac{\lvk}{2\pi}\int A\wedge dA,
$$ 
where $A$ is the $U(1)$ gauge field.
It is well-known \cite{Wen} that the Hilbert space $\mathcal{H}_{U(1)}$ of the ground states of the theory has a $\lvk$-fold degeneracy\footnote{Its ground-state degeneracy is $(\lvk)^g$ on a genus-$g$ Riemann surface instead \cite{WenNiu}.}, which we will denote by $\ket{a}$ ($a=0,\dots,\lvk-1$). Let $\alpha$ and $\beta$ be two fundamental $1$-cycles of $T^2$, where $\alpha$ corresponds to a loop at constant $x_2$, with $x_1$ varying from $0$ to $1$, and $\beta$ corresponds to a similar loop at constant $x_1$ with $x_2$ varying from $0$ to $1$. 
Consider the Wilson loop operators 
\begin{equation}
W_1\equiv\exp\left(i\oint_\alpha A\right)\quad \text{and}\quad W_2\equiv\exp\left(i\oint_\beta A\right).
\end{equation}
Their action on the ground states is given by the ``clock'' and ``shift'' matrices:
\begin{equation}\label{eqn:W}
W_1\ket{a} = e^{2\pi i a/\lvk}\ket{a}\,,\qquad
W_2\ket{a}=\ket{a+1}\,.
\end{equation}
We will need the action of large diffeomorphisms, generated by $\csT$ and $\csS$,
\begin{equation}
\label{eqn:csST}
\csS\ket{a} = \frac{1}{\sqrt{\lvk}}\sum_{b=0}^{\lvk-1}e^{-2\pi i a b/\lvk}\ket{b}\,,\qquad
\csT\ket{a} = e^{-i\pi/12}e^{\pi i a^2/\lvk}\ket{a}\,.
\end{equation}
Up to the constant phase $e^{-i\pi/4}$, this can be checked by making sure that the commutation relations $\csS^{-1}W_i\csS$ and $\csT^{-1}W_i\csT$ are as they should be (for $i=1,2$), given the geometrical interpretation of $\csT$ and $\csS$ as torus MCG generators. The phase $e^{-i\pi/4}$ is determined by requiring $\left(\csS\csT\right)^3=\csS^2$, which is the charge conjugation operator, so that we obtain a {\it linear} instead of a {\it projective} $\SL(2,\Z)$ representation. It can be derived more systematically by explicitly writing the ground-state wavefunctions as a function of holonomies of the gauge fields \cite{EMSS,axelrod1991geometric}, or by recalling the connection between the $U(1)$ Chern-Simons theory and the 2d CFT of a free chiral boson \cite{Witten:1988hf}. %The states $\ket{a}$ can be associated with Virasoro characters of primary states, and $\csS$ and $\csT$ act by modular transformations \cite{Witten:1988hf}. The factor $e^{-i\pi/12}$ is related to the shift in ground-state energy by $-c/24$ where $c=1$ is the central charge.
Note that the equation for $\csT$ is ill-defined for odd $\lvk$ (because it is inconsistent with $\ket{a}=\ket{a+\lvk}$). In that case only even powers of $\csT$ are well-defined.

We now set $\lvk\equiv q$.
Up to an $e^{\pi i/4}$ phase, the (complex conjugate of) quadratic Gauss sum appearing on the RHS of \eqref{eqn:LSrel} can then be written as
\be\label{eqn:QGtr}
\frac{1}{\sqrt{q}}\sum_{n=0}^{q-1}e^{-2\pi i n^2 p/q}
= e^{(p+1)\pi i/12}
\sum_{n=0}^{q-1}\bra{n}\csS\csT^{2+2p}\ket{n}
=e^{(p+1)\pi i/12}\text{Tr}_{\mathcal{H}_{U(1)}}\left(\csS\csT^{2+2p}\right).
\ee
We will see in \secref{subsec:QGSPF} that, up to a phase, \eqref{eqn:QGtr} is also the partition function $\PF$ that we calculated in \eqref{eqn:PFFPFsumRed}, for $\matMx=\genSx\genTx^{q+2}$. In \secref{subsec:Berry} we will calculate the phase of the partition function in a T-dual formulation and observe that it receives a contribution from a Berry phase that depends on the details of how the complex structure varies with time. We also note that a much deeper analysis of the partition function of Chern-Simons theory on a mapping torus with $SU(2)$ gauge group has been carried out by Jeffrey (see \S4 of \cite{Jeffrey:1992tk} and \S4 of \cite{Jeffrey:2012}), where the result\footnote{With a particular ``canonical'' choice of {\it 2-framing} of the mapping torus $M_3$, i.e., the choice of a homotopy equivalence class of trivialization of of $TM_3\oplus TM_3$ \cite{Atiyah}, so that the theory is free of framing anomaly.} is similarly given by a trace of the action of $\matMx$ on the Hilbert space and yields a quadratic Gauss sum.

To move from the special case of \eqref{eqn:LSrel} to the general case \eqref{eqn:PFFPFsumRed}, we need $U(1)^\tpowNum$ Chern-Simons theory on $T^2$.
We first recall some basic facts, and then we explain in \secref{subsec:isoCSMT} why it is related to \eqref{eqn:PFFPFsumRed}. The Chern-Simons coupling constants are given by a symmetric matrix, which for $\tpowNum>2$ takes the tri-diagonal form with corners\footnote{$U(1)^\tpowNum$ Chern-Simons on more general 3-manifolds, such as plumbed 3-manifolds, has been considered in \cite{Gadde:2013sca,Gukov:2016gkn}.}
\be
\label{eqn:Kcpl}
\matKcpl\equiv
\begin{pmatrix}
\tpow_1 & -1 & 0 &  & -1 \\
-1 & \ddots & \ddots & \ddots & \\
0 & \ddots & \ddots & \ddots & 0 \\
&\ddots  & \ddots & \ddots & -1 \\
-1 &  & 0 & -1 & \tpow_\tpowNum \\
\end{pmatrix}\,,
\ee
and for $\tpowNum=2$ takes the form
\be
\label{eqn:KcplTwo}
\matKcpl\equiv
\left(\begin{array}{rr}
\tpow_1 & -2 \\
-2 &  \tpow_2 \\
\end{array}\right)\,.
\ee
(It is conventionally called {\it the $K$-matrix} in condensed matter literature\footnote{The ground-state degeneracy of this Chern-Simons theory on a genus-$g$ Riemann surface is $|\det K|^g$ \cite{WenZee}.}, and is used to describe the $\tpowNum$-component abelian fractional quantum Hall effect \cite{WenZee}.)
The Hilbert space of $U(1)^\tpowNum$ Chern-Simons theory on $T^2$ with coupling constant matrix $\matKcpl$ has a basis of states $\ket{\yV}$ parametrized by $\yV\in\Z^\tpowNum$ such that
$\ket{\yV}=\ket{\yU}$ if $\yV-\yU=\matKcpl\vNCS$ for some $\vNCS\in\Z^\tpowNum$.
Define the lattice
\begin{equation}
\label{eq:sublattice}
\latWv\simeq\matKcpl\left(\Z^\tpowNum\right)
\equiv\left\{
\matKcpl\yW:\yW\in\Z^\tpowNum
\right\}\subset\Z^\tpowNum\,.
\end{equation}
So $\latWv$ is the sublattice of $\Z^\tpowNum$ that is generated by the columns of the matrix $\matKcpl$. The coset $\Z^\tpowNum/\latWv$ is a finite abelian group.
The Hilbert space of $U(1)^\tpowNum$ Chern-Simons theory on $T^2$ with coupling constant matrix $\matKcpl$ has a basis of states which can be identified with elements of $\Z^\tpowNum/\latWv$. Pick a nontrivial generator of $\pi_1(T^2)$, and consider the corresponding $\tpowNum$ Wilson loops acting on the Hilbert space. They form an abelian group which can be identified with $\Z^\tpowNum/\latWv$.

Next, we define the action of $\SL(2,\Z)$ on the Hilbert space, as a physical realization of \eqref{eqn:WeilRMT}. From the Chern-Simons perspective, this is the action of the MCG of $T^2$.
We assume that $\matKcpl$ is even, i.e., all $\tpow_i\in2\Z$ ($i=1,\dots,\tpowNum$).
The generators act on states as: 
\be
\label{eqn:TSCSK}
\mcS\ket{\yV} = 
\frac{1}{\sqrt{|\Z^\tpowNum/\latWv|}}\sum_{\yU\in\Z^\tpowNum/\latWv}
e^{-2\pi i\yU^t\matKcpl^{-1}\yV}\ket{\yV}
\,,\quad
\mcT\ket{\yV} =
e^{-i\phaseW}e^{\pi i\yV^t\matKcpl^{-1}\yV}\ket{\yV},
\ee
generalizing the previous $\csS$ and $\csT$ actions in $U(1)$ Chern-Simons theory, given by \eqref{eqn:csST}.
The phase $\phaseW$ equals $\pi i\tpowNum/12$ if all $\tpow_i>0$, and in general it is $\pi i/12$ times the signature of $\matKcpl$.
Note that $\mcT$ is well-defined when $\matKcpl$ defines an even bilinear form.
We will now show how the Chern-Simons picture is isomorphic to the mapping-torus picture.

% -------------------------------------------------------------
\subsection{Isomorphism between the Chern-Simons and mapping torus descriptions}
\label{subsec:isoCSMT}

We take the powers $\tpow_1,\dots,\tpow_\tpowNum$ to match the powers in one of the (non-unique) decompositions of $\matMx\in\SL(2,\mathbb{Z})$ into $\genSx$ and $\genTx$ generators:
$$
\matMx = \genSx\genTx^{\tpow_1}\genSx\genTx^{\tpow_2}\cdots\genSx\genTx^{\tpow_\tpowNum}.
$$
Then, it can be shown, using elementary row and column operations, that there exist $\smithP,\smithQ\in \PSL(\tpowNum,\Z)$ (unimodular) so that
$$
\smithP\matKcpl\smithQ = 
\begin{pmatrix}
\matMx-\Id &  &  &  &  \\
& 1 &  &  & \\
& & 1 & &  \\
& && \ddots &  \\
&  &  &  & 1 \\
\end{pmatrix}
$$
is a unique block-diagonal matrix. In other words, $\matMx-\Id$ and $\matKcpl$ have the same {\it Smith normal form}\footnote{Here we adopt the uncommon convention that on the diagonal, each lower-right element divides the element upper-left to it, opposite to that in \cite{Ganor:2014pha}.}, i.e., according to \eqref{eqn:latZvdef} and \eqref{eq:sublattice}, $\Lambda/\mathbb{Z}^2$ is isomorphic to $\mathbb{Z}^{\overline{\mathbf{r}}}/\Lambda'$.
Now let us construct an explicit isomorphism between them. Define the $(2\tpowNum)\times 2$ matrix
$$
\matJ\equiv
\begin{pmatrix}1 & 0 \\ 0 & 1 \\ 0 & 0 \\ \vdots & \vdots\\ 0 & 0\\ \end{pmatrix}\in
\Hom\left(\Z^2,\Z^\tpowNum\right).
$$
Now, suppose $\xV$ satisfies $(\matMx-\Id)\xV\in\Z^2$ and define
$$
\widetilde{\varphi}(\xV)=\matKcpl\smithQ\matJ\xV
=\smithP^{-1}
\begin{pmatrix}
\matMx-\Id &  &  &  &  \\
& 1 &  &  & \\
& & 1 & &  \\
& && \ddots &  \\
&  &  &  & 1 \\
\end{pmatrix}
\begin{pmatrix}\xV\\ 0 \\ 0 \\ \vdots \\ 0 \\ \end{pmatrix}
=\smithP^{-1}
\matJ(M-\Id)\xV\in\Z^{\tpowNum},
$$
implying that $\widetilde{\varphi}(\xV)\in\Z^{\tpowNum}$ if $(\matMx-\Id)\xV\in\Z^2$, so $\widetilde{\varphi}$ is surjective. Its injectivity is due to both $P$ and $M-\Id$ being invertible, and $\matJ^t\matJ=\Id$.
Therefore $\widetilde{\varphi}$ is an isomorphism between finite abelian groups $\widetilde{\varphi}:\latZv/\Z^2\cong\Z^\tpowNum/\latWv$ with $\widetilde{\varphi}(v)\equiv\tilde{v}$ (see Appendix A in \cite{Ganor:2014pha} by the authors of this paper for more details).

We take \eqref{eqn:TSCSK} as the definition of the action of $\SL(2,\Z)$ on the states. In the basis $\ket{v}$ of \secref{subsec:MT} we define the $\SL(2,\Z)$ action using the isomorphism $\widetilde{\varphi}$.
We conjecture that this definition is the same as formulae \eqref{eqn:SyMT} and \eqref{eqn:TyMT} for the $\mathscr{T}$ and $\mathscr{S}$ generators [with a suitable choice of $\pm$ signs in the matrix elements of \eqref{eqn:SyMT} and \eqref{eqn:TyMT}], and we prove it explicitly in the case $\tpowNum=2$ in Appendix \ref{app:proof1}, with numerical evidence for the $\tpowNum=3$ cases in \appref{app:proof2}.

% ==============================================================

\section{Quadratic Reciprocity from double-Janus \texorpdfstring{$\sigma$}{}-models}
\label{sec:QRDJSM}

The partition function of the double-Janus $\sigma$-model that we discussed in \secref{sec:PF} reduces to a sum \eqref{eqn:PFFPFsumRed} over the finite abelian group $\GrSt$ defined in \eqref{eqn:GrStDef}.
We will now show how, in a special case, $\PF$ reduces to the Gauss sum defined in \secref{sec:QRbasics}, and we will follow with a discussion on the general case.
We will give a Hilbert space interpretation of \eqref{eqn:PFFPFsumRed}, with $\xWS_2$ identified as ``time'', and we will subsequently argue that the Landsberg-Schaar identity \eqref{eqn:LSS} can be understood by switching the role of ``time'' from $\xWS_2$ to $\xWS_1$.

% --------------------------------------------------------------
\subsection{Quadratic Gauss sum as a special case of \texorpdfstring{$\PF(M,\widetilde{M})$}{}}
\label{subsec:QGSPF}

Taking the special case \eqref{eqn:MyMxSpecial}:
$$
\matMx=\genSx\genTx^{q+2}=\begin{pmatrix}
0 & -1 \\ 1 & q+2
\end{pmatrix},\qquad
\matMy=\genSy\genTy^{2+2p}=\begin{pmatrix}
0 & -1 \\ 1 & 2p+2
\end{pmatrix},
$$
we can identify $\GrSt$ with $\Z_q$ as follows.
Setting
$$
\vN=\begin{pmatrix}
\Nx \\ \Ny \\ 
\end{pmatrix},\qquad
\vK=\begin{pmatrix}
\Kx \\ \Ky \\ 
\end{pmatrix},
$$
the identification $\vN\sim\vN+(\Id-\matMx)\vK$ [see \eqref{eqn:vNi}] becomes
$$
\Nx\sim\Nx+\Kx+\Ky,\qquad\Ny\sim\Ny-\Kx-\Ky-q\Ky.
$$
We define $\Nz\equiv\Nx+\Ny$, and use the freedom to choose $\Kx=-\Nx-\Ky$ to set $\Nx=0$.
Then, $\Nz\sim\Nz-q\Ky$, and so $\Nz$ can be identified with an element of $\Z_q$. Setting $\tpowy=2+2p$, we then calculate
%$$
%i\pi(\tpowy-2)\vN^t\matEps(\Id-\matMx)^{-1}\vN=\frac{2\pi i p\Nz^2}{q}
%%$$
\be\label{eqn:ZasQGS}
\PF=\frac{1}{\sqrt{q}}\sum_{\vN\in\GrSt}
\exp\left\lbrack
i\pi(2-\tpowy)\vN^t\matEps(\Id-\matMx)^{-1}\vN
\right\rbrack
=\frac{1}{\sqrt{q}}\sum_{\Nz=0}^{q-1}\exp\left(-\frac{2\pi i p\Nz^2}{q}\right),
\ee
as the standard quadratic Gauss sum.

% -------------------------------------------------------------
\subsection{Berry phase}
\label{subsec:Berry}

The Landsberg-Schaar identity \eqref{eqn:LSrel} relates Gauss sum of $(4p)^{\text{th}}$ roots of unity to Gauss sum of $q^{\text{th}}$ roots of unity.
The partition function $\PF$, derived in \eqref{eqn:PFFPFsumRed}, reproduces the latter sum, as we saw in \secref{subsec:QGSPF}, and this form is closely associated with a Hilbert space interpretation whereby time is identified with $\xWS_2$.
We will now show that the other side of the Landsberg-Schaar identity can be interpreted in terms of a different Hilbert space, with time identified with $\xWS_1$. In such a Hilbert space interpretation the duality wall at $\xWS_1=0$ contributes a trace of $\matMx$, which has to be multiplied by a Berry phase resulting from the variation of complex structure $\tauCS$ (which we can take to be adiabatic).\footnote{A similar Berry phase for fermions (either Dirac or Majorana) on a 3d mapping torus under an adiabatic diffeomorphism appeared in \cite{Witten:2015aba}.}
We will see that the phase of $e^{\pi i/4}$ that appears in \eqref{eqn:LSrel} can be reproduced by a combination of the representation of $\matMx$ on the low-energy Hilbert space and the Berry phase. In calculating the Berry phase, the form of the profile $\tauCS(\xWS_1)$ is important, and we will see that when $\tauCS(\xWS_1)$ takes values along a semi-circle in its upper half-plane, as in \secref{subsec:MoreDJSols}, the correct phase is reproduced.

In \secref{sec:sigmaM} we have chosen the radii of the $\xWS_1$ and $\xWS_2$ circles to be equal (and given by $1/2\pi$).
This was mostly to avoid a cumbersome notation, and indeed, we can easily allow the radii to be different. The partition function $\PF$ is independent of the radii, thanks to supersymmetry.
We will now take the limit that the $\xWS_2$ direction is much smaller than the $\xWS_1$ circle.
In this limit, we can study the Hilbert space of the problem at a fixed $\xWS_1$, reduce to ground states, and then introduce the wall at $\xWS_1=0$ by inserting the operator representing $\matMx$ on the subspace of ground states.
For $0<\xWS_1<1$, the most relevant terms in the action are given by the single-derivative terms \eqref{eqn:Ibprime}, and we can set $\vX_0=\vX_1$, since the $\xWS_2$ circle is assumed to be small, and momentum modes along it are suppressed.
Writing $\vXR(\xWS_1)$ instead of both $\vX_0$ and $\vX_1$, we are therefore left with
\be\label{eqn:IbprimeReduced}
I_b'=
\frac{i(2-\tpowy)}{4\pi}\int\vXR^t\matEps d\vXR,
\ee
where the integral is over $\xWS_1$ and $\vXR$ is independent of $\xWS_2$.
We set $\levely\equiv\tpowy-2$, and assume $\levely>0$.
$\vXR$ describes a coordinate on $T^2$, and the Hilbert space of ground states can be identified with that of geometric quantization at level $\levely$ (i.e., with a symplectic form $\displaystyle{\frac{\levely}{4\pi} d\vXR^t\matEps\wedge d\vXR=\frac{\levely}{2\pi} d\xf^1\wedge d\xf^2}$).

% -------------------------------------------------------------
%\subsection{Theta functions, geometric quantization, and the Landau problem}

Quantization on a torus at level $\levely$ is one of the simplest examples of geometric quantization \cite{Haldane1,theta,geometry}.
Our $T^2$ is parametrized by $(\xf^1,\xf^2)$ with $0\le \xf^1,\xf^2<2\pi$.
It is convenient to add the kinetic term to get the Lagrangian of a Landau problem (a particle in a uniform and constant magnetic field) on $T^2$:
\be\label{eqn:LLag}
I_L=\frac{1}{2\pi}
\int\left(\frac{1}{2}\massL G_{IJ}\dot{\xf}^I\dot{\xf}^J - \frac{i\levely}{2}\epsilon_{IJ}\xf^I\dot{\xf}^J\right)dt\,,
\ee
where for $G_{IJ}$ we take the metric $ds^2=\frac{1}{\tauLL_2}|\tauLL d\xf^2-d\xf^1|^2$.
The ground states of \eqref{eqn:LLag} are normalized Landau wavefunctions\footnote{They are essentially $1/\levely$ Laughlin states on a torus as proposed in \cite{Haldane1}.} (independent of $\massL$ and the area of the $T^2$):
\bear
\wfF_{j,\levely}\left(\xf^1,\xf^2\right) &=&
\frac{1}{2\pi}
\left(2 \levely \tauLL_2 \right)^{\frac{1}{4} } 
e^{-\frac{i\levely\xf^1\xf^2}{4\pi}}
\sum_{n=-\infty}^\infty e^{i\left(n\levely + j\right)\xf^1
+\pi i\levely\tauLL\left(n+\frac{j}{\levely}-\frac{\xf^2}{2\pi}\right)^2}
\nn\\
&=&
\frac{1}{2\pi}
\left(2 \levely \tauLL_2 \right)^{\frac{1}{4} } 
e^{-\frac{i\levely\xf^1\xf^2}{4\pi}}
e^{ \pi i \tauLL\levely \left( \tfrac{ \xf^2 }{2 \pi}  \right)^2}  \Theta _{j,\levely} \left(\tfrac{\xf^1 -\tauLL\xf^2}{2\pi}; \tauLL \right)
\label{eqn:Lwf}
\eear
where the $\Theta$-function is defined as
\be\label{eqn:Thetafn}
\Theta_{j,\levely}(u,\tauLL)\equiv
\sum_{n=-\infty}^\infty e^{\pi i\levely\tauLL\left(n+\frac{j}{\levely}\right)^2 
	+ 2\pi i\levely\left(n + \frac{j}{\levely}\right)u}
\ee
and is holomorphic in $u$ and $\tauLL$.
In the low-energy limit, the kinetic term in \eqref{eqn:LLag} can be dropped, and as is well-known, we are left with the Lagrangian that describes a noncommutative $T^2$ with a symplectic form $\omega=\displaystyle{\frac{\levely}{2\pi}d\xf^1\wedge d\xf^2}$.

% -------------------------------------------------------------
%\subsection{The Berry Phase}
%\label{subsec:Berry}

Defining the Berry connection in a standard way as
$$
(\ABerry_\tauLL)_{l j} = i\bra{\wfF_{j,\levely}}\partial_\tauLL\ket{\wfF_{l,\levely}}
\,,\qquad
(\ABerry_\btauLL)_{l j} = i\bra{\wfF_{j,\levely}}\partial_\btauLL\ket{\wfF_{l,\levely}},
$$
we calculate
$$
(\ABerry)_{l j} = (\ABerry_\tauLL)_{l j} d\tauLL + (\ABerry_\btauLL)_{l j} d\btauLL 
= -\frac{1}{4\tauLL_2}\delta_{l j}d\tauLL_1.
$$
We need to calculate the Berry phase along the path that $\tauLL$ takes, as $\xWS_1$ varies from $0$ to $1$. We parametrize the arc of the semicircle described at the end of \secref{subsec:MoreDJSols} (see \cite{Gaiotto:2008sd,Ganor:2014pha,Assel:2018vtq}) in terms of the variable $\GWpsi$ and parameters $\GWa$ and $\GWD$, introduced in \cite{Gaiotto:2008sd}:
$$
\tauGW = \GWa + 4\pi\GWD e^{2 i\GWpsi},
$$
where
$$
\GWa = \frac{\xa-\xd}{2\xc}\,,\qquad
4\pi\GWD=\frac{\sqrt{(\xa+\xd)^2-4}}{2|\xc|}\,,
$$
Defining $\GWpsi_0$ and $\GWpsi_1$ as the values of the phase $\GWpsi$ at the start and end of the arc, i.e., 
$$
\tauCS(0) = \GWa + 4\pi\GWD e^{2 i\GWpsi_0},\qquad
\tauCS(1) = 
\frac{\xa\tauCS(0)+\xb}{\xc\tauCS(0)+\xd}=
\GWa + 4\pi\GWD e^{2 i\GWpsi_1},
$$
we calculate the phase difference as \cite{Ganor:2014pha}:
$$
e^{i(\GWpsi_1-\GWpsi_0)}=
\sgn(\xa+\xd)
\frac{|\xc\tauCS(0)+\xd|}{\xc\tauCS(0)+\xd}\,,
$$
and the total Berry phase is easily calculated to be
\be\label{eqn:BerryPhaseValue}
e^{i\int_0^1\ABerry} = 
e^{\frac{i}{2}(\GWpsi_1-\GWpsi_0)}
=\left\lbrack
\sgn(\xa+\xd)
\frac{|\xc\tauCS(0)+\xd|}{\xc\tauCS(0)+\xd}
\right\rbrack^{1/2}.
\ee
The sign of the square root is determined so that $-\frac{\pi}{2}<\frac{1}{2}(\GWpsi_1-\GWpsi_0)<\frac{\pi}{2}$.

% --------------------------------------------------------------
\subsection{Modular transformations of the Landau wavefunctions}
\label{subsec:modularWF}

To introduce the $\matMx$-duality wall, we need to examine the behavior of the wavefunction under an $\SL(2,\Z)$ transformation that acts as
$$
\xf^1\rightarrow \xa\xf^1 +\xb\xf^2\,,\qquad
\xf^2\rightarrow \xc\xf^1 + \xd\xf^2\,,\qquad
\tauLL\rightarrow\frac{\xa\tauLL+\xb}{\xc\tauLL+\xd}\,.
$$

The general $\SL(2,\Z)$ transformation can be composed from the $\genSx$ and $\genTx$ generators, which act on wavefunctions as
\bear
S:\quad
\wfF_{j,\levely}\left(-\xf^2,\xf^1;-\tfrac{1}{\tauLL}\right) &=&
e^{-\frac{\pi i}{4}}
\left(\frac{\tauLL}{|\tauLL|}\right)^{\frac{1}{2}}
\frac{1}{\sqrt{\levely}}
\sum_{l=0}^{\levely-1}
e^{-\frac{2\pi i}{\levely} j l}
\wfF_{l,\levely}(\xf^1,\xf^2;\tauLL)
\,,\label{eqn:wfFS}
\eear
and for even $\levely$ we have
\be\label{eqn:wfFT2}
T:\quad
\wfF_{j,\levely}\left(\xf^1+\xf^2,\xf^2;\tauLL+1\right) =
e^{\frac{\pi i j^2}{\levely}} \wfF_{j,\levely}\left(\xf^1,\xf^2;\tauLL\right).
\ee
We note that for any $\levely\in\Z$ we have
\bear
\wfF_{j,\levely}\left(\xf^1+\xf^2,\xf^2;\tauLL+1\right) &=&
e^{\frac{i\pi j^2}{\levely}}
\frac{1}{2\pi}
\left(2 \levely \tauLL_2 \right)^{\frac{1}{4}} 
e^{-\frac{i\levely\xf^1\xf^2}{4\pi}}
\sum_{n=-\infty}^\infty 
e^{i(\levely n+j)\xf^1+ \pi i\levely\tauLL\left(n+\frac{j}{\levely}-\frac{\xf^2}{2\pi}\right)^2}
(-1)^{\levely n}.
\nn\\ &&
\label{eqn:wfFTOdd}
\eear
This is well-defined on the $\levely$-dimensional Hilbert space for $\levely\in 2\Z$, since the RHS is a linear combination of the $\wfF_{j,\levely}$'s, but for odd $\levely\in\Z$ only the square $\genTx^2$ is a well-defined operator on the Hilbert space.

% -----------------------------------------------------------
%\subsection{Fermionic contribution to the phase}
%We now discuss the reduction of the fermionic fields $\vPsi$ and $\bvPsi$ to 0+1D, and show that it does not contribute any nontrivial phase.

% -----------------------------------------------------------
\subsection{Recovering the Landsberg-Schaar relation}
\label{subsec:RecoverLS}
For $\matMx=\genSx\genTx^{q+2}$ and $\levely=2p$ we combine the two modular transformations \eqref{eqn:wfFS}-\eqref{eqn:wfFT2} to get
\be\label{eqn:wfFAfterM}
\wfF_{j,2p}\rightarrow
\left(\frac{\tauLL}{|\tauLL|}\right)^{\frac{1}{2}}
\frac{e^{-\frac{\pi i}{4}}}{\sqrt{2p}}
e^{\frac{\pi i (q+2)j^2}{2p}}
\sum_{l=0}^{2p-1}
e^{-\frac{\pi i}{p} j l}\wfF_{l,2p}\,.
\ee
In this expression $\tauLL$ is a shorthand for $\tauCS(0)+q+2$ [since this is the value of $\tauCS$ after $\genTx^{q+2}$ acts on $\tauCS(0)$], and $\wfF_{l,2p}$ on the RHS is a shorthand for $\wfF_{l,2p}\left(\xf^2+(q+2)\xf^1,-\xf^1; \tauCS(0)\right)$. The RHS of \eqref{eqn:wfFAfterM} represents the wavefunction at $\xWS_1=1$ (not including the Berry phase yet) after the action by $\matMx$. To complete the calculation of the partition function, we must multiply the RHS of \eqref{eqn:wfFAfterM} by $e^{-I''}$ [with $I''$ given in \eqref{eqn:IppAlt}], take its inner product with $\wfF_{j,2p}$, sum over $j$, and multiply by the Berry phase \eqref{eqn:BerryPhaseValue}.
Note that the role of $I''$ is to ensure periodicity in $\left(\xf^1,\xf^2\right)$, since the arguments in $\wfF_{l,2p}$ correspond to $\vX$ at $\xWS_1=1$, while those in $\wfF_{j,2p}$ correspond to $\vX$ at $\xWS_1=0$, and they can differ by $\vN$, as defined in \eqref{eqn:vNdef}. Moreover, since $\wfF_{j,2p}$ is periodic only up to a gauge transformation (in the language of the Landau problem of a particle in a uniform magnetic field), periodicity in $\left(\xf^1,\xf^2\right)$ can only be restored by including $e^{-I''}$ which plays the role of a gauge factor.
The resulting partition function is
$$
\PF' = 
\frac{e^{-\frac{\pi i}{4}}}{\sqrt{2p}}
\sum_{j=0}^{2p-1}e^{\frac{\pi i q j^2}{2p}} =
\frac{e^{-\frac{\pi i}{4}}}{\sqrt{2p}}\DualQGaussSum{q}{p}\,.
$$
Equating $\PF$ calculated in \eqref{eqn:PFFPFsumRed} with $\PF'$, we recover the (complex conjugate of the) basic Landsberg-Schaar relation \eqref{eqn:LSS}.

% ==========================================
\section{Identities for generalizations of Gauss sums}
\label{sec:Generalizations}

In previous sections we saw how the Landsberg-Schaar identity \eqref{eqn:LSrel} is recovered for duality twists of the form $\matMx=\genSx\genTx^{q+2}$ and $\matMy=\genSy\genTy^{2+2p}$, as in \secref{subsec:QGSPF}. We can get more complicated identities by looking at $\SL(2,\Z)$ elements which are expressed as longer words, with more $\genSx$ generators in $\matMx$ or $\genSy$ generators in $\matMy$.
In all cases, the identities that we get are of the schematic form
$$
\text{Tr}_{\mathcal{H}_{\matMx}}\left(\matMy\right)=e^{i\varphi}\text{Tr}_{\mathcal{H}_{\matMy}}(\matMx),
$$
where $\text{Tr}_{\mathcal{H}_{\matMx}}$ is a trace over the $|\det(\Id-\matMx)|$-dimensional Hilbert space $\mathcal{H}_{\matMx}$ of ground states of the $\matMx$-twisted circle compactification [whose states correspond to the finite abelian group $\GrSt$ defined in \eqref{eqn:GrStDef}], and $\text{Tr}_{\mathcal{H}_{\matMy}}$ is a similar trace over the $|\det(\Id-\matMy)|$-dimensional Hilbert space $\mathcal{H}_{\matMy}$ of ground states of the $\matMy$-twisted circle compactification, and $\varphi$ is a phase correction (arising from the Berry phase\footnote{The first place where a Berry phase appears as a multiplicative factor in the partition function is in \cite{Haldane2}, where each ``hedgehog'' defect, a singular configuration, of a spin field in a $(2+1)$d antiferromagnet, described by an $O(3)$ NLSM, carries a Berry phase. It is also directly related to the Wess-Zumino term \cite{Altland-Simons}} as in \secref{subsec:Berry}).

In some cases we will be able to rewrite the sum \eqref{eqn:PFFPFsumRed} explicitly, which requires identifying the abelian group $\GrSt$ as a direct sum of cyclic groups, and turns out to be of the form $\Z_{d_1}\oplus\Z_{d_2}$.
To achieve this we need to calculate the ``Smith normal form'' of the matrix $\Id-\matMx$, i.e., to find matrices $\smithP,\smithQ\in \SL(2,\Z)$ and unique integers $d_1, d_2\in\Z$ such that
$$
\Id-\matMx=\smithP\begin{pmatrix} d_1 & 0 \\ 0 & d_2 \\ \end{pmatrix}\smithQ.
$$
We will present a few examples below.

% -------------------------------------------------------------
\subsection{\texorpdfstring{$\matMx$}{} generated by \texorpdfstring{$\genSx$}{} and \texorpdfstring{$\genTx^2$}{} and \texorpdfstring{$\matMy=\genSy\genTy^{2+\levely}$}{}}
\label{subsec:MST2gen}

Let us assume that $\matMx$ can be expanded as the word
\be\label{eqn:MxST2}
\matMx = 
\genSx\genTx^{2\thalfpowx_1}
\genSx\genTx^{2\thalfpowx_2}
\cdots
\genSx\genTx^{2\thalfpowx_\tpowNumx},
\qquad
\text{with $\thalfpowx_1,\ldots,\thalfpowx_\tpowNumx$ nonzero integers.}
\ee
We will also assume that $\matMy=\genSy\genTy^{2+\levely}$, with $\levely$ even.
We recall that in $\SL(2,\Z)$ there are no relations among $\genSx$ and $\genTx^2$, other than those that follow from inserting an even number of $\genSx^2=-\Id$ in expressions, and therefore if $\matMx$ is of the form \eqref{eqn:MxST2}, the decomposition is unique. In fact the subgroup of $\SL(2,\Z)$ freely generated by $\genSx$ and $\genTx^2$ is isomorphic to the {\it Hecke congruence subgroup} $\Gamma_0(2)$. (See Example 3.7 of \cite{KeithConrad}.)

We now get an identity that equates the partition function $\PF$ given in the form \eqref{eqn:PFFPFsumRed} (with $\tpowy=\levely-2$), to a partition function calculated by combining the modular transformations of the $\levely$ ground-state wavefunctions and the Berry phase, as in \secref{subsec:Berry}-\secref{subsec:modularWF}.
The result of the latter is
$$
\left(\frac{e^{\frac{\pi i}{4}}}{\sqrt{\levely}}\right)^{\tpowNumx}
\sum_{l_1,\ldots,l_\tpowNumx=0}^{\levely-1}
\exp\left\{-\frac{2\pi i}{\levely}\left(
\sum_{j=1}^\tpowNumx\thalfpowx_j l_j^2
+\sum_{j=1}^{\tpowNumx-1} l_j l_{j+1} + l_1 l_\tpowNumx\right)
\right\}.
$$
The generalized identity for Gauss sums then takes the form
\be\label{eqn:MxManyTSMyTS}
\frac{1}{\sqrt{|\GrSt|}}
\sum_{\vN\in\GrSt}
\exp\left[
-i\pi\levely\vN^t\matEps(\Id-\matMx)^{-1}\vN
\right]
=
\frac{e^{\frac{\pi i\tpowNumx}{4}}}{\levely^{\tpowNumx/2}}
\sum_{l_1,\ldots,l_\tpowNumx=0}^{\levely-1}
e^{-\frac{2\pi i}{\levely}\left(
\sum_{j=1}^\tpowNumx\thalfpowx_j l_j^2
+\sum_{j=1}^{\tpowNumx-1} l_j l_{j+1} + l_1 l_\tpowNumx\right)
}\,.
\ee
For example, we take $\tpowNumx=2$ and
$$
\matMx = \genSx\genTx^{2\thalfpowx_1}
\genSx\genTx^{2\thalfpowx_2}
=\begin{pmatrix}
-1\quad  & -2\thalfpowx_2 \\
2\thalfpowx_1\quad & 4\thalfpowx_1\thalfpowx_2-1 \\
\end{pmatrix}.
$$
For simplicity, we assume that $\thalfpowx_1,\thalfpowx_2\ge 1$.
Then
$$
|\GrSt|=|\det(\Id-\matMx)|=4(\thalfpowx_1\thalfpowx_2-1).
$$
The Smith normal form of $\Id-\matMx$ is given by
$$
\Id-\matMx = 
\begin{pmatrix}
1\quad & \thalfpowx_2 \\
-2\thalfpowx_1\quad & 1-2\thalfpowx_1\thalfpowx_2 \\
\end{pmatrix}
\begin{pmatrix}
2(1-\thalfpowx_1\thalfpowx_2)\quad & 0 \\
0\quad & 2 \\
\end{pmatrix}
\begin{pmatrix}
1\quad & 0 \\
\thalfpowx_1\quad & 1 \\
\end{pmatrix}.
$$
[Note that the leftmost matrix on the RHS is in $\SL(2,\Z)$.]
An element of $\GrSt\cong\Z^2/(\Id-\matMx)(\Z^2)$ can then be parametrized as
$$
\vN=
\begin{pmatrix}
1\quad & \thalfpowx_2 \\
-2\thalfpowx_1\quad & 1-2\thalfpowx_1\thalfpowx_2 \\
\end{pmatrix}
\begin{pmatrix}
\nS \\
\aS\\
\end{pmatrix}
=
\begin{pmatrix}
\nS+\thalfpowx_2\aS \\
(1-2\thalfpowx_1\thalfpowx_2)\aS-2\thalfpowx_1\nS\\
\end{pmatrix}
\,,
$$
with
$$
\aS=0,1\,,\qquad
\nS=0, \dots, 2(\thalfpowx_1\thalfpowx_2-1)-1.
$$
And then we calculate
\be\label{eqn:NteN}
\vN^t\matEps(\Id-\matMx)^{-1}\vN=
\frac{\thalfpowx_1\nS^2}{2(\thalfpowx_1\thalfpowx_2-1)}
-\frac{\thalfpowx_2\aS^2}{2},
\ee
so the LHS of \eqref{eqn:MxManyTSMyTS} can be expressed as
$$
\frac{1}{\sqrt{|\GrSt|}}
\sum_{\vN\in\GrSt}
\exp\left[
-i\pi\levely\vN^t\matEps(\Id-\matMx)^{-1}\vN
\right]
=
\frac{1+i^{-\levely\thalfpowx_2}}{2\sqrt{\thalfpowx_1\thalfpowx_2-1}}
\sum_{\nS=0}^{2(\thalfpowx_1\thalfpowx_2-1)-1}
\exp\left(\frac{i\pi\levely\thalfpowx_1\nS^2}{2(\thalfpowx_1\thalfpowx_2-1)}\right).
$$
Setting $a=\thalfpowx_1$, $b=\thalfpowx_2$, and taking the complex conjugate, we find that \eqref{eqn:MxManyTSMyTS} becomes

%\be\label{eqn:LSgen1}
%\frac{1+i^{\levely\thalfpowx_2}}{2\sqrt{\thalfpowx_1\thalfpowx_2-1}}
%\sum_{\nS=0}^{2(\thalfpowx_1\thalfpowx_2-1)-1}
%\exp\left(-\frac{i\pi\levely\thalfpowx_1\nS^2}{2(\thalfpowx_1\thalfpowx_2-1)}\right)
%=
%-\frac{i}{\levely}
%\sum_{l_1,l_2=0}^{\levely-1}
%e^{\frac{2\pi i}{\levely}(\thalfpowx_1 l_1^2+2\thalfpowx_1\thalfpowx_2 +\thalfpowx_2 l_2^2)}
%\,.
%\ee

\be
\label{eqn:LSgen1}
\boxed{
-\frac{i}{\levely}
\sum_{m,n=0}^{\levely-1}
e^{\frac{2\pi i}{\levely}(a m^2 + b n^2 -2m n)}
=
\frac{1+i^{\levely b}}{2\sqrt{a b -1}}
\sum_{\nS=0}^{2a b -3}
\exp\left(
-\frac{\pi i\levely a\nS^2}{2(a b -1)}
\right)
}
\ee
for $a, b\in\Z, \, ab>1$, and $\levely\in 2\Z_{+}$. This is our first concrete generalization of the Landsberg-Schaar relation, whose proof we include in Appendix \ref{app:proof}.
Identity \eqref{eqn:LSgen1} is actually a special case of a collection of generalizations of the basic Landsberg-Schaar identity derived by Krazer in 1912 \cite{Krazer:1912} and other authors from then onwards, and the requirement for even $\levely$ also appears there. In our case it is a requirement that appeared at the end of \secref{sec:DTwists}.
We will discuss Krazer's and others' work in \secref{subsec:KrazerEtAl}.
We also note that double quadratic Gauss sums with denominators [$\levely$ in \eqref{eqn:LSgen1}] that are powers of a prime have been evaluated in \cite{Alaca:2016} in terms of the Legendre symbol.

% ---------------------------------------------------------------------------------
\subsection{More generalizations}
\label{subsec:MoreGens}

We can obtain more identities by allowing $\matMy$ to take the more general form
\be\label{eqn:MyST2}
\matMy = 
\genSy\genTy^{2\thalfpowy_1}
\genSy\genTy^{2\thalfpowy_2}
\cdots
\genSy\genTy^{2\thalfpowy_\tpowNumy},
\ee
with $\thalfpowy_1,\ldots,\thalfpowy_\tpowNumx$ nonzero integers.

We recall that $\genSy$ and $\genTy^2$ generate an index-$3$ congruence subgroup of $\SL(2,\Z)$, called the {\it theta subgroup}\footnote{However, any subgroup $\langle S,T^m\rangle$ with $m>2$ does not have a finite index inside $\SL(2,\mathbb{Z})$.}, and is isomorphic to the Hecke congruence subgroup $\Gamma_0(2)$ \cite{KeithConrad,Wang1}.
%\footnote{It is easy to check that 
%\begin{equation*}
%    \langle S,T^2\rangle=\left\{A\in \SL(2,\mathbb{Z})\bigg|A\equiv\begin{pmatrix}
%    1 & 0\\0 &1
%   \end{pmatrix}\text{ or }
%    \begin{pmatrix}
%    0 & 1\\1 & 0
%    \end{pmatrix}\text{ mod }2\right\},
%\end{equation*}
%and the conjugation class
%\begin{equation*}
%    \Gamma_0(2)=\begin{pmatrix}
%    1 & 0\\1 & 1
%    \end{pmatrix}\Lambda
%    \begin{pmatrix}
%    1 & 0\\1 & 1
%    \end{pmatrix}^{-1}.
%\end{equation*}}
Inserting the $\matMy$-twist in the $\xWS_2$ direction amounts to inserting $\tpowNumy$ duality walls of the type \eqref{eqn:Ibprime}, one for each $\genSy\genTy^{2\thalfpowy_j}$ factor ($j=1,\dots,\tpowNumy$).
The combined phase factor of modular transformations and the Berry phase, as in \secref{subsec:Berry}-\secref{subsec:modularWF}, now comes out to $\exp(\frac{i\pi}{4}\tpowNumx\tpowNumy)$.

If we repeat the analysis of \secref{subsec:Berry} for the system with $\tpowNumy$ duality walls, we get instead of \eqref{eqn:IbprimeReduced}, a reduced (0+1)d system that describes geometric quantization of $T^{2\tpowNumy}$ with an action given by
$$
-\frac{i}{4\pi}\int \sum_{\iKy,\jKy=1}^{\tpowNumy}\matKcply_{\iKy\jKy}\vXR_\iKy^t\matEps d\vXR_\jKy,
$$
where $\matKcply$ is an $\tpowNumy\times\tpowNumy$ integer coupling constant matrix given by an expression similar to \eqref{eqn:Kcpl}-\eqref{eqn:KcplTwo}, which for $\tpowNumy$ takes the form
\be\label{eqn:Kcply}
\matKcply\equiv
\begin{pmatrix}
2\thalfpowy_1 & -1 & 0 &  & -1 \\
-1 & \ddots & \ddots & \ddots & \\
0 & \ddots & \ddots & \ddots & 0 \\
&\ddots  & \ddots & \ddots & -1 \\
-1 &  & 0 & -1 & 2\thalfpowy_\tpowNumy \\
\end{pmatrix}\,,
\ee
and $\vXR_\iKy$ ($\iKy=1,\dots,\tpowNumy$) are coordinates on the $\iKy^{th}$ $T^2$ factor.
For $\tpowNumy=2$, $\matKcply$ takes the form
\be\label{eqn:KcplyTwo}
\matKcply\equiv
\begin{pmatrix}
2\thalfpowy_1 & -2 \\
-2 & 2\thalfpowy_2 \\
\end{pmatrix}\,.
\ee
For example, for $\tpowNumx=\tpowNumy=2$ we obtain the identity
\begin{subequations}
\begin{empheq}[box=\widefbox]{align}
\label{eqn:2v2}
&\frac{1}{pq-1}
\left[ 1+(-1)^{sq}\right]\sum_{m,n=0}^{2pq-3}
e^{-\frac{\pi i p}{pq-1}(s m^2+t n^2-2mn)}
\\
&= -\frac{1}{st-1}
\bigl\lbrack 1+(-1)^{tp}\bigr\rbrack
\sum_{m,n=0}^{2st-3}
e^{\frac{\pi i s}{st-1}(p m^2+q n^2-2mn)}\,,\qquad
\text{for $tq\in 2\Z$.}
\end{empheq}
\end{subequations}
where we have set $p=\thalfpowy_1$, $q=\thalfpowy_2$, $s=\thalfpowx_1$, $t=\thalfpowx_2$.
The requirement that $tq$ must be even arises as follows.
The abelian group $\GrSt$ defined in \eqref{eqn:GrStDef} turns out to be isomorphic to $\Z_{st-1}\oplus\Z_2^2$ in this case, and the sum over the $\Z_2^2$ factor produces a factor of $\bigl\lbrack 1+(-1)^{tq}\bigr\rbrack
\bigl\lbrack 1+(-1)^{sq}\bigr\rbrack$ on the LHS of \eqref{eqn:2v2}, while the RHS receives a factor of
$\bigl\lbrack 1+(-1)^{tq}\bigr\rbrack\bigl\lbrack 1+(-1)^{tp}\bigr\rbrack$ instead.
We cancelled the common factor of $\bigl\lbrack 1+(-1)^{tq}\bigr\rbrack$ by assuming $tq\in2\Z$. For $t$ and $q$ both odd, the relation in \eqref{eqn:2v2} is not generally correct, for example for $s=p=2$ and $t=q=1$ the two sides differ by a $(-)$ sign, and for $s=1$, $p=3$, $t=q=1$, the LHS is $8$ while the RHS is $0$.

% ----------------------------------------------------------------------------------
\subsection{Relation to Krazer's, Jeffrey's, Deloup's, and Turaev's reciprocity formulae}
\label{subsec:KrazerEtAl}

Let us now briefly discuss the relationship among the identities we found in \secref{subsec:MST2gen}-\secref{subsec:MoreGens} and a few known results in the mathematical literature spanning centuries.
In the late 19$^{\text{th}}$ century, a univariate formula which slightly generalizes the Landsberg-Schaar identity \eqref{eqn:LSrel} was discovered independently by Cauchy, Dirichlet, and Kronecker \cite{Turaev}:
\begin{equation}
\label{eqn:CDK}
|b|^{-1/2}\sum_{x\in\mathbb{Z}/b\mathbb{Z}}e^{\frac{\pi ia}{b}(x+\omega)^2}=e^{\frac{\pi i}{4}\text{sign}(ab)}|a|^{-1/2}\sum_{x\in\mathbb{Z}/a\mathbb{Z}}e^{-\frac{\pi ib}{a}x^2-2\pi i\omega x}\,,
\end{equation}
where $a,\,b$ are nonzero integers and $\omega\in\mathbb{Q}$ such that $ab+2a\omega\in2\mathbb{Z}$.
Later, a version of \eqref{eqn:CDK} for multivariate Gauss sums was obtained around 1912 by A. Krazer \cite{Krazer:1912,Braun, Siegel}:
\be
\label{eqn:Krazer}
d^{-\frac{m}{2}}\sum_{x\in(\mathbb{Z}/d\mathbb{Z})^m}e^{\frac{\pi ix^tAx}{d}}=\frac{d^{\frac{m-r}{2}}e^{\frac{\pi i}{4}\sigma(A)}}{|\det A|^{\frac{1}{2}}}\sum_{y\in\mathbb{Z}/A'\mathbb{Z}^m}e^{-\pi idy^tA'^{-1}y},
\ee
where $d$ is again a nonzero integer,  $A$ is a symmetric $m\times m$ matrix with integer entries, $\sigma(A)\in\mathbb{Z}$ is the signature of $A$ (i.e., the difference between numbers of positive and negative eigenvalues), and either $d$ or $A$ is {\it even} (i.e., all diagonal entries of $A$ are even). The $r\times r$ symmetric invertible matrix $A'$ with integer entries is determined from $A$ by finding a unimodular matrix $P$ such that $P^tAP=A'\oplus(0_{m-r})$, where $0_{m-r}$ is the zero matrix of size $m-r$.
Equation \eqref{eqn:Krazer} generalizes the case $\omega=0$ of \eqref{eqn:CDK} by replacing one of the numbers $a$ and $b$ in the exponents there by an integer-valued quadratic form given by $A$. Note that the input to the identity is a single bilinear form $A$ and an integer $d$, because $A'$ is determined by $A$.

In 1992, Jeffrey studied the semiclassical expansion of $SU(2)$ Chern-Simons partition functions on Lens spaces and torus bundles \cite{Jeffrey:1992tk}, and discovered a generalization %(Proposition 4.3 therein)
(which was slightly corrected by Deloup and Turaev \cite{Turaev2} in 2005):
\be
\label{eqn:Jeffrey}
    \text{vol}\left(\Lambda^*\right)\sum_{\lambda\in\Lambda/r\Lambda}e^{i\pi \langle\lambda,B\lambda/r\rangle+2\pi i\langle\lambda,\psi\rangle}=\left(\frac{\det B}{r^l}\right)^{-1/2}
    e^{i\pi\sigma(g)/4}
    \sum_{\mu\in \Lambda^*/B\Lambda^*}e^{-i\pi\langle\mu+\psi,rB^{-1}(\mu+\psi)\rangle},
\ee
where $\Lambda$ is a lattice of finite rank $l$ with $\Lambda^*$ being its dual, $\langle\cdot,\cdot\rangle$ is the inner product on the real vector space $\Lambda_{\mathbb{R}}=\Lambda\otimes_{\mathbb{Z}}\mathbb{R}$, $\psi\in \Lambda_{\mathbb{R}}$, $r\in\mathbb{Z}_{>0}$, and $B$ is a self-adjoint automorphism on $\Lambda_{\mathbb{R}}$ (i.e., a bilinear form). The volume $\text{vol}(\Lambda^*)$ is the absolute value of the determinant of a matrix obtained by expanding a basis of $\Lambda^*$ in terms of an orthonormal basis of $\Lambda_{\mathbb{R}}$.
%Because $\Lambda$ and $\Lambda^*$ are dual lattices, there is only one single quadratic form, not two, i.e., \textbf{A}$ and $\textbf{B}$ in Deloup's formula. 
%This formula was later corrected by Deloup and Turaev in 2005 \cite{Turaev2}, and the corrected version can be obtained from either Deloup's formula \cite{Deloup} or Turaev's formula \cite{Turaev} as a special case. 
The symmetric bilinear form $g:\Lambda\times \Lambda\rightarrow \mathbb{Z}$ is defined by $g(x,y)=\langle x,B(y)\rangle$ for all $x,y\in \Lambda$, and $\sigma(g)$ is the signature of a diagonal matrix presenting the bilinear extension $\Lambda_{\mathbb{R}}\times \Lambda_{\mathbb{R}}\rightarrow\mathbb{R}$ of $g$. Formula \eqref{eqn:Jeffrey} extends Krazer's formula in that the lattice to be summed over is now arbitrary, and on both sides there are additional linear terms in $\psi$ in the exponents, whose significance will be discussed in the next paragraph. Notice that there is still only {\it one single} bilinear form $B$.

In 1996, independently of Jeffrey, Deloup \cite{Deloup,Deloup3} geometrically generalized Krazer's formula \eqref{eqn:Krazer} (as well as Jeffrey's) by essentially replacing {\it both} integers $a$ and $b$ in \eqref{eqn:CDK} with bilinear forms, and applied his result to calculate topological invariants of $3$-manifolds, such as Witten-Reshetikhin-Turaev (WRT) invariants (i.e., Chern-Simons partition functions). For integral quadratic forms determined by {\it two} invertible, even, symmetric matrices $\matADeloup$ and $\matBDeloup$, Deloup's reciprocity theorem relates a Gauss sum with bilinear form $\matADeloup\otimes\matBDeloup^{-1}$ to another Gauss sum with bilinear form $-\matADeloup^{-1}\otimes\matBDeloup$.\footnote{For an $m\times m$ matrix $\matADeloup$ and $n\times n$ matrix $\matBDeloup^{-1}$, the $(mn)\times(mn)$ matrix $\matADeloup\otimes\matBDeloup^{-1}$ denotes the tensor (Kronecker) product.}
This identity appears as Theorem 3 in \cite{Deloup}, and we do not present it here, since it would require quite a few new notations and definitions.
In the context of our \secref{subsec:MoreGens}, $\matADeloup$ can be identified with the coupling constant matrix $\matKcply$ (derived from $\matMy$) of abelian Chern-Simons theory, while $\matBDeloup$ can be identified with a similar but independent matrix derived from $\matMx=ST^{2v_1}\cdots ST^{2v_s}$:
\be\label{eqn:Kcplx}
\matKcplx\equiv
\begin{pmatrix}
2\thalfpowx_1 & -1 & 0 &  & -1 \\
-1 & \ddots & \ddots & \ddots & \\
0 & \ddots & \ddots & \ddots & 0 \\
&\ddots  & \ddots & \ddots & -1 \\
-1 &  & 0 & -1 & 2\thalfpowx_\tpowNumx \\
\end{pmatrix}\,.
\ee
We note that in a Gauss sum with a bilinear form governed by $\matKcplx^{-1}$, the sum would be over the finite abelian group $\Z^n/\matKcplx(\Z^n)$ [where $\matKcplx(\Z^n)$ is the sublattice of $\Z^n$ generated by the columns of $\matKcplx$]. This abelian group is equivalent to $\GrSt$ in \eqref{eqn:GrStDef}, as shown in Appendix A of \cite{Ganor:2014pha}.
Deloup's reciprocity relation is actually more general, allowing non-even $\matADeloup$ and $\matBDeloup$ by introducing arithmetic ``Wu classes''. For a quadratic form $\xDeloup^t\matADeloup\xDeloup$ (with $\xDeloup\in\Z^n$), a {\it Wu class} is realized by a constant vector $\WuClass\in\Z^n$ such that $\xDeloup^t\matADeloup\xDeloup+\WuClass^t\xDeloup\in2\Z$ for all $\xDeloup\in\Z^n$ \cite{Deloup}. By adding such linear terms (also similarly in Jeffrey's), one overcomes the ambiguity in the definition of Gauss sums in \eqref{eqn:Krazer} for a non-even $\matADeloup$.
In our context, this suggests a possible extension to twists $\matMx$ and $\matMy$ beyond the $\SL(2,\mathbb{Z})$ form given by \eqref{eqn:MyST2}, by inserting operators linear in the bosonic field $\vX$ introduced in \secref{subsec:MinSUSYDJ}, which would correspond to ``vertex operators'', but we will not explore this possibility in the present paper.

% The identity is shown as Theorem 3 in his paper, which we reproduce here:

In 1998, Turaev further generalized \cite{Turaev} Deloup's formula to capture an arbitrary ``rational Wu class'' (which means that the Gauss sums are sums of exponentials in quadratic forms on a lattice plus linear terms with rational coefficients that ensure that the exponents are well-defined up to $2\pi i$).
Overall, the place of our construction and generalization in \secref{subsec:MST2gen}-\secref{subsec:MoreGens} is somewhere in between Krazer's/Jeffrey's formula and Deloup's theorem -- for example, if we set the duality twists to be  $\matMx=\genSx\genTx^{2\thalfpowx_1}
\genSx\genTx^{2\thalfpowx_2}$ and $\matMy=\genSy\genTy^{2\thalfpowy_1}
\genSy\genTy^{2\thalfpowy_2}$, we get bivariate quadratic forms on both sides of the identity \eqref{eqn:2v2}; the result is a special case of Deloup's formula, but beyond Krazer's formula.
In other words, our physical system is able to accommodate two independent bilinear forms, and the identities obtained in \eqref{eqn:LSgen1} and \eqref{eqn:2v2} are slightly more general than Jeffrey's formula \eqref{eqn:Jeffrey}.

%If we continue to add arbitrary numbers of copies of $\genSx\genTx^{2\thalfpowx_i}$ and $\genSy\genTy^{2\thalfpowy_j}$ on both twists $\matMx$ and $\matMy$, we will reach the limit of our construction. Our proof in Appendix \ref{app:proof} is specific to \eqref{eqn:LSgen1}, and the most general proof of Deloup's formula involves arithmetic ``Wu classes'' \cite{Deloup2} and is beyond classical harmonic analysis.

Recently, the corrected formula \eqref{eqn:Jeffrey} has been extensively used in studying the categorification of WRT invariants \cite{Gukov:2016gkn,Zedhat,Cheng:2018vpl,Chun:2019mal}. It plays an essential role in the derivation of $\hat{Z}$-invariants, or ``{\it homological blocks}'', of plumbed 3-manifolds $M_3$. The topology of $M_3$ is encoded in its plumbing graph, hence the linking matrix of the link corresponding to the graph. The colored Jones polynomials of this link, or equivalently the WRT invariant of $M_3$, is a Laurent polymonial $J[q,q^{-1}]$ in variable\footnote{$h^{\vee}$ is the dual Coxeter number of the gauge group of the pure Chern-Simons theory defined on $M_3$.} $\displaystyle{q\equiv \exp\left(\frac{2i\pi}{\lvk+h^{\vee}}\right)}$, which are raised to powers dictated by linking matrix elements. Formula \eqref{eqn:Jeffrey} then basically converts $J[q,q^{-1}]$ into a linear combination of homological blocks, i.e., a summation over the lattice defined by the linking matrix, whose cokernel determines $H^1(M_3)$.

Finally, for a more comprehensive and technical historical account of the long sequence of Quadratic Reciprocity formulae up to the early 20$^{\text{th}}$ century [i.e., just prior to Krazer's formula \eqref{eqn:Krazer}], consult Chapter 1 of \cite{Doyle} or Chapter 4 of \cite{Lemmermeyer}.

% ==============================================================
% ==============================================================
% ==============================================================

\section{Discussion and outlook}
\label{sec:Discussion}

We have constructed a supersymmetric double-Janus configuration for a 2d $\sigma$-model with $T^2$ target space, where all the moduli are allowed to vary along both coordinates, and we focused on a particular solution of the SUSY conditions whereby the complex structure varies along one direction and the K\"ahler structure varies along the other. We then placed the 2d double-Janus configuration on $T^2$ with periodic boundary condition that include $\SL(2,\Z)$-duality walls. We discovered a nontrivial interaction at the intersection of the duality walls, and we calculated the partition function and showed that it can be expressed as a quadratic Gauss sum. The fermionic and bosonic modes generally describe what might be called "a second quantized supersymmetric quantum mechanics'', where the single-particle energy levels are those of a 0+1d supersymmetric system with an arbitrary periodic superpotential. The fermionic and bosonic determinants cancel each other, leaving only a number-theoretic quadratic Gauss sum, which we could compute in two different ways, verifying the Landsberg-Schaar relation and obtaining generalizations. Our derivation of the Landsberg-Schaar relation contains somewhat similar ingredients to a method introduced by Armitage and Rogers \cite{ArmitageRogers}, where quantum mechanics with a toroidal phase space was also considered, although in their approach the physical time was quantized.

%As we have seen in Sections 3.1-3.3, our sigma-model is valid for target space $T^N$, and the correspondence between F-strings on a mapping torus and Chern-Simons on $T^2\times\mathbb{R}$ is established through the Hilbert spaces of the respective systems.

%{\bf (If we use the Chern-Simons on $T^4\times\mathbb{R}$, and again only consider $\\SL(2,\Z)$ twists on two boundaries, then we should obtain some identity resembling the quartic reciprocity.} 

Our work here is a prelude to the problem of an \SUSY{4} Super-Yang-Mills theory with nonabelian $U(n)$ gauge group compactified on a closed double-Janus configuration with an $\SL(2,\Z)$-duality twist. As suggested in \cite{Ganor:2014pha}, such a system can be studied by realizing it in terms of a stack of D$3$-branes and then mapping it to a system of weakly-coupled strings in Type-IIA on a mapping torus. That system can be compactified on another mapping torus, and the work in this paper provides the partition function of a sector of the weakly-coupled Type-IIA strings.
It suggests interesting connections between number theory and partition functions with $\SL(2,\Z)$-duality walls.

\section*{Acknowledgements}
We are grateful to Richard E. Borcherds, Chi-Ming Chang, Zhu-Xi Luo, Ryan Thorngren, Alessandro Tomasiello, Daniel Waldram, and Zhenghan Wang for helpful conversations and discussions. OJG is grateful to E.~Witten and the organizers of the workshop ``Master Lectures on Mathematical Physics: Edward Witten'' that took place at Tsinghua Sanya International Mathematics Forum (TSIMF), in December 2015, where part of this work was presented. This research was supported by the Berkeley Center of Theoretical Physics. The research of N.~R.~Torres-Chicon was supported by the National Science Foundation Graduate Research Fellowship Program under Grant No. DGE-11-06400.

\appendix

% ==============================================================

\section{Janus compactifications and mapping tori}
\label{app:JanusAndMT}

We begin with the 6d $(2,0)$ theory on a spacetime with metric given by
$$
ds^2=-dt^2+dx_1^2+dx_2^2+dy^2+R_4(y)^2 dx_4^2 +R_5^2 dx_5^2,
$$
where $R_5$ is a constant and $R_4$ varies with periodicities $x_4\sim x_4+2\pi$ and $x_5\sim x_5+2\pi$, and noncompact $-\infty<y<\infty$.
Note that direction $x_4$ forms an $S^1$-fibration over $\R$, the direction $y$, whose total space is a noncompact Riemann surface. The 6d $(2,0)$-theory compactification therefore falls within the class of theories studied by Gaiotto in \cite{Gaiotto:2009we}.
To preserve $8$ supersymmetries we introduce, as in \cite{Gaiotto:2009we}, an R-symmetry twist that matches the holonomy of the Riemann surface. This is done with a background gauge field for an $SO(2)$ subgroup of the $\Spin(5)$ R-symmetry group. Dimensional reduction along directions $x_4,x_5$ gives rise to a Janus configuration with $\tauYM(y)=iR_5/R_4$ varying with $y$ and taking values along the imaginary axis. Now, apply an $\SL(2,\R)$ transformation that acts as
\be\label{eqn:tauYMSL2R}
h:\tauYM\rightarrow\frac{\alpha\tauYM+\beta}{\gamma\tauYM+\delta},
\qquad
h=\begin{pmatrix}
\alpha & \beta \\
\gamma & \delta \\
\end{pmatrix}\in \SL(2,\R),
\ee
and linearly transforms $x_4,x_5$ to obtain the metric
\begin{equation}
\label{eq:comp}
ds^2=dt^2+dx_1^2+dx_2^2
+dy^2+R_4(y)^2\left(\delta dx_4'-\beta dx_5'\right)^2 
+R_5^2 \left(\alpha dx_5'-\gamma dx_4'\right)^2.
\end{equation}
We also modify the periodicity condition so that the new coordinates $x_4',x_5'$ are periodic with periodicity $2\pi$. The $3$-manifold in directions $y,x_4',x_5'$ still has an $SO(2)$ holonomy, but dimensional reduction along $x_4',x_5'$ now produces a 4d gauge theory with a complex structure $\tauYM(y)$ that varies along a semicircle in the upper half-plane. The semicircle is the image of the imaginary axis under \eqref{eqn:tauYMSL2R}.
The R-symmetry twist creates additional mass terms for the R-charged fields, as discovered in \cite{Gaiotto:2008sd}.

In the following we would like to take the further step of compactifying the $y$ direction as well, say $y\sim y+1$. This would be possible if we could find an $\SL(2,\Z)$ MCG transformation acting on $x_4',x_5'$ that relates $\tauYM(y)$ to $\tauYM(y+1)$. The $\SL(2,\Z)$ element will then be identified with $\matMx$, \textcolor{blue}{the S-duality twist in \eqref{eq:sdual}}, and one of the eigenvalues of $\matMx$ has to be $R_4(1)/R_4(0)$ [and the other eigenvalue will be $R_4(0)/R_4(1)$]. Note, however, that the area of the $T^2$ in directions $x_4',x_5'$, which is $(2\pi)^2R_4(y) R_5$, is not generally a periodic function of $y$, and so we can only compactify $y$ after the limit $R_4,R_5\rightarrow 0$ has been taken to reach the 4d SYM low-energy limit.

From the 4d SYM theory we can proceed to 2d by compactifying directions $x_1,x_2$ on another $T^2$ and taking the low-energy limit.
If the gauge group is $U(1)$ then we obtain a $\sigma$-model with $T^2$ target space from the gauge fields, whose complex structure is $\tauYM(y)$.
We can now proceed in a parallel fashion to \eqref{eq:comp} and replace the so far untouched 3d part of the metric with
\begin{equation}
dt^2+dx_1^2+dx_2^2
\rightarrow
dt^2+R_1(t)^2\left(\delta' dx_1'-\beta' dx_2'\right)^2 
+R_2^2 \left(\alpha' dx_2'-\gamma' dx_1'\right)^2,
\end{equation}
where
$$
\begin{pmatrix}
\alpha' & \beta' \\
\gamma' & \delta' \\
\end{pmatrix}\in\SL(2,\R).
$$
In the low-energy limit, the $U(1)$ gauge field will reduce to a 2d $\sigma$-model with $T^2$ target space whose complex structure varies with $y$ and whose K\"ahler structure varies with $t$. We can then compactify $t$ by imposing the periodicity $t\sim t+1$ together with an associated linear transformation on $(x_1',x_2')$ given by $\matMy\in\SL(2,\Z)$. This is possible in the limit $R_1,R_2\rightarrow 0$ provided that $R_1(0)/R_1(1)$ is one of the eigenvalues of $\matMy$ [the other eigenvalue being $R_1(1)/R_1(0)$].

We stress that the above construction is only a motivation for the $\sigma$-model that we studied in the main paper, which is a supersymmetric 2d $\sigma$-model with varying parameters.

% ==============================================================================
\section{Locality and action \texorpdfstring{$I''$}{} at the intersection of the \texorpdfstring{$\SL(2,\mathbb{Z})$}{} duality walls}
\label{app:Locality}

The bosonic part of the duality wall corresponding to $\matMy=\genS\genT^\tpow$ was given by the sum of $I_b'$ and $I''$ from \eqref{eqn:Ibprime} and \eqref{eqn:Ipp}:
\be\label{eqn:IbpIppAlt}
I_b'+I'' =
\left.\frac{i}{4\pi}\int_0^1
\bigl(2\vX_1^t\matEps d\vX_0-\tpow\vX_0^t\matEps d\vX_0\bigr)
\right\rvert_{\xWS_2=0}
+\frac{i}{4\pi}\left(\tpow-2\right)\vX(1,0)^t\matEps\matMx\vX(0,0)
\,.
\ee
We mentioned below \eqref{eqn:Ipp} that $I''$ is necessary to incorporate the periodic identification \eqref{eqn:vXvKshift} in a local way.
Let us now explain in what sense \eqref{eqn:IbpIppAlt} is a local action.
For one, it is crucial that the coefficient in \eqref{eqn:Ipp} is $\left(2-\tpow\right)$.
The key point is that $\vX_0$ and $\vX_1$ are only one set of representative functions on $\R^2$ that represent a map to the coset $T^2\simeq\R^2/\Z^2$. 
In order to analyze the locality of the expression, let us divide the range $0\le \xWS_1< 1$ into $\Nr$ segments labeled by $\Ir=0,\dots,\Nr-1$ with endpoints 
$$
0=\xWS_{1,0}<\xWS_{1,1}<\cdots<\xWS_{1,\Nr-1}<\xWS_{1,\Nr}=1\,.
$$
In each segment $\left[\xWS_{1,\Ir},\xWS_{1,\Ir+1}\right]$, we define fields $\vX_{0,\Ir}(\xWS_1)$, with $\xWS_{1,\Ir}\le\xWS_1\le\xWS_{1,\Ir+1}$. Similarly, at $\xWS_2=1$ we have another sequence of vector functions $\left\{\vX_{1,\Ir}\right\}_{\Ir=1}^\Nr$. Thanks to the equivalence \eqref{eqn:vXvKshift}, there is an infinite number of ways to pick $\vX_{0,\Ir}$ and $\vX_{1,\Ir}$ in each segment, but different choices differ by a constant vector $\vK$. To incorporate that ambiguity in a local way, we allow the boundary conditions at the point were $\xWS_{1,\Ir+1}$, where one segment touches the next one, to include a shift by $\vN_\Ir\in\Z^2$ [which plays the role of $\vK$ in \eqref{eqn:vXvKshift}]:
\be\label{eqn:segmentBC}
\vX_{i,\Ir}(\xWS_{1,\Ir+1})
=\matMx^{\matMpow_\Ir}\vX_{i,\Ir+1}\left(\xWS_{1,\Ir+1}\right)+2\pi\vN_\Ir
\,,\qquad
i=0,1.
\ee
In \eqref{eqn:segmentBC}, we also allow a twist by an integer power $\matMpow_\Ir\in\Z$ of $\matMx$ between the $\Ir^{th}$ and $(\Ir+1)^{th}$ segment. In order to comply with \eqref{eqn:vXMZ}, we require that the net twist be $\matMx$, so that:
$$
\sum_{\Ir=0}^{\Nr-1}\matMpow_\Ir = 1.
$$
Our convention is that $\Nr\sim 0$ so that $\vN_\Nr=\vN_0$.
We require that the action be a sum of local expressions, and we start by replacing the first two terms in \eqref{eqn:IbpIppAlt} with
\be\label{eqn:tIbp}
\tI_b'\equiv
-\frac{i}{4\pi}\sum_{\Ir=0}^{\Nr-1}\int_{\xWS_{1,\Ir}}^{\xWS_{1,\Ir+1}}
\left(
\tpow\vX_{0,\Ir}^t\matEps d\vX_{0,\Ir}
-2\vX_{1,\Ir}^t\matEps d\vX_{0,\Ir}
\right)
\,.
\ee
Next, we require local gauge invariance. The gauge transformations are labeled by $\Nr$ vectors $\vK_\Ir\in\Z^2$ and act as
\be\label{eqn:LocalK}
\vX_{i,\Ir}\rightarrow\vX_{i,\Ir}+2\pi\vK_\Ir\,\quad(i=0,1)\,,\qquad
\vN_\Ir\rightarrow\vN_\Ir+\vK_\Ir-\matMx^{\matMpow_\Ir}\vK_{\Ir+1}
\,,\qquad
\Ir=0,\dots,\Nr-1.
\ee
Under this gauge transformation the action transforms as
\bear
\tI_b' &\rightarrow&
\tI_b'
-\frac{i}{2}\left(\tpow-2\right)\sum_{\Ir=0}^{\Nr-1}
\vK_\Ir^t\matEps\left[
\matMx^{\matMpow_\Ir}\vX_{0,\Ir+1}(\xWS_{1,\Ir+1})
-\vX_{0,\Ir}(\xWS_{1,\Ir})\right]
-i\left(\tpow-2\right)\pi\sum_{\Ir=0}^{\Nr-1}\vK_\Ir^t\matEps\vN_\Ir
\,.\qquad
\label{eqn:tIbpRArrow}
\eear
Since we assume that $\tpow\in 2\Z$, the last term is in $2\pi\Z$ and we can drop it, since the path integral is over $e^{-I}$.
For the remaining terms, we can rearrange the sum to read
$$
\sum_{\Ir=0}^{\Nr-1}
	\vK_\Ir^t\matEps\left[\matMx^{\matMpow_\Ir}\vX_{0,\Ir+1}(\xWS_{1,\Ir+1})
	-\vX_{0,\Ir}(\xWS_{1,\Ir})\right]
=
\sum_{\Ir=1}^{\Nr}
\left[\vK_{\Ir-1}^t\left(\matMx^{-\matMpow_{\Ir-1}}\right)^t -\vK_\Ir^t
\right]
\matEps\vX_{0,\Ir}(\xWS_{1,\Ir}),
$$
where we used the identity
$
\left(\matMx^{\matMpow_{\Ir-1}}\right)^t\matEps\matMx^{\matMpow_{\Ir-1}}
=(\det\matMx)^{\matMpow_{\Ir-1}}\matEps=\matEps\,.
$
Thus, \eqref{eqn:tIbpRArrow} can be rewritten as
$$
\tI_b'\rightarrow
\tI_b'
-\frac{i}{2}\left(\tpow-2\right)\sum_{\Ir=1}^{\Nr}
\left[\vK_{\Ir-1}^t\left(\matMx^{-\matMpow_{\Ir-1}}\right)^t -\vK_\Ir^t
\right]
\matEps\vX_{0,\Ir}(\xWS_{1,\Ir})
\qquad
{\pmod{2\pi\Z}}\,.
$$
So, $\tI_b'$ is not gauge invariant on its own.
To fix it, we can add to $\tI_b'$ the term
\be\label{eqn:tIpp}
\tI''\equiv
\frac{i}{2}\left(\tpow-2\right)\sum_{\Ir=0}^{\Nr-1}
\vN_{\Ir-1}^t\left(\matMx^{-\matMpow_{\Ir-1}}\right)^t\matEps\vX_{0,\Ir}(\xWS_{1,\Ir}).
\ee
We can now check that $\tI_b'+\tI''$ is invariant,
using \eqref{eqn:LocalK} to calculate
$$
\vN_\Ir\rightarrow
\vN_\Ir+\vK_\Ir-\matMx^{\matMpow_\Ir}\vK_{\Ir+1}
\,,\qquad
\vN_{\Ir-1}^t\rightarrow
\vN_{\Ir-1}^t+\vK_{\Ir-1}^t-\vK_{\Ir}^t\left(\matMx^{\matMpow_{\Ir-1}}\right)^t
\,,
$$
and thus
$$
\vN_{\Ir-1}^t\left(\matMx^{-\matMpow_{\Ir-1}}\right)^t\rightarrow
\vN_{\Ir-1}^t\left(\matMx^{-\matMpow_{\Ir-1}}\right)^t
+\vK_{\Ir-1}^t\left(\matMx^{-\matMpow_{\Ir-1}}\right)^t-\vK_{\Ir}^t
\,.
$$
Now, if we set
$$
\matMpow_{-1}\equiv\matMpow_{\Nr-1}=1\,,\qquad
\matMpow_0 = \matMpow_1 = \cdots = \matMpow_{\Nr-2}=0
\,,
$$
and
$$
\vN_{-1}\equiv\vN_{\Nr-1}=\vN\,,\qquad
\vN_0=\vN_1 = \cdots = \vN_{\Nr-2}=0
\,,
$$
then we recover the action \eqref{eqn:IbpIpp}.

%\noindent\hrulefill
%
%To see this, note that \eqref{eqn:tIpp} now becomes
%$$
%\tI''=
%\frac{i}{2}\left(\tpow-2\right)
%\vN^t\left(\matMx^{-1}\right)^t\matEps\vX_{0,0}(\xWS_{1,0})
%=\frac{i}{2}\left(\tpow-2\right)
%\vN^t\matEps\matMx\vX(0,0),
%$$
%where we have used the relation $\left(\matMx^{-1}\right)^t\matEps=\matMx\matEps$ for $\matMx\in \SL(2,\R)$.
%According to \eqref{eqn:segmentBC}, with $\Ir=-1$,
%\be\label{eqn:vXMinus1}
%\vX_{i,-1}(\xWS_{1,0})
%=\matMx\vX_{i,0}(\xWS_{1,0})+2\pi\vN
%=\matMx\vX_{i,0}(0)+2\pi\vN
%\,,\qquad
%i=0,1.
%\ee
%Since $\xWS_{1,0}=0$ (which is identified with $\xWS_1=1$, and since $\vX_{i,-1}$ is defined to be $\vX_{i,\Nr-1}$, we translate $\vX_{i,-1}(\xWS_{1,0})$ to $\vX_i(1,0)$ and \eqref{eqn:vXMinus1} becomes
%$$
%\vX_i(1,0)
%=\matMx\vX_i(0,0)+2\pi\vN
%\,,\qquad
%i=0,1.
%$$
%Which agrees with \eqref{eqn:vNdef}.
%We then calculate
%\bear
%\tI'' &=&
%\frac{i}{2}\left(\tpow-2\right)\vN^t\matEps\matMx\vX(0,0)
%\frac{i}{4\pi}\left(\tpow-2\right)[\vX(1,0)-\matMx\vX(0,0)]^t\matEps\matMx\vX(0,0)
%\nn\\
%&=&
%\frac{i}{4\pi}\left(\tpow-2\right)\vX(1,0)^t\matEps\matMx\vX(0,0)
%-\frac{i}{4\pi}\left(\tpow-2\right)\vX(0,0)^t\matMx^t\matEps\matMx\vX(0,0)
%\,,
%\nn
%\eear
%and since $\matMx^t\matEps\matMx = (\det\matMx)\matEps = \matEps$ and $\vX(0,0)^t\matMx^t\matEps\matMx\vX(0,0) = 0$ by antisymmetry of $\matEps$, we finally arrive at
%$$
%\tI'' =
%\frac{i}{4\pi}\left(\tpow-2\right)\vX(1,0)^t\matEps\matMx\vX(0,0)
%$$
%as in \eqref{eqn:IbpIpp}.

% ==============================================================================
\section{On the equivalence between \texorpdfstring{$2\qfMT(\cdot)$}{} and \texorpdfstring{$\xV^t\matEps\matMx\xV$}{} in \texorpdfstring{\eqref{eqn:2qfMT}}{}}
\label{app:Proof2qfMT}

\subsection{Proof for \texorpdfstring{$\matMx=\genSx\genTx^{2\thalfpowx_1}$}{} and \texorpdfstring{$\matMx=\genSx\genTx^{2\thalfpowx_1}\genSx\genTx^{2\thalfpowx_2}$}{}}
\label{app:proof1}

We now show that \eqref{eqn:2qfMT} holds for the cases $\matMx=\genSx\genTx^{2\thalfpowx_1}$ and $\matMx=\genSx\genTx^{2\thalfpowx_1}\genSx\genTx^{2\thalfpowx_2}$.
Equation \eqref{eqn:2qfMT} states that the quadratic form $2\qfMT(\xV)$ can be written as $\xV^t\matEps\matMx\xV$ up to an integer. 

The case $\matMx=\genSx\genTx^{2\thalfpowx_1}$ is simple. The group $\GrSt$, defined in \eqref{eqn:GrStDef}, is isomorphic to the cyclic group $\Z_{(2\thalfpowx_1-1)}$, and an element of $(\matMx-\Id)^{-1}(\Z^2)/\Z^2$ can be expressed as 
$$
\xV=\frac{\nS}{2\thalfpowx_1-2}\begin{pmatrix}
1 \\
-1 \\
\end{pmatrix},\qquad
\text{for $\nS=0,\dots,2(\thalfpowx_1-1)-1.$}
$$
Then (half) the RHS of \eqref{eqn:2qfMT} is equal to
$$
\tfrac{1}{2}\xV^t\matEps\matMx\xV=\frac{\nS^2}{4(\thalfpowx_1-1)}.
$$
The coupling constant matrix $\matKcpl$ is $1\times 1$ and given by $(2\thalfpowx_1-2)$ which results in the same expression. Thus, in this case $\qfMT(\xV)$ equals $\frac{1}{2}\xV^t\matEps\matMx\xV$.

For the case $\matMx=\genSx\genTx^{2\thalfpowx_1}\genSx\genTx^{2\thalfpowx_2}$, (half) the RHS of \eqref{eqn:2qfMT} was calculated in \eqref{eqn:NteN} as
\be\label{eqn:xVteMxV}
\tfrac{1}{2}\xV^t\matEps\matMx\xV=
\frac{\thalfpowx_1\nS^2}{4(\thalfpowx_1\thalfpowx_2-1)}
-\frac{\thalfpowx_2\aS^2}{4}
\ee
using an identification of $\GrSt$ with $\Z_{2(\thalfpowx_1\thalfpowx_2-1)}\oplus\Z_2$ via
$$
\xV\mapsto\begin{pmatrix}\nS\\ \aS\\ \end{pmatrix},
\qquad
\aS=0,1\,,\qquad
\nS=0, \dots, 2(\thalfpowx_1\thalfpowx_2-1)-1.
$$
[The relation $\xV=(\matMx-\Id)^{-1}\vN$ is needed to relate \eqref{eqn:NteN} to \eqref{eqn:xVteMxV}.]

To calculate $\qf(\cdot)$ we note that
$$
\matKcpl=\begin{pmatrix}
2\thalfpowx_1 & -2\\
-2 & 2\thalfpowx_2 \\
\end{pmatrix}
=\begin{pmatrix}
0 & -1 \\
1 & \thalfpowx_2 \\
\end{pmatrix}
\begin{pmatrix}
2 (\thalfpowx_1 \thalfpowx_2-1) & 0 \\
0 & 2 \\
\end{pmatrix}
\begin{pmatrix}
1 & 0 \\
-\thalfpowx_1 & 1\\
\end{pmatrix}.
$$
Thus, we can define the isomorphism $\isomMTCS$ by
$$
\isomMTCS(\xV)=
\mcV=
\begin{pmatrix}
0 & -1 \\
1 & \thalfpowx_2 \\
\end{pmatrix}
\begin{pmatrix}
\nS \\
\aS\\
\end{pmatrix},
$$
and use \eqref{eqn:qfMTisomqf} to calculate 
\be\label{eqn:qfMTxV}
\qfMT(\xV)=\qf(\isomMTCS(\xV))
=
\tfrac{1}{2}\mcV^t K^{-1}\mcV
=\frac{\thalfpowx_2}{4}\aS^2+\frac{\thalfpowx_1}{4(\thalfpowx_1\thalfpowx_2-1)}\nS^2+\frac{1}{2}\aS\nS.
\ee
Comparing \eqref{eqn:qfMTxV} to \eqref{eqn:xVteMxV} we see that in general
$$
\qfMT(\xV)\neq\tfrac{1}{2}\xV^t\matEps\matMx\xV{\pmod\Z}
$$
but
$$
2\qfMT(\xV)\equiv\xV^t\matEps\matMx\xV{\pmod\Z},
$$
as stated.

% ==============================================================================

% ----------------------------------------------------------------------------------
%\section{Examples of ``exotic'' \texorpdfstring{$\mathcal{T}$}{} and \texorpdfstring{$\mathcal{S}$}{} matrices from strings on a mapping torus in \secref{subsec:MT}}
%In \secref{subsec:MT}, we have encountered 
%According to definitions 

\subsection{Example for \texorpdfstring{$M$}{} with three \texorpdfstring{$ST^{\#}$}{} factors}
\label{app:proof2}
Although we did not prove the equivalence of $2\qf(\xV)$ and $\xV^t\matEps\matMx\xV$ (mod $1$) for the case with three factors of $ST^{\#}$ inside the decomposition of $M$, we here present a numerical example to support our conjecture that signs could be added by hand to the {\it naive}\footnote{In this section, by ``naive'' we mean that we do not consider any $\pm$ sign shown in \eqref{eqn:TyMT}, or the phase $\phi$ predicted by the signature $\sig$ of the quadratic form $\xV^t\matEps\matMx\xV$ in \eqref{eqn:2qfMT}.} $\widetilde{\mathscr{T}}$ to satisfy $(\mathscr{S}\widetilde{\mathscr{T}})^3=\mathscr{S}^2$. The naive $\widetilde{\mathscr{T}}$ is defined as
$$
\widetilde{\mathscr{T}}\ket{v}\equiv e^{-\pi iv^t\epsilon Mv}\ket{v}, \quad v\in \GrSt.
$$

Let us consider
\begin{equation}
\label{eq:example}
M=ST^2ST^2ST^4=\begin{pmatrix}
-2 & -7\\3 & 10
\end{pmatrix},
\end{equation}
which yields a cyclic group $\GrSt\cong\mathbb{Z}_6$ with an ordered basis
$$
\left\{\left(\frac{5}{6},\frac{1}{2}\right),\left(\frac{2}{3},0\right),\left(\frac{1}{2},\frac{1}{2}\right),\left(\frac{1}{3},0\right),\left(\frac{1}{6},\frac{1}{2}\right),\left(0,0\right)\right\}.
$$

Here we do not calculate $\mathscr{S}$ according to the definition \eqref{eqn:SyMT}. The reason is as follows: without knowing the correct $\pm$ sign in front of each matrix element beforehand, if one were to choose all signs to be, say, $+$, then it is a numerical observation that almost always one would obtain a singular $\mathscr{S}$ where at least two rows are identical, and furthermore $\mathscr{S}^2$ would not be the charge conjugation operator. Hence, we start by using the unsymmetrized\footnote{The exponent is unsymmetrized (no symmetrization or antisymmetrization whatsoever), but $\mathscr{S}$ is still a symmetric matrix.}/naive definition:
\begin{equation}
\label{eq:unsymm}
\widetilde{\mathscr{S}}\ket{\xV} \equiv \frac{1}{\sqrt{|\GrSt|}}\sum_{\xU\in\latZv/\Z^2}
e^{2\pi i\xV^t\matEps(\matMx-\Id)\xU}\ket{\xU}, \quad u,v\in \GrSt.
\end{equation}

Then we calculate
$$
\widetilde{\mathscr{T}}=
\left(\begin{array}{cccccc}
    e^{\frac{5i\pi}{6}}   & &  & &  \\
     &    e^{-\frac{2i\pi}{3}}   &     &     &  \\ 
    &     &  -i   &     &  \\
          &  &  &  e^{\frac{i\pi}{3}}   &  \\
       &   &          &       &  e^{\frac{5i\pi}{6}}&  \\
       &   &          &       &  & 1 \\
  \end{array}\right),
$$
and
$$
\widetilde{\mathscr{S}}=
\frac{1}{\sqrt{6}}\left(\begin{array}{cccccc}
    e^{\frac{i\pi}{3}}   & e^{\frac{2i\pi}{3}}& -1 &e^{-\frac{2i\pi}{3}} & e^{-\frac{i\pi}{3}} &1 \\
    e^{\frac{2i\pi}{3}} &    e^{-\frac{2i\pi}{3}}   &  1   &   e^{\frac{2i\pi}{3}}  & e^{-\frac{2i\pi}{3}} & 1 \\ 
    -1 &  1   &  -1   &   1  & -1 & 1 \\
    e^{-\frac{2i\pi}{3}}  & e^{\frac{2i\pi}{3}} & 1 &  e^{-\frac{2i\pi}{3}}   & e^{\frac{2i\pi}{3}} & 1\\
    e^{-\frac{i\pi}{3}}   & e^{-\frac{2i\pi}{3}}  & -1  &   e^{\frac{2i\pi}{3}}    &  e^{\frac{i\pi}{3}} &  1\\
     1  &  1 &  1  &  1 & 1 & 1 \\
  \end{array}\right).
$$
We can easily check that $\widetilde{\mathscr{S}}^2$ is the Sylvester's ``shift'' matrix in the canonical basis
$$
\left(
\begin{array}{cccccc}
 0 & 0 & 0 & 0 & 1 & 0 \\
 0 & 0 & 0 & 1 & 0 & 0 \\
 0 & 0 & 1 & 0 & 0 & 0 \\
 0 & 1 & 0 & 0 & 0 & 0 \\
 1 & 0 & 0 & 0 & 0 & 0 \\
 0 & 0 & 0 & 0 & 0 & 1 \\
\end{array}
\right),
$$
i.e., the charge conjugation sending $\ket{v}$ to $\ket{-v}$, and $\widetilde{\mathscr{S}}^4=\Id$.

We found four solutions to $\big|(\widetilde{\mathscr{S}}\mathscr{T})^3\big|=\widetilde{\mathscr{S}}^2$, and below is the list of phases\footnote{Up to multiplications by $e^{\frac{2\pi i}{3}}$ and $e^{\frac{4\pi i}{3}}$, due to the three distinct ways of lifting a projective representation of $SL(2,\mathbb{Z})$ to a linear representation
(i.e., three distinct  ``linearizations''), see for example \cite{Wang1,TenerWang}.} required to multiply the entire $\mathscr{T}$ to possibly become the {\it refined} $\mathscr{T}$ as in \eqref{eqn:TyMT} satisfying $(\widetilde{\mathscr{S}}\mathscr{T})^3=\widetilde{\mathscr{S}}^2$, along with positions on the diagonal of $\widetilde{\mathscr{T}}$ where signs were flipped in each solution:
\begin{itemize}
    \item $\,$phase: $e^{-\frac{i\pi}{12}}$, positions: 1, 2, 3, 5, 6;
    \item $\,$phase: $e^{\frac{i\pi}{12}}$, positions: 1, 3, 4, 5;
    \item $\boxed{\text{phase:}\,\, e^{-\frac{i\pi}{4}}, \,\,\text{positions:}\,\, 2, 6};$
    \item $\,$phase: $e^{\frac{i\pi}{4}}$, position: 4.
\end{itemize}
We see that the third sign configuration yields a phase matching $\phi=e^{-i\pi\sigma(K)/12}$ in $\mcT$ in \eqref{eqn:WeilR}, since the signature of the $K$-matrix \eqref{eqn:KcplM} in this example 
\begin{equation}
    \begin{pmatrix}
    2 & -1 & 0\\
    -1 & 2 & -1\\
    0 & -1 & 4
    \end{pmatrix}
\end{equation}
is 3 [because exponents in $ST^{\#}$ factors of $M$ in \eqref{eqn:WeilR} are all positive].

Our final comment is that the naive $\widetilde{\mathscr{T}}$ and unsymmetrized $\widetilde{\mathscr{S}}$ matrices here also enable us to find the correct sign configuration on the refined $\mathscr{T}$ in \eqref{eqn:TyMT}, with a phase matching the signature of $K$-matrix for cases with two $ST^{\#}$ factors in $M$, as expected from \appref{app:proof1}. We also conjecture that the correct sign configuration on $\mathcal{S}$ in \eqref{eqn:STmatrixElements} and $\mathscr{S}$ in \eqref{eqn:SyMT} is exactly given by $\widetilde{\mathscr{S}}$ on the face value.

\subsection{Concrete exercises for \texorpdfstring{$M$}{} with two \texorpdfstring{$ST^{\#}$}{} factors}
The purpose of this subsection is to provide examples for \appref{app:proof1}. Our first example is 
$$
M=ST^2ST^4=\begin{pmatrix}
-1 & -4\\2 & 7
\end{pmatrix},
$$
which generate $\GrSt$ as $\Z_2\oplus\Z_2$, the Klein 4-group\footnote{Incidentally, we note that
$$
M=ST^{-1}ST^{-3}ST=\begin{pmatrix}
3 & 4\\2 & 3
\end{pmatrix}
$$
also give the same $\mathbb{Z}_2\oplus\mathbb{Z}_2$ with the same basis.}. Since everything is mod 1 here, we choose the ordered basis $\left\{(1/2,1/2),(1/2,0),(0,1/2),(0,0)\right\}$. Then its naive $\widetilde{\mathscr{T}}$ matrix is
$$
\mathscr{T}=\begin{pmatrix}
-i & 0 & 0& 0\\
0 & i & 0& 0\\
0 & 0 & -1& 0\\
0 & 0 & 0& 1\\
\end{pmatrix},
$$

If we were to following the customs to symmetrize the quadratic form on the exponent in \eqref{eqn:SyMT} to be a quadratic refinement (a symmetric bilinear form) \cite{Moore, Anyon}, then we would get a candidate ``$\mathcal{S}$-matrix'' with \textit{symmetrized} exponents $e^{-\pi i(\xV^t\matEps\matMx\xU+\xU^t\matEps\matMx\xV)}$:
$$
\mathcal{S}'=\frac{1}{2}\begin{pmatrix}
-1 & -1 & 1& 1\\
-1 & -1 & 1& 1\\
1 & 1 & 1& 1\\
1 & 1 & 1& 1\\
\end{pmatrix}.
$$
However, as we have claimed in \appref{app:proof2}, it is singular and its square is not a charge conjugation matrix.

On the other hand, another candidate ``$\mathcal{S}$-matrix'' with \textit{antisymmetrized} exponents $e^{-\pi i(\xV^t\matEps\matMx\xU-\xU^t\matEps\matMx\xV)}$ is
\begin{equation}
    \label{eq:antisymm}
\mathcal{S}''=\frac{1}{2}\begin{pmatrix}
1 & -1 & -1& 1\\
-1 & 1 & -1& 1\\
-1 & -1 & 1& 1\\
1 & 1 & 1& 1\\
\end{pmatrix},
\end{equation}
which now squared to be the identity, but one cannot achieve $\big|(\mathcal{S}''\widetilde{\mathscr{T}})^3\big|=\mathcal{S}''^2$ by flipping signs on elements in $\widetilde{\mathscr{T}}$, and the best one could possibly do is to flip the lower right entry in $\widetilde{\mathscr{T}}$ so that
$$
\left[\mathcal{S}''\begin{pmatrix}
-i & 0 & 0& 0\\
0 & i & 0& 0\\
0 & 0 & -1& 0\\
0 & 0 & 0& \boxed{-1}\\
\end{pmatrix}\right]^3=\begin{pmatrix}
1 & 0 & 0& 0\\
0 & 1 & 0& 0\\
0 & 0 & -1& 0\\
0 & 0 & 0& -1\\
\end{pmatrix},
$$
which is fine because we are working with (mod $1$) for MCG actions in \eqref{eqn:SyMT} and \eqref{eqn:TyMT}. However, in the upcoming example we will not be so lucky with antisymmetrization, which would fail there.

Now let us consider 
$$
M=ST^4ST^6=\begin{pmatrix}
-1 & -6\\4 & 23
\end{pmatrix},
$$
which generate $\GrSt$ with the basis of the following order 
\begin{equation}
\begin{split}
&\left\{\left(0,0\right),\left(\frac{7}{10},\frac{1}{10}\right),\left(\frac{2}{5},\frac{1}{5}\right),\left(\frac{1}{10},\frac{3}{10}\right),\left(\frac{4}{5},\frac{2}{5}\right),\left(\frac{1}{2},\frac{1}{2}\right),\left(\frac{1}{5},\frac{3}{5}\right),\left(\frac{9}{10},\frac{7}{10}\right),\right.\\ &\,\,\,\,\left(\frac{3}{5},\frac{4}{5}\right),\left(\frac{3}{10},\frac{9}{10}\right),\left(\frac{9}{10},\frac{1}{5}\right),\left(\frac{3}{5},\frac{3}{10}\right),\left(\frac{3}{10},\frac{2}{5}\right),\left(0,\frac{1}{2}\right),\left(\frac{7}{10},\frac{3}{5}\right),\\
&\,\,\,\,\left. \left(\frac{4}{5},\frac{9}{10}\right),\left(\frac{1}{10},\frac{4}{5}\right),\left(\frac{1}{2},0\right),\left(\frac{2}{5},\frac{7}{10}\right),\left(\frac{1}{5},\frac{1}{10}\right)\right\},
\end{split}
\end{equation}
so its naive $\widetilde{\mathscr{T}}$ matrix is
\begin{equation}
\begin{split}
\widetilde{\mathscr{T}}=
    \text{diag}&\left(1,e^{-\frac{3\pi}{10}i},e^{\frac{4\pi}{5}i},e^{-\frac{7\pi}{10}i},e^{-\frac{4\pi}{5}i},i,e^{-\frac{4\pi}{5}i},e^{-\frac{7\pi}{10}i},e^{\frac{4\pi}{5}i},e^{-\frac{3\pi}{10}i}, e^{-\frac{\pi}{5}i},e^{-\frac{3\pi}{10}i},e^{\frac{\pi}{5}i},-i,\right.\\
    &\,\,\,\left.e^{\frac{\pi}{5}i},e^{\frac{7\pi}{10}i},e^{-\frac{\pi}{5}i},-1,e^{\frac{3\pi}{10}i},e^{\frac{7\pi}{10}i}\right).
    \end{split}
\end{equation}

The antisymmatrization proposal similar to \eqref{eq:antisymm} will yield a candidate
$$
\mathcal{S}''=\scalemath{0.72}{\frac{1}{2\sqrt{5}}\left(
\begin{array}{cccccccccccccccccccc}
 1 & 1 & 1 & 1 & 1 & 1 & 1 & 1 & 1 & 1 & 1 & 1 & 1 & 1 & 1 & 1 & 1 & 1 & 1 & 1 \\
 1 & 1 & 1 & 1 & 1 & 1 & 1 & 1 & 1 & 1 & -1 & -1 & -1 & -1 & -1 & -1 & -1 & -1 & -1 & -1 \\
 1 & 1 & 1 & 1 & 1 & 1 & 1 & 1 & 1 & 1 & 1 & 1 & 1 & 1 & 1 & 1 & 1 & 1 & 1 & 1 \\
 1 & 1 & 1 & 1 & 1 & 1 & 1 & 1 & 1 & 1 & -1 & -1 & -1 & -1 & -1 & -1 & -1 & -1 & -1 & -1 \\
 1 & 1 & 1 & 1 & 1 & 1 & 1 & 1 & 1 & 1 & 1 & 1 & 1 & 1 & 1 & 1 & 1 & 1 & 1 & 1 \\
 1 & 1 & 1 & 1 & 1 & 1 & 1 & 1 & 1 & 1 & -1 & -1 & -1 & -1 & -1 & -1 & -1 & -1 & -1 & -1 \\
 1 & 1 & 1 & 1 & 1 & 1 & 1 & 1 & 1 & 1 & 1 & 1 & 1 & 1 & 1 & 1 & 1 & 1 & 1 & 1 \\
 1 & 1 & 1 & 1 & 1 & 1 & 1 & 1 & 1 & 1 & -1 & -1 & -1 & -1 & -1 & -1 & -1 & -1 & -1 & -1 \\
 1 & 1 & 1 & 1 & 1 & 1 & 1 & 1 & 1 & 1 & 1 & 1 & 1 & 1 & 1 & 1 & 1 & 1 & 1 & 1 \\
 1 & 1 & 1 & 1 & 1 & 1 & 1 & 1 & 1 & 1 & -1 & -1 & -1 & -1 & -1 & -1 & -1 & -1 & -1 & -1 \\
 1 & -1 & 1 & -1 & 1 & -1 & 1 & -1 & 1 & -1 & 1 & -1 & 1 & -1 & 1 & -1 & 1 & 1 & -1 & -1 \\
 1 & -1 & 1 & -1 & 1 & -1 & 1 & -1 & 1 & -1 & -1 & 1 & -1 & 1 & -1 & 1 & -1 & -1 & 1 & 1 \\
 1 & -1 & 1 & -1 & 1 & -1 & 1 & -1 & 1 & -1 & 1 & -1 & 1 & -1 & 1 & -1 & 1 & 1 & -1 & -1 \\
 1 & -1 & 1 & -1 & 1 & -1 & 1 & -1 & 1 & -1 & -1 & 1 & -1 & 1 & -1 & 1 & -1 & -1 & 1 & 1 \\
 1 & -1 & 1 & -1 & 1 & -1 & 1 & -1 & 1 & -1 & 1 & -1 & 1 & -1 & 1 & -1 & 1 & 1 & -1 & -1 \\
 1 & -1 & 1 & -1 & 1 & -1 & 1 & -1 & 1 & -1 & -1 & 1 & -1 & 1 & -1 & 1 & -1 & -1 & 1 & 1 \\
 1 & -1 & 1 & -1 & 1 & -1 & 1 & -1 & 1 & -1 & 1 & -1 & 1 & -1 & 1 & -1 & 1 & 1 & -1 & -1 \\
 1 & -1 & 1 & -1 & 1 & -1 & 1 & -1 & 1 & -1 & 1 & -1 & 1 & -1 & 1 & -1 & 1 & 1 & -1 & -1 \\
 1 & -1 & 1 & -1 & 1 & -1 & 1 & -1 & 1 & -1 & -1 & 1 & -1 & 1 & -1 & 1 & -1 & -1 & 1 & 1 \\
 1 & -1 & 1 & -1 & 1 & -1 & 1 & -1 & 1 & -1 & -1 & 1 & -1 & 1 & -1 & 1 & -1 & -1 & 1 & 1 \\
\end{array}
\right)},
$$
whose square and the fourth power are nothing close to a charge conjugation operator or the identity matrix.

Following the unsymmetrized definition \eqref{eq:unsymm}, we have the $\widetilde{\mathscr{S}}$ matrix
$$
\widetilde{\mathscr{S}}=\scalemath{0.72}{\frac{1}{2\sqrt{5}}\left(
\begin{array}{cccccccccccccccccccc}
 1 & 1 & 1 & 1 & 1 & 1 & 1 & 1 & 1 & 1 & 1 & 1 & 1 & 1 & 1 & 1 & 1 & 1 & 1 & 1 \\
 1 & e^{\frac{3 i \pi }{5}} & e^{-\frac{4 i \pi}{5}} & e^{-\frac{i \pi }{5}} & e^{\frac{2 i \pi }{5}} & -1 & e^{-\frac{2 i \pi }{5}} & e^{\frac{i \pi }{5}} & e^{\frac{4 i \pi }{5}} & e^{-\frac{3 i \pi }{5}} & e^{\frac{i \pi }{5}} & e^{\frac{4 i \pi }{5}} & e^{-\frac{3 i \pi}{5}} & 1 & e^{\frac{3 i \pi }{5}} & e^{\frac{2 i \pi }{5}} & e^{-\frac{i \pi }{5}} & -1 & e^{-\frac{4 i \pi }{5}} & e^{-\frac{2 i \pi}{5}} \\
 1 & e^{-\frac{4 i \pi}{5}} & e^{\frac{2 i \pi }{5}} & e^{-\frac{2 i \pi}{5}} & e^{\frac{4 i \pi }{5}} & 1 & e^{-\frac{4 i \pi}{5}} & e^{\frac{2 i \pi }{5}} & e^{-\frac{2 i \pi}{5}} & e^{\frac{4 i \pi }{5}} & e^{\frac{2 i \pi }{5}} & e^{-\frac{2 i \pi }{5}} & e^{\frac{4 i \pi }{5}} & 1 & e^{-\frac{4 i \pi}{5}} & e^{\frac{4 i \pi }{5}} & e^{-\frac{2 i \pi}{5}} & 1 & e^{\frac{2 i \pi }{5}} & e^{-\frac{4 i \pi}{5}} \\
 1 & e^{-\frac{i \pi}{5}} & e^{-\frac{2 i \pi}{5}} & e^{-\frac{3 i \pi}{5}} & e^{-\frac{4 i \pi }{5}} & -1 & e^{\frac{4 i \pi }{5}} & e^{\frac{3 i \pi }{5}} & e^{\frac{2 i \pi }{5}} & e^{\frac{i \pi }{5}} & e^{\frac{3 i \pi }{5}} & e^{\frac{2 i \pi }{5}} & e^{\frac{i \pi }{5}} & 1 & e^{-\frac{i \pi}{5}} & e^{-\frac{4 i \pi}{5}} & e^{-\frac{3 i \pi }{5}} & -1 & e^{-\frac{2 i \pi}{5}} & e^{\frac{4 i \pi }{5}} \\
 1 & e^{\frac{2 i \pi }{5}} & e^{\frac{4 i \pi }{5}} & e^{-\frac{4 i \pi }{5}} & e^{-\frac{2 i \pi}{5}} & 1 & e^{\frac{2 i \pi }{5}} & e^{\frac{4 i \pi }{5}} & e^{-\frac{4 i \pi}{5}} & e^{-\frac{2 i \pi}{5}} & e^{\frac{4 i \pi }{5}} & e^{-\frac{4 i \pi}{5}} & e^{-\frac{2 i \pi}{5}} & 1 & e^{\frac{2 i \pi }{5}} & e^{-\frac{2 i \pi}{5}} & e^{-\frac{4 i \pi}{5}} & 1 & e^{\frac{4 i \pi }{5}} & e^{\frac{2 i \pi }{5}} \\
 1 & -1 & 1 & -1 & 1 & -1 & 1 & -1 & 1 & -1 & -1 & 1 & -1 & 1 & -1 & 1 & -1 & -1 & 1 & 1 \\
 1 & e^{-\frac{2 i \pi}{5}} & e^{-\frac{4 i \pi }{5}} & e^{\frac{4 i \pi }{5}} & e^{\frac{2 i \pi }{5}} & 1 & e^{-\frac{2 i \pi}{5}} & e^{-\frac{4 i \pi}{5}} & e^{\frac{4 i \pi }{5}} & e^{\frac{2 i \pi }{5}} & e^{-\frac{4 i \pi }{5}} & e^{\frac{4 i \pi }{5}} & e^{\frac{2 i \pi }{5}} & 1 & e^{-\frac{2 i \pi}{5}} & e^{\frac{2 i \pi }{5}} & e^{\frac{4 i \pi }{5}} & 1 & e^{-\frac{4 i \pi}{5}} & e^{-\frac{2 i \pi}{5}} \\
 1 & e^{\frac{i \pi }{5}} & e^{\frac{2 i \pi }{5}} & e^{\frac{3 i \pi }{5}} & e^{\frac{4 i \pi }{5}} & -1 & e^{-\frac{4 i \pi}{5}} & e^{-\frac{3 i \pi}{5}} & e^{-\frac{2 i \pi}{5}} & e^{-\frac{i \pi}{5}} & e^{-\frac{3 i \pi}{5}} & e^{-\frac{2 i \pi}{5}} & e^{-\frac{i \pi}{5}} & 1 & e^{\frac{i \pi }{5}} & e^{\frac{4 i \pi }{5}} & e^{\frac{3 i \pi }{5}} & -1 & e^{\frac{2 i \pi }{5}} & e^{-\frac{4 i \pi}{5}} \\
 1 & e^{\frac{4 i \pi }{5}} & e^{-\frac{2 i \pi}{5}} & e^{\frac{2 i \pi }{5}} & e^{-\frac{4 i \pi}{5}} & 1 & e^{\frac{4 i \pi }{5}} & e^{-\frac{2 i \pi}{5}} & e^{\frac{2 i \pi }{5}} & e^{-\frac{4 i \pi}{5}} & e^{-\frac{2 i \pi}{5}} & e^{\frac{2 i \pi }{5}} & e^{-\frac{4 i \pi}{5}} & 1 & e^{\frac{4 i \pi }{5}} & e^{-\frac{4 i \pi}{5}} & e^{\frac{2 i \pi }{5}} & 1 & e^{-\frac{2 i \pi}{5}} & e^{\frac{4 i \pi }{5}} \\
 1 & e^{-\frac{3 i \pi}{5}} & e^{\frac{4 i \pi }{5}} & e^{\frac{i \pi }{5}} & e^{-\frac{2 i \pi}{5}} & -1 & e^{\frac{2 i \pi }{5}} & e^{-\frac{i \pi}{5}} & e^{-\frac{4 i \pi}{5}} & e^{\frac{3 i \pi }{5}} & e^{-\frac{i \pi}{5}} & e^{-\frac{4 i \pi}{5}} & e^{\frac{3 i \pi }{5}} & 1 & e^{-\frac{3 i \pi}{5}} & e^{-\frac{2 i \pi}{5}} & e^{\frac{i \pi }{5}} & -1 & e^{\frac{4 i \pi }{5}} & e^{\frac{2 i \pi }{5}} \\
 1 & e^{\frac{i \pi }{5}} & e^{\frac{2 i \pi }{5}} & e^{\frac{3 i \pi }{5}} & e^{\frac{4 i \pi }{5}} & -1 & e^{-\frac{4 i \pi}{5}} & e^{-\frac{3 i \pi}{5}} & e^{-\frac{2 i \pi}{5}} & e^{-\frac{i \pi}{5}} & e^{\frac{2 i \pi }{5}} & e^{\frac{3 i \pi }{5}} & e^{\frac{4 i \pi }{5}} & -1 & e^{-\frac{4 i \pi}{5}} & e^{-\frac{i \pi}{5}} & e^{-\frac{2 i \pi}{5}} & 1 & e^{-\frac{3 i \pi}{5}} & e^{\frac{i \pi }{5}} \\
 1 & e^{\frac{4 i \pi }{5}} & e^{-\frac{2 i \pi}{5}} & e^{\frac{2 i \pi }{5}} & e^{-\frac{4 i \pi}{5}} & 1 & e^{\frac{4 i \pi }{5}} & e^{-\frac{2 i \pi }{5}} & e^{\frac{2 i \pi }{5}} & e^{-\frac{4 i \pi}{5}} & e^{\frac{3 i \pi }{5}} & e^{-\frac{3 i \pi}{5}} & e^{\frac{i \pi }{5}} & -1 & e^{-\frac{i \pi}{5}} & e^{\frac{i \pi }{5}} & e^{-\frac{3 i \pi}{5}} & -1 & e^{\frac{3 i \pi }{5}} & e^{-\frac{i \pi}{5}} \\
 1 & e^{-\frac{3 i \pi}{5}} & e^{\frac{4 i \pi }{5}} & e^{\frac{i \pi }{5}} & e^{-\frac{2 i \pi }{5}} & -1 & e^{\frac{2 i \pi }{5}} & e^{-\frac{i \pi}{5}} & e^{-\frac{4 i \pi}{5}} & e^{\frac{3 i \pi }{5}} & e^{\frac{4 i \pi }{5}} & e^{\frac{i \pi }{5}} & e^{-\frac{2 i \pi}{5}} & -1 & e^{\frac{2 i \pi }{5}} & e^{\frac{3 i \pi }{5}} & e^{-\frac{4 i \pi}{5}} & 1 & e^{-\frac{i \pi}{5}} & e^{-\frac{3 i \pi}{5}} \\
 1 & 1 & 1 & 1 & 1 & 1 & 1 & 1 & 1 & 1 & -1 & -1 & -1 & -1 & -1 & -1 & -1 & -1 & -1 & -1 \\
 1 & e^{\frac{3 i \pi }{5}} & e^{-\frac{4 i \pi}{5}} & e^{-\frac{i \pi}{5}} & e^{\frac{2 i \pi }{5}} & -1 & e^{-\frac{2 i \pi}{5}} & e^{\frac{i \pi }{5}} & e^{\frac{4 i \pi }{5}} & e^{-\frac{3 i \pi}{5}} & e^{-\frac{4 i \pi}{5}} & e^{-\frac{i \pi}{5}} & e^{\frac{2 i \pi }{5}} & -1 & e^{-\frac{2 i \pi}{5}} & e^{-\frac{3 i \pi }{5}} & e^{\frac{4 i \pi }{5}} & 1 & e^{\frac{i \pi }{5}} & e^{\frac{3 i \pi }{5}} \\
 1 & e^{\frac{2 i \pi }{5}} & e^{\frac{4 i \pi }{5}} & e^{-\frac{4 i \pi}{5}} & e^{-\frac{2 i \pi }{5}} & 1 & e^{\frac{2 i \pi }{5}} & e^{\frac{4 i \pi }{5}} & e^{-\frac{4 i \pi}{5}} & e^{-\frac{2 i \pi }{5}} & e^{-\frac{i \pi}{5}} & e^{\frac{i \pi }{5}} & e^{\frac{3 i \pi }{5}} & -1 & e^{-\frac{3 i \pi }{5}} & e^{\frac{3 i \pi }{5}} & e^{\frac{i \pi }{5}} & -1 & e^{-\frac{i \pi}{5}} & e^{-\frac{3 i \pi}{5}} \\
 1 & e^{-\frac{i \pi}{5}} & e^{-\frac{2 i \pi }{5}} & e^{-\frac{3 i \pi}{5}} & e^{-\frac{4 i \pi }{5}} & -1 & e^{\frac{4 i \pi }{5}} & e^{\frac{3 i \pi }{5}} & e^{\frac{2 i \pi }{5}} & e^{\frac{i \pi }{5}} & e^{-\frac{2 i \pi}{5}} & e^{-\frac{3 i \pi}{5}} & e^{-\frac{4 i \pi}{5}} & -1 & e^{\frac{4 i \pi }{5}} & e^{\frac{i \pi }{5}} & e^{\frac{2 i \pi }{5}} & 1 & e^{\frac{3 i \pi }{5}} & e^{-\frac{i \pi}{5}} \\
 1 & -1 & 1 & -1 & 1 & -1 & 1 & -1 & 1 & -1 & 1 & -1 & 1 & -1 & 1 & -1 & 1 & 1 & -1 & -1 \\
 1 & e^{-\frac{4 i \pi}{5} } & e^{\frac{2 i \pi }{5}} & e^{-\frac{2 i \pi}{5}} & e^{\frac{4 i \pi }{5}} & 1 & e^{-\frac{4 i \pi}{5}} & e^{\frac{2 i \pi }{5}} & e^{-\frac{2 i \pi}{5}} & e^{\frac{4 i \pi }{5}} & e^{-\frac{3 i \pi}{5}} & e^{\frac{3 i \pi }{5}} & e^{-\frac{i \pi}{5}} & -1 & e^{\frac{i \pi }{5}} & e^{-\frac{i \pi}{5}} & e^{\frac{3 i \pi }{5}} & -1 & e^{-\frac{3 i \pi}{5}} & e^{\frac{i \pi }{5}} \\
 1 & e^{-\frac{2 i \pi}{5}} & e^{-\frac{4 i \pi}{5} } & e^{\frac{4 i \pi }{5}} & e^{\frac{2 i \pi }{5}} & 1 & e^{-\frac{2 i \pi }{5}} & e^{-\frac{4 i \pi}{5}} & e^{\frac{4 i \pi }{5}} & e^{\frac{2 i \pi }{5}} & e^{\frac{i \pi }{5}} & e^{-\frac{i \pi}{5}} & e^{-\frac{3 i \pi}{5}} & -1 & e^{\frac{3 i \pi }{5}} & e^{-\frac{3 i \pi }{5}} & e^{-\frac{i \pi }{5}} & -1 & e^{\frac{i \pi }{5}} & e^{\frac{3 i \pi }{5}} \\
\end{array}
\right)},
$$
which squared to be the charge conjugation operator -- the ``shift'' matrix in a non-canonical basis.

It is not hard to use \textit{Mathematica} to find 8 solutions to $\big|(\widetilde{\mathscr{S}}\mathscr{T})^3\big|=\widetilde{\mathscr{S}}^2$, out of $2^{20}=1,048,576$ possible sign configurations of $\widetilde{\mathscr{T}}$'s diagonal elements. Below for each solution, we list positions of entries in $\widetilde{\mathscr{T}}$ whose signs are  flipped, as well as the phase needs to be added by hand to convert the naive $\widetilde{\mathscr{T}}$ matrix into the refined $\mathscr{T}$ so that $(\widetilde{\mathscr{S}}\mathscr{T})^3=\widetilde{\mathscr{S}}^2$:
\begin{itemize}
    \item $\,$phase: $e^{\frac{i\pi}{6}}$ or $e^{\frac{5i\pi}{6}}$ or $e^{-\frac{i\pi}{2}}$, positions: 1, 2, 3, 4, 5, 6, 7, 8, 9, 10, 11, 13, 15, 17, 18;
    \item $\,$phase: $1$ or $e^{\frac{2i\pi}{3}}$ or $e^{\frac{4i\pi}{3}}$, positions 1, 2, 3, 4, 5, 6, 7, 8, 9, 10, 12, 14, 15, 19, 20;
    \item $\boxed{\text{phase:}\,\, e^{\frac{i\pi}{2}} \,\,\text{or}\,\, e^{\frac{7i\pi}{6}} \,\,\text{or}\,\, e^{-\frac{i\pi}{6}}, \,\,\text{positions: 1, 3, 5, 7, 9, 11, 12, 13, 14, 15, 16, 17, 18, 19, 20}}$;
    \item $\,$phase: $1$ or $e^{\frac{2i\pi}{3}}$ or $e^{\frac{4i\pi}{3}}$, positions: 1, 3, 5, 7, 9;
    \item $\,$phase: $-1$ or $e^{\frac{i\pi}{3}}$ or $e^{-\frac{i\pi}{3}}$, positions: 2, 4, 6, 8, 10, 11, 12, 13, 14, 15, 16, 17, 18, 19, 20;
    \item $\,$phase: $e^{\frac{i\pi}{6}}$ or $e^{\frac{5i\pi}{6}}$ or $e^{-\frac{i\pi}{2}}$, positions: 2, 4, 6, 8, 10;
    \item $\,$phase: $-1$ or $e^{\frac{i\pi}{3}}$ or $e^{-\frac{i\pi}{3}}$, positions: 11, 13, 15, 17, 18;
    \item $\boxed{\text{phase:}\,\, e^{\frac{i\pi}{2}} \,\,\text{or}\,\, e^{\frac{7i\pi}{6}} \,\,\text{or}\,\, e^{-\frac{i\pi}{6}}, \,\,\text{positions: 12, 14, 16, 19, 20}}$.
\end{itemize}

Now in this case, the $K$-matrix defined as \eqref{eqn:KcplM2} is
\begin{equation}
    \begin{pmatrix}
    4 & -2\\
    -2 & 6
    \end{pmatrix},
\end{equation}
whose signature is $2$, meaning that the third and the last solutions yield the right phase as dictated by comments following \eqref{eqn:WeilR} and \eqref{eqn:TSCSK}.

\subsection{A concrete exercise for \texorpdfstring{$M$}{} with 4 factors of \texorpdfstring{$ST^{\#}$}{}}
In fact, our above procedure goes beyond 3 factors of $ST^{\#}$ in $M$. To support this claim, we finally consider 
\begin{equation}
M=ST^2ST^3ST^2ST^3=\begin{pmatrix}
-5 & -12\\8 & 9
\end{pmatrix}
\end{equation}
with four $ST^{\#}$ factors. 

In the ordered basis of $\GrSt$
\begin{equation}
\begin{split}
&\left\{(0,0),\left(0,\frac{1}{2}\right),\left(\frac{1}{2},\frac{2}{3}\right),\left(\frac{1}{2},\frac{1}{6}\right),\left(0,\frac{1}{3}\right),\left(0,\frac{5}{6}\right),\left(\frac{1}{2},0\right),\left(\frac{1}{2},\frac{1}{2}\right),\left(0,\frac{2}{3}\right),\left(0,\frac{1}{6}\right),\right.\\ &\left.\,\,\,\left(\frac{1}{2},\frac{1}{3}\right),\left(\frac{1}{2},\frac{5}{6}\right)\right\},
\end{split}
\end{equation}
the naive $\widetilde{\mathscr{T}}$ matrix is
\begin{equation}
    \widetilde{\mathscr{T}}=\text{diag}\left(1,-1,e^{-\frac{2 i \pi }{3}},e^{\frac{i \pi }{3}},e^{-\frac{2 i \pi}{3}},e^{\frac{i \pi }{3}},1,-1,e^{-\frac{2 i \pi}{3}},e^{\frac{i \pi }{3}},e^{-\frac{2 i \pi}{3}},e^{\frac{i \pi }{3}}\right).
\end{equation}

The unsymmetrized $\widetilde{\mathscr{S}}$ matrix is
    \begin{equation}
    \widetilde{\mathscr{S}}=\scalemath{0.85}{\frac{1}{2\sqrt{3}}\left(
\begin{array}{cccccccccccc}
 1 & 1 & 1 & 1 & 1 & 1 & 1 & 1 & 1 & 1 & 1 & 1 \\
 1 & 1 & -1 & -1 & 1 & 1 & -1 & -1 & 1 & 1 & -1 & -1 \\
 1 & -1 & e^{-\frac{2 i \pi}{3}} & e^{\frac{i \pi }{3}} & e^{\frac{2 i \pi }{3}} & e^{-\frac{i \pi }{3}} & 1 & -1 & e^{-\frac{2 i \pi}{3}} & e^{\frac{i \pi }{3}} & e^{\frac{2 i \pi }{3}} & e^{-\frac{i \pi }{3}} \\
 1 & -1 & e^{\frac{i \pi }{3}} & e^{-\frac{2 i \pi}{3}} & e^{\frac{2 i \pi }{3}} & e^{-\frac{i \pi}{3}} & -1 & 1 & e^{-\frac{2 i \pi}{3}} & e^{\frac{i \pi }{3}} & e^{-\frac{i \pi}{3}} & e^{\frac{2 i \pi }{3}} \\
 1 & 1 & e^{\frac{2 i \pi }{3}} & e^{\frac{2 i \pi }{3}} & e^{-\frac{2 i \pi}{3}} & e^{-\frac{2 i \pi}{3}} & 1 & 1 & e^{\frac{2 i \pi }{3}} & e^{\frac{2 i \pi }{3}} & e^{-\frac{2 i \pi}{3}} & e^{-\frac{2 i \pi }{3}} \\
 1 & 1 & e^{-\frac{i \pi}{3}} & e^{-\frac{i \pi }{3}} & e^{-\frac{2 i \pi}{3}} & e^{-\frac{2 i \pi}{3}} & -1 & -1 & e^{\frac{2 i \pi }{3}} & e^{\frac{2 i \pi }{3}} & e^{\frac{i \pi }{3}} & e^{\frac{i \pi }{3}} \\
 1 & -1 & 1 & -1 & 1 & -1 & 1 & -1 & 1 & -1 & 1 & -1 \\
 1 & -1 & -1 & 1 & 1 & -1 & -1 & 1 & 1 & -1 & -1 & 1 \\
 1 & 1 & e^{-\frac{2 i \pi}{3}} & e^{-\frac{2 i \pi}{3}} & e^{\frac{2 i \pi }{3}} & e^{\frac{2 i \pi }{3}} & 1 & 1 & e^{-\frac{2 i \pi}{3}} & e^{-\frac{2 i \pi}{3}} & e^{\frac{2 i \pi }{3}} & e^{\frac{2 i \pi }{3}} \\
 1 & 1 & e^{\frac{i \pi }{3}} & e^{\frac{i \pi }{3}} & e^{\frac{2 i \pi }{3}} & e^{\frac{2 i \pi }{3}} & -1 & -1 & e^{-\frac{2 i \pi}{3}} & e^{-\frac{2 i \pi}{3}} & e^{-\frac{i \pi}{3}} & e^{-\frac{i \pi}{3}} \\
 1 & -1 & e^{\frac{2 i \pi }{3}} & e^{-\frac{i \pi}{3}} & e^{-\frac{2 i \pi}{3}} & e^{\frac{i \pi }{3}} & 1 & -1 & e^{\frac{2 i \pi }{3}} & e^{-\frac{i \pi}{3}} & e^{-\frac{2 i \pi}{3}} & e^{\frac{i \pi }{3}} \\
 1 & -1 & e^{-\frac{i \pi}{3}} & e^{\frac{2 i \pi }{3}} & e^{-\frac{2 i \pi}{3}} & e^{\frac{i \pi }{3}} & -1 & 1 & e^{\frac{2 i \pi }{3}} & e^{-\frac{i \pi}{3}} & e^{\frac{i \pi }{3}} & e^{-\frac{2 i \pi}{3}} \\
\end{array}.
\right)}
\end{equation}

We find 8 solutions to $(\widetilde{\mathscr{S}}\mathscr{T})^3=\widetilde{\mathscr{S}}^2$ by flipping signs and multiplying phases on the diagonal of $\mathscr{T}$:
\begin{itemize}
    \item $\,$phase: $e^{\frac{i\pi}{3}}$, positions: 1, 2, 3, 5, 6, 7, 9, 10, 11;
    \item $\boxed{\text{phase:}\,\, e^{\frac{-i\pi}{3}}, \,\,\text{positions: 1, 2, 4, 5, 6, 8, 9, 10, 12}}$;
    \item $\,$phase: $e^{\frac{i\pi}{3}}$, positions: 1, 3, 4, 5, 7, 8, 9, 11, 12;
    \item $\,$phase: $e^{\frac{i\pi}{3}}$, positions: 2, 3, 4, 6, 7, 8, 10, 11, 12;
    \item $\boxed{\text{phase:}\,\, e^{\frac{-i\pi}{3}}, \,\,\text{positions: 1, 5, 9}}$;
    \item $\boxed{\text{phase:}\,\, e^{\frac{-i\pi}{3}}, \,\,\text{positions: 1, 3, 4, 5, 7, 8, 9, 11, 12}}$;
    \item $\,$phase: $e^{\frac{i\pi}{3}}$, positions: 1, 2, 4, 5, 6, 8, 9, 10, 12;
    \item $\boxed{\text{phase:}\,\, e^{\frac{-i\pi}{3}}, \,\,\text{positions: 1, 2, 3, 5, 6, 7, 9, 10, 11}}$.
\end{itemize}

The $K$-matrix of $M$ is
\begin{equation}
    \begin{pmatrix}
    2 & -1 & 0 & -1\\
    -1 & 3 & -1 & 0\\
    0 & -1 & 2 & -1\\
    -1 & 0 & -1 & 3
    \end{pmatrix}
\end{equation}
with signature 4, and again all phases are up to multiplications by $e^{\frac{2i\pi}{3}}$ and $e^{-\frac{2i\pi}{3}}$, so we see that the second, the fifth, the sixth and the last solutions yield the correct results.

Finally, we propose the following conjecture:

\noindent
\fbox{
\begin{minipage}{5.75 in}
\textbf{Conjecture} For any $M\in\SL(2,\Z)$ built from a finite number of $ST^{\#}$ factors with any integer power $\#$, we perform the following steps:
\begin{itemize}
    \item Compute the unsymmetrized $\widetilde{\mathscr{S}}$ and the naive $\widetilde{\mathscr{T}}$ \`a la \eqref{eqn:STmatrixElements}, or \`a la \eqref{eqn:SyMT} and \eqref{eqn:TyMT}; \item Flip signs on $\widetilde{\mathscr{T}}$'s diagonal to get $\mathscr{T}$ satisfying $\big|(\widetilde{\mathscr{S}}\mathscr{T})^3\big|=\widetilde{\mathscr{S}}^2$; \item Multiply $\mathscr{T}$ with a necessary phase $\phi$ to satisfy $(\widetilde{\mathscr{S}}\mathscr{T})^3=\widetilde{\mathscr{S}}^2$. 
    \end{itemize}
    
    We claim that the resulting $\widetilde{\mathscr{S}}$ and $\mathscr{T}$ always exist, and they give the correct sign configurations for $\mathcal{S}$ and $\mathcal{T}$ in \eqref{eqn:STmatrixElements}; as well as $\mathscr{S}$ in \eqref{eqn:SyMT} and $\mathscr{T}$ in \eqref{eqn:TyMT}. Moreover, $\phi$ equals $e^{-i\pi \sigma(K)/12}$ as mentioned below \eqref{eqn:WeilR} and \eqref{eqn:TSCSK}, where $K$ is defined by \eqref{eqn:KcplM} together with \eqref{eqn:KcplM2}.
    \end{minipage}
    }

\section{Proof of the bivariate Landsberg-Schaar relation in \texorpdfstring{\eqref{eqn:LSgen1}}{}}
\label{app:proof}
%Here we copy (\ref{eqn:LSgen1}). 
The identity to be proved is:
\be
\label{1}
\boxed{-\frac{i}{q}
\sum_{m,n=0}^{q-1}
e^{-\frac{2\pi i}{q}\left(2 m n -a m^2-b n^2\right)}
=
\left(\frac{1+i^{q b}}{2\sqrt{a b -1}}\right)
\sum_{n=0}^{2a b -3}
\exp\left(
-\frac{\pi i q a n^2}{2(a b -1)}
\right)
}\,,
\ee
for $q\in 2\Z_{+},\, a, b\in\Z,\, ab>1.$

In the following subsections, we will prove this by discussing two separate situations, in which $q=4r$ and $q=4r+2$, respectively, where $r$ is an odd integer.

\subsection{Analysis for \texorpdfstring{$q=4r$}{}}

To prove this identity \eqref{1} we start with the Poisson resummation in two variables:
$$
\sum_{m,n\in\Z} f(m,n) = \sum_{x,y\in\Z}\hat{f}(x,y),
$$
where the Fourier transform is defined with normalization as:
$$
\hat{f}(x,y)\equiv\int_{-\infty}^\infty\int_{-\infty}^\infty e^{-2\pi i (m x + n y)}f(u,v)du dv$$
We choose the function to be
$$
f(u,v) \equiv e^{2\pi i \tau\left(a u^2 + b v^2 - 2 u v\right)}\,,\qquad
{\text{Im}}\tau>0,\qquad a,b>1,
$$
then
\begin{equation}
\begin{split}
	\hat{f}(x,y) &=
	\int_{-\infty}^\infty\int_{-\infty}^\infty e^{2\pi i \left(a\tau x^2 + b\tau y^2 -2\tau x y - u x - v y\right)}dx dy\\
&=-\frac{1}{2i\sqrt{a b-1}}
\exp\left[
-\frac{\pi i\left(b u^2 + a v^2 + 2 u v\right)}{2(a b-1)\tau}
\right].
\end{split}
\end{equation}
So, we have
\begin{equation}
\label{eq:2}
-\frac{1}{2i\sqrt{a b-1}}
\sum_{m,n\in\Z}\exp\left[
-\frac{\pi i\left(b m^2 + a n^2 + 2 m n\right)}{2(a b-1)\tau}
\right]
=
\sum_{m,n\in\Z}e^{2\pi i\tau \left(a m^2 + b n^2 - 2 m n\right)}
\,.
\end{equation}
%by changing variables $u$ and $v$ into $m$ and $n$, respectively.
Now, we set 
\begin{equation}
    \label{eq:limit}
\tau = \frac{1}{q}+i\epsilon,\quad q=4r,
\end{equation}
and take the limit $\epsilon\rightarrow 0$, so that\footnote{This is similar to the standard complex-analytical technique in proving the basic univariate Landsberg-Schaar relation \eqref{eqn:LSrel}, shown for example in \cite{Analysis}.}
\begin{equation}
\label{eq:limit2}
-\frac{1}{\tau} = 
-q + i q^2\epsilon + O\left(\epsilon^2\right).
\end{equation}
The expression on the RHS of \eqref{eq:2} can be expanded by setting
$$
m \equiv m_1 q + m_0,\qquad
n \equiv n_1 q + n_0,\qquad
m_0,n_0 = 0,\dots,q-1,\qquad
m_1, n_1\in\Z,
$$
so that with \eqref{eq:limit},
$$
\exp\left[2\pi i\tau \left(a m^2+bn^2-2mn\right)\right]
\approx
\exp\left[
\frac{2\pi i}{q}\left(a m_0^2 + b n_0^2 - 2 m_0 n_0\right)
\right]
e^{-2\pi q^2\epsilon\left(a m_1^2 + b n_1^2 - 2 m_1 n_1\right)},
$$
where in the exponent of the second factor on the RHS, we have ignored terms $2am_0m_1q$, $bm_0^2$, $2bn_0n_1q$, $bn_0^2$, $-2m_1n_0q$, $-2m_0n_1q$, and $-2m_0n_0$, which are proportional to either $q$ or 1, because both $m_0$ and $n_0$ are only comparable to $q$ at most.

Then
\begin{equation}
\label{eq:inter1}
\sum_{m,n\in\Z}e^{2\pi i\tau \left(a m^2 + b n^2 - 2 m n\right)}
=
\left(
\sum_{m,n=0}^{q-1}e^{\frac{2\pi i}{q}\left(a m^2 + b n^2 - 2 m n\right)}
\right)
\sum_{m,n\in\Z}e^{-2\pi q^2\epsilon\left(a m^2 + b n^2 - 2 m n\right)}.
\end{equation}
In the limit $\epsilon\rightarrow0$, the leftmost sum on RHS of \eqref{eq:inter1} can be evaluated by converting it into an integral with a change of variables $u\equiv m\sqrt{\epsilon}$ and $v\equiv n\sqrt{\epsilon}$:
$$
\lim_{\epsilon\rightarrow 0}\sum_{m,n\in\Z}e^{-2\pi q^2\epsilon\left(a m^2 + b n^2 - 2 m n\right)}
\approx 
\frac{1}{\epsilon}\int_{-\infty}^{\infty}du\int_{-\infty}^{\infty}dv
e^{-2\pi q^2\left(a m^2 + b n^2 - 2 m n\right)}
=\frac{1}{2\epsilon q^2\sqrt{a b-1}}.
$$
So, we have
\begin{equation}
\label{eq:double2}
\lim_{\epsilon\rightarrow 0}\sum_{m,n\in\Z}e^{2\pi i\tau \left(a m^2 + b n^2 - 2 m n\right)}
\approx
\frac{1}{2\epsilon q^2\sqrt{a b-1}}
\sum_{m,n=0}^{q-1}e^{\frac{2\pi i}{q}\left(a m^2 + b n^2 - 2 m n\right)}.
\end{equation}

Now to approximate the LHS of \eqref{eq:2}, we need to perform a similar manipulation on the double sum
\begin{equation}
\label{eq:double}
\sum_{m,n\in\Z}\exp\left\lbrack
-\frac{\pi i\left(b m^2 + a n^2 + 2 m n\right)}{2(a b-1)\tau}
\right\rbrack\,.
\end{equation}
It would help to know the Smith Normal Form of the matrix
$\begin{pmatrix}
a & -1 \\
-1 & b \\
\end{pmatrix}$,
which is related to the inverse of the quadratic form in the exponent of \eqref{eq:double} by
$$
\begin{pmatrix}
a & -1 \\
-1 & b \\
\end{pmatrix}
=(ab-1)
\begin{pmatrix} 	
b & 1 \\
1 & a \\
\end{pmatrix}^{-1}\,.
$$

Its decomposition into the Smith Normal Form is:
$$
\begin{pmatrix}
a & -1 \\
-1 & b \\
\end{pmatrix} =
\begin{pmatrix}
1 & -a \\
0 & 1 \\
\end{pmatrix}
\begin{pmatrix}
a b -1 & 0 \\
0 & 1 \\
\end{pmatrix}
\begin{pmatrix}
0 & 1 \\
-1 & b \\
\end{pmatrix}.
$$
We want to convert the sum over $(m,n)\in\Z^2$ in \eqref{eq:double} into a sum over a finite number of points in the fundamental cell generated by the columns of $\begin{pmatrix}
a & -1 \\
-1 & b \\
\end{pmatrix}$, times a sum over the lattice points generated by these columns.
So, we write $(m,n)$ as
$$
\begin{pmatrix} m \\ n \\ \end{pmatrix}
=
\begin{pmatrix} m_0 \\ n_0 \\ \end{pmatrix}
+
\begin{pmatrix}
a & -1 \\
-1 & b \\
\end{pmatrix}
\begin{pmatrix} m_1 \\ n_1 \\ \end{pmatrix}
=
\begin{pmatrix} m_0 + a m_1 - n_1 \\ n_0 + b n_1 - m_1 \\ \end{pmatrix}
\,,
$$
where $(m_0, n_0)$ take $a b -1$ possible integer values.

Replacing
$$
\begin{pmatrix} m_1 \\ n_1 \\ \end{pmatrix}
\rightarrow
\begin{pmatrix}
0 & 1 \\
-1 & b \\
\end{pmatrix}
\begin{pmatrix} m_1 \\ n_1 \\ \end{pmatrix},
$$
we can write
$$
\begin{pmatrix} m \\ n \\ \end{pmatrix}
=
\begin{pmatrix} m_0 \\ n_0 \\ \end{pmatrix}
+
\begin{pmatrix}
1 & -a \\
0 & 1 \\
\end{pmatrix}
\begin{pmatrix}
a b -1 & 0 \\
0 & 1 \\
\end{pmatrix}
\begin{pmatrix} m_1 \\ n_1 \\ \end{pmatrix}
\,.
$$
Setting
$$
\begin{pmatrix} m_0 \\ n_0 \\ \end{pmatrix}
=
\begin{pmatrix}
1 & -a \\
0 & 1 \\
\end{pmatrix}
\begin{pmatrix} j \\ 0 \\ \end{pmatrix}
\,,\qquad
j=0,\dots, ab-2\,,
$$
we find
$$
\begin{pmatrix} m \\ n \\ \end{pmatrix}
=
\begin{pmatrix} j + m_1(ab-1)-n_1 a \\ n_1 \\ \end{pmatrix}
\,.
$$
So, we set $n\equiv n_1$ and $m \equiv j + (ab-1)k -n a$, where $j=0,\dots,ab-2$.
Then,
$$
b m^2 + a n^2 + 2 m n 
= b j^2
+(ab-1)\left[
a(b k-n)^2-b(k-j)^2+2(k-j) n+b j^2
\right]
\,,
$$
and \eqref{eq:double} becomes

\bear
\sum_{j=0}^{a b -2}
\left(
e^{-\frac{\pi i b j^2}{2(a b-1)\tau}}
\sum_{n,k\in\Z}
\exp\left\lbrack
-\frac{\pi i\left[ a(b k-n)^2-b(k-j)^2+2(k-j) n+b j^2\right]}{2\tau}
\right\rbrack
\right)
\nn
\eear
In the limit $\tau=1/q+i\epsilon$ \eqref{eq:limit2} with $\epsilon\rightarrow 0$, we can approximate the exponent
\begin{eqnarray}
\label{3}
\lefteqn{
	-\frac{\pi i}{2\tau}\left[ a(b k-n)^2-b(k-j)^2+2(k-j) n+b j^2\right]}
\nn\\
&\approx&
-\frac{\pi i}{2}q\left[ a(b k-n)^2-b(k-j)^2+2(k-j) n+b j^2\right]
-\frac{\pi}{2} q^2\epsilon\left[ a(b k-n)^2-bk^2+2k n\right].
\nn\\
\end{eqnarray}
Because $4\lvert q$, the first term on the last line in \eqref{3} is an integer multiple of $2\pi i$ and can be dropped, so we are left with
\begin{equation}
\label{eq:4|q}
\begin{aligned}
\lefteqn{
	\sum_{m,n\in\Z}\exp\left\lbrack
	-\frac{\pi i\left(b m^2 + a n^2 + 2 m n\right)}{2(a b-1)\tau}
	\right\rbrack
}\\
&=&
\left(
\sum_{j=0}^{a b -2}
e^{-\frac{\pi iq  b j^2}{2(a b-1)}}
\right)
\lim_{\epsilon\rightarrow 0}
\sum_{n,k\in\Z}
e^{-\frac{\pi}{2} q^2\epsilon\left[ a(b k-n)^2-bk^2+2k n\right]}.
\end{aligned}
\end{equation}
The limit $\epsilon\rightarrow 0$ can be evaluated by converting \eqref{eq:4|q} into an integral
with a change of variables $u\equiv k\sqrt{\epsilon}$ and $v\equiv n\sqrt{\epsilon}$:
\bear
\lefteqn{
	\lim_{\epsilon\rightarrow 0}
	\sum_{n,k\in\Z}
	e^{-\frac{\pi}{2} q^2\epsilon\left[ a(b k-n)^2-bk^2+2k n\right]}
}\nn\\
&=&
\frac{1}{\epsilon}\int_{-\infty}^{\infty}du\int_{-\infty}^{\infty}dv
e^{-\frac{\pi}{2} q^2\left[ a(b u-v)^2-bu^2+2u v\right]}
=\frac{2}{q^2(ab-1)\epsilon}.
\nn
\eear

Finally, we need to show that in \eqref{eq:4|q},
\begin{equation}
\label{eq:sum}
\sum_{j=0}^{a b -2}
e^{-\frac{\pi iq  b j^2}{2(a b-1)}}=\sum_{j=ab-1}^{2a b -3}
e^{-\frac{\pi iq  b j^2}{2(a b-1)}}.
\end{equation}
In order to achieve this, we consider the pairings between exponents of forms
$$
-\frac{qbj^2}{ab-1}\frac{\pi i}{2} \quad \text{and}\quad -\frac{qb(ab-1+j)^2}{ab-1}\frac{\pi i}{2},\quad j=0,\dots, ab-2,
$$
then the difference between $j^2$ and $(ab-1+j)^2$ is $a^2b^2-2ab+1+2j(ab-1)=(ab-1)^2+2j(ab-1)$, so the difference between two exponents is $(ab-1+2j)qb\pi i/2$, which is an integer multiple of $2\pi i$ because $4\lvert q$. So the summands in \eqref{eq:sum} can be paired using the one-to-one correspondence between $j$ and $ab-1+j$, and \eqref{eq:sum} is true.

At this point, we have results \eqref{eq:2}, \eqref{eq:double2}, and the newly proved:
\begin{align}
&\lim_{\epsilon\rightarrow0}\sum_{m,n\in\Z}\exp\left\lbrack
-\frac{\pi i\left(b m^2 + a n^2 + 2 m n\right)}{2(a b-1)\tau}
\right\rbrack\\
&\approx
\frac{2}{q^2(ab-1)\epsilon}
\sum_{j=0}^{a b -2}
e^{-\frac{\pi iq  b j^2}{2(a b-1)}}=\frac{2}{q^2(ab-1)\epsilon}
\sum_{j=0}^{2a b -3}
e^{-\frac{\pi iq  b j^2}{2(a b-1)}}.
\end{align}
Combining these three equations, by inspection, \eqref{1} holds if $4\lvert q$.

\subsection{Analysis for \texorpdfstring{$q=4r+2=2s$}{}}
\subsubsection{The case of odd \texorpdfstring{$b$}{}}
\label{sec:oddb}
In this case, the RHS of \eqref{1} is obviously zero. Hence we need to show that the LHS of \eqref{1} also vanishes, i.e., for fixed $q$ and $m$, 
$$
\sum_{n=0}^{q-1}
e^{-\frac{2\pi i}{q}\left(2 m n -a m^2-b n^2\right)}=0.
$$
We need to show that with $m$ fixed, for every $n$, there is always a single $n'=n+t$ such that $am^2+bn^2-2mn$ and $am^2+b(n+t)^2-2m(n+t)$ differ by an odd multiple of $s$, hence the pairing $\displaystyle{e^{\frac{\pi i}{s}\left(am^2+bn^2-2 m n\right)}+e^{\frac{\pi i}{s}\left(am^2+bn'^2-2 m n'\right)}}$ is 0. 

We calculate this difference $am^2+b(n+t)^2-2m(n+t)-\left(am^2+bn^2-2mn\right)=bt^2+2bnt-2mt$, and set it to be $-\delta s$. Then we only need to show that $\delta$ can be odd for every $n$. We start from the quadratic equation in $t$:
\begin{equation}
\label{eq:quadratic}
bt^2+2(bn-m)t+\delta s=0,
\end{equation}
with solutions $\displaystyle{t_{\pm}=-n+\frac{m\pm\sqrt{(bn-m)^2-b\delta s}}{b}}$. We set the discriminant $\Delta$ to be $x^2$, $x\in\mathbb{Z}$, then it follows that
\be
\label{4}
(bn-m-x)(bn-m+x)=b\delta s.
\ee
We also denote $\displaystyle{\frac{m\pm \sqrt{\Delta}}{b}\equiv y}$, leading to $m=by\pm x$. Then \eqref{4} becomes 
$$
(n-y)(bn-by\pm2x)=\delta s.
$$
In order to establish the pairing between $n$ and $n'$ for all $n$ and $n'$, we have to require $-n+y$, a solution to \eqref{eq:quadratic}, be exactly $s=q/2$ [so that the pairings are all ``diagonal'' in the irregular $q$-sided polygon on the complex plane, whose vertices are $e^{-\frac{2\pi i}{q}\left(2mn-am^2-bn^2\right)}$]. So we have
$$
bn-by\pm2x=-bs\pm2x=-\delta.
$$
Since both $b$ and $s$ are odd, $\delta$ has to be odd, as expected. Hence, for $q=4r+2$ and odd $b$, both sides of \eqref{1} are zero.

\subsubsection{The case of even \texorpdfstring{$b$}{} and odd \texorpdfstring{$a$}{}}
\label{sec:evenb}
In this case, we do not construct pairings as before, but resort to an analytic method. The first term in \eqref{3} expands as
$$
s\pi i\left[ab^2k^2-2abnk+an^2-bk^2+2bjk+2(k-j)n\right].
$$
Since all terms in the square brackets, except for $an^2$, are even, the overall contributing exponent provided by \eqref{3} is 
$$
-\frac{\pi}{2}q^2\epsilon\left[a(bk-n)^2-bk^2+2kn\right]-asn^2\pi i.
$$
So the sum concerning $n$ and $k$ following \eqref{3} becomes 
\begin{equation}
\label{eq:evenb}
\lim_{\epsilon\rightarrow 0}
\sum_{n,k\in\Z}
e^{-\frac{\pi}{2} q^2\epsilon\left[ a(b k-n)^2-bk^2+2k n\right]}e^{-asn^2\pi i},
\end{equation}
where the second factor is just $(-1)^n$. We evaluate the first factor again by converting it into a single integral via a change of variables $u\equiv k\sqrt{\epsilon}$ and $v\equiv n\sqrt{\epsilon}$:
\be
\label{5}
\frac{1}{\epsilon}\int_{-\infty}^{\infty}du\int_{-\infty}^{\infty}dv
e^{-\frac{\pi}{2} q^2\left[ a(b u-v)^2-bu^2+2u v\right]}dudv
=\frac{1}{q}\sqrt{\frac{2}{b(ab-1)\epsilon}} \sum_{n \in\Z}e^{-\frac{\pi q^2\epsilon n^2}{2b}},
\ee
where we only performed the $u$-integral and have converted $v$ back to $n$ after the equal sign. Then we consider the remaining infinite sum in \eqref{eq:evenb}:
$$
\frac{1}{q}\lim_{\epsilon\rightarrow 0}\sqrt{\frac{2}{b(ab-1)\epsilon}}\sum_{n \in\Z}(-1)^ne^{-\frac{\pi q^2\epsilon n^2}{2b}},
$$
which is zero because this is an alternating Riemann sum, and the decay in the exponent is $\sim\epsilon\rightarrow0$, while the denominator of here goes as $\sim 1/\sqrt{\epsilon}$.

Notice that the result here agrees with the symmetry between $a$ and $b$ which is manifest on the LHS of \eqref{1}, as well as the fact that both sides vanishes for an odd $b$ as shown in \secref{sec:oddb}.

\subsection{If both \texorpdfstring{$a$}{} and \texorpdfstring{$b$}{} are even}
Then again the first term in \eqref{3} is an integer multiple of $2\pi i$, and the remaining argument conincide with that in the previous subsection \secref{sec:evenb}.
\qed

% ----------------------------------------------------------------------------------
% ----------------------------------------------------------------------------------

% ==============================================================
% ==============================================================
% ==============================================================
% ==============================================================

\bibliographystyle{my-h-elsevier}

\end{document}